\documentclass[preprint,pre,showpacs,preprintnumbers,amsmath,amssymb]{revtex4}

\usepackage{graphicx}
\usepackage{dcolumn}
\usepackage{bm}
\usepackage{braket}
\usepackage{graphicx}
\usepackage{epstopdf}
\usepackage{mwe}    
\usepackage{subfig}
\usepackage[T1]{fontenc}
\usepackage{caption} 
\usepackage{amssymb}
\usepackage{amsmath}
\usepackage{booktabs}
\usepackage{tabularx}
\usepackage{multirow}

\newcommand{\tensr}[1]{\bm{\mathsf{#1}}} 

\usepackage{xcolor}

\begin{document}

\preprint{PREPRINT}

\title{Galilean Invariant Preconditioned Central Moment Lattice Boltzmann Method without Cubic Velocity Errors for Efficient Steady Flow Simulations}

\author{Farzaneh  Hajabdollahi}
\email{farzaneh.hajabdollahi-ouderji@ucdenver.edu}
\affiliation{Department of Mechanical Engineering, University of Colorado Denver, 1200 Larimer street, Denver, CO  80124, U.S.A.\\}

\author{Kannan N. Premnath}
\email{kannan.premnath@ucdenver.edu}
\affiliation{Department of Mechanical Engineering, University of Colorado Denver, 1200 Larimer street, Denver, CO  80124, U.S.A.\\}


\date{\today}

\begin{abstract}
Lattice Boltzmann (LB) models used for the computation of fluid flows represented by the Navier-Stokes (NS) equations on standard
lattices can lead to non Galilean invariant (GI) viscous stress involving cubic velocity errors. This arises from the dependence
of their third order diagonal moments on the first order moments for standard lattices, and strategies have recently been
introduced to restore GI without such errors using a modified collision operator involving either corrections to the relaxation
times or to the moment equilibria. Convergence acceleration in the simulation of steady flows can be achieved by solving the
preconditioned NS equations, which contain a preconditioning parameter that can be used to tune the effective sound speed,
and thereby alleviating the numerical stiffness. In the present study, we present a GI formulation of the preconditioned cascaded
central moment LB method used to solve the preconditioned NS equations, which is free of cubic velocity errors on a standard
lattice, for steady flows. A Chapman-Enskog analysis reveals the structure of the spurious non-GI defect terms
and it is demonstrated that the anisotropy of the resulting viscous stress is dependent on the preconditioning parameter,
in addition to the fluid velocity. It is shown that partial correction to eliminate the cubic velocity defects is achieved by scaling
the cubic velocity terms in the off-diagonal third-order moment equilibria with the square of the preconditioning parameter. Furthermore,
we develop additional corrections based on the extended moment equilibria involving gradient terms with coefficients dependent locally on
the fluid velocity and the preconditioning parameter. Such parameter dependent corrections eliminate the remaining truncation
errors arising from the degeneracy of the diagonal third-order moments and fully restores GI without cubic defects for the
preconditioned LB scheme on a standard lattice. Several conclusions are drawn from the analysis of the structure of the non-GI
errors and the associated corrections, with particular emphasis on their dependence on the preconditioning parameter. The new
GI preconditioned central moment LB method is validated for a number of complex flow benchmark problems and its effectiveness
to achieve convergence acceleration and improvement in accuracy is demonstrated.
\end{abstract}

\maketitle
\section{\label{1}Introduction}
The lattice Boltzmann (LB) method has now been established as a powerful kinetic scheme based computational fluid dynamics
approach (\cite{Succi2001},~\cite{Guo2013},~\cite{Kruger2016}). It is a mesoscopic method based on local conservation and discrete symmetry principles, and may be derived as a special discretization of the Boltzmann equation. Algorithmically, it involves the streaming of the particle distribution functions as a perfect shift advection step along the lattice directions and followed by a local collision step as a relaxation process towards an equilibria, and accompanied by special strategies for the implementation of impressed forces. The hydrodynamic fields characterizing the fluid motion are then obtained via the various kinetic moments of the evolving distribution functions and its consistency to the Navier-Stokes (NS) equations may be established by a Chapman-Enskog expansion or Taylor series expansions under appropriate scaling between the discrete space step and time step. As such,
the LB method has been applied for the computation of a wide range of complex flows including turbulence, multiphase and multicomponent flows, particulate flows and microflows~(\cite{Chen1998},~\cite{Aidun2010}). Its various appealing features, including its inherent parallelism, natural framework to incorporate kinetic models for complex flows and the ease of boundary conditions has made it an unique and efficient approach for computational fluid
dynamics (CFD). During the last decade, many efforts were made to further improve its numerical stability, accuracy and efficiency. In particular, sophisticated collision models based on multiple relaxation times and involving raw moments, central moments or cumulants, and entropic formulations have significantly expanded the capabilities of the LB method. The significant achievements of these developments and their applications to a variety of complex flow problems have been discussed, for example, in~\cite{Aidun2010,luo2010lattice,geller2013turbulent,karlin2014gibbs,frapolli2015entropic,succi2015lattice,mazloomi2015entropic,Ning2015,krafczyk2015dns,yang2016intercomparison,dorschner2016entropic,hajabdollahi2018central}.

There exist various additional aspects in the LB approach that require further attention and present scope for improvements.
In particular, the finiteness of the lattice can introduce certain truncation errors that manifest as non-Galilean invariant
viscous stress, i.e. fluid velocity dependent viscosity. This lack of Galilean invariance (GI) arises due to the fact that
the diagonal terms in the third-order moments are not independently supported by the standard tensor-product lattices (i.e.
D2Q9 and D3Q27). More specifically, for example,
\begin{equation*}
\widehat{\kappa}^{'}_{xxx} = \sum_\alpha e_{\alpha x}^3 f_\alpha = \sum_\alpha e_{\alpha x} f_\alpha = \widehat{\kappa}^{'}_{x}.
\end{equation*}
Here, and in the following, the primed quantities denote raw moments. In other words, there is a degeneracy of the third-order
diagonal (longitudinal) moments that results in a deviation between the emergent macroscopic equations derived by the
Chapman-Enskog expansion and the Navier-Stokes (NS) equations. Such cubic-velocity truncation errors are grid independent
and persist in finer grids especially under high shear and flow velocity. Moreover, such emergent anisotropic viscous stress
may have a negative impact on numerical stability as a result of a negative dependence of the emergent viscosity on the fluid
velocity. In order to overcome this shortcoming, various attempts have been made.

One possibility is to consider a lattice with a larger particle velocity set, such as the D2Q17 lattice in two-dimensions~\cite{Qian1998},
which was pursued after~\cite{Qian1993} pointed out nonlinear, cubic-velocity deviations of the emergent
equations of the LB models with standard lattice sets from the NS equations. This involved the use of higher order velocity terms
in the equilibrium distribution. However,~\cite{Hazi2006} showed that the specific equilibria adopted in~\cite{Qian1998} does not fully eliminate the cubic-velocity errors. Moreover, the use of non-standard lattice stencils with
larger number of particle velocities increases the computational cost and propensity of the numerical instability at grid scales.
On the other hand, it was shown more recently by various others (\cite{Wagner1999},~\cite{Hazi2006},~\cite{Keating2007}) that partial corrections to the GI errors on the standard lattice (i.e. D2Q9 lattice) may
be achieved by adopting special forms of the off-diagonal, third-order moments in the equilibria. That is,
\begin{equation*}
\widehat{\kappa}^{eq'}_{xxy}=c_s^2\rho u_y+\underline{\rho u_x^2u_y}, \quad \widehat{\kappa}^{eq'}_{xyy}=c_s^2\rho u_x+\underline{\rho u_xu_y^2}.
\end{equation*}
Here, $c_s$ is the speed of sound and the particular choices of the cubic-velocity terms that are underlined are crucial to partially restore
GI for the above identified moments. Here, we also point out that the above forms of the off-diagonal, third-order raw moment equilibria that
allow such partial GI corrections naturally arise in the central moment LB formulations, when the equilibrium central moment components are
set to zero and and then rewritten in terms of their corresponding raw moments. However, since $\widehat{\kappa}^{eq'}_{xxx}=\widehat{\kappa}^{eq'}_{x}$
and $\widehat{\kappa}^{eq'}_{yyy}=\widehat{\kappa}^{eq'}_{y}$ due to the degeneracy of the third-order longitudinal moments, which is inherent to
the standard tensor-product lattices, additional corrections are required to restore GI completely free of cubic-velocity errors. In this
regard, in order to compensate the terms which violate GI on standard lattices, LB schemes with single relaxation time models were augmented with finite
difference expressions~\cite{prasianakis2007lattice,prasianakis2008lattice,prasianakis2009lattice}. On the other hand, more recently,~\cite{Dellar2014} introduced small intentional anisotropies into a matrix collision operator that corrects the anisotropy in the resulting viscous stress tensor thereby addressing the above mentioned issue. In addition, independently,~\cite{Geier2015} introduced additional corrections involving velocity gradients to the equilibria that achieved equivalent results. These two studies provided strategies to represent the Navier-Stokes equations in LB models on standard lattices completely free of cubic-velocity errors. In addition,~\cite{Dubois17} presented finite difference based corrections to the method proposed in~\cite{Asinari12} to reduce the resulting spurious velocity dependent viscosity effects on standard lattices.

While the LB schemes have found applications to a wide range of fluid flow problems, there has also been considerable interest to an important class of problems related to low Reynolds number steady state flows. They include analysis and design optimization of a variety of Stokes flows through capillaries, porous media flows, heat transfer problems under stationary conditions, and since the LB methods are explicit marching in nature, efficient solution techniques need to devised to accelerate their convergence (see e.g.~\cite{verberg1999simulation, bernaschi2001accelerated, tolke2002multigrid, Guo2004, mavriplis2006multigrid, pingen2008parallel, Izquierdo2008, Premnath2009a, hubner2010efficient, patil2014multigrid, hu2015simulation, atanasov2016steady, Hajabdollahi2017, de2017preconditioned, Hajabdollahi2017b}). A recent review of the literature in the LB approach for such problems can be found in~\cite{Hajabdollahi2017, de2017preconditioned}. Generally, multigrid and preconditioning techniques can be devised to improve the steady state convergence of the LB scheme. A comparison of a multigrid LB formulation with the conventional solvers showed significant improvement in efficiency~\cite{patil2014multigrid}. At low Mach numbers, the convergence can be further accelerated by means of preconditioning for both the traditional single grid LB methods~\cite{Guo2004, Izquierdo2008, Premnath2009a, Hajabdollahi2017, de2017preconditioned} and multigrid LB scheme~\cite{Hajabdollahi2017b}. The present work addresses a further refinement to the LB techniques for steady state flows, viz., improving the accuracy of the acceleration strategy based on the preconditioned LB formulation without GI cubic velocity and parameter dependent errors.

Thus, it is clear that another aspect of the LB method, similar to certain schemes based on the classical CFD, is its slow convergence to steady state at low Mach numbers. In such conditions, there is a relatively large disparity between the sound speed and the convection speed of the fluid motion resulting in higher eigenvalue stiffness and larger number of iterations for convergence. This stiffness can be alleviated and convergence can be significantly improved by preconditioning. Reference~\cite{Guo2004} presented a preconditioned LB method based on a single relaxation time model by modifying
the equilibrium distribution function by using a preconditioning parameter. Then,~\cite{Izquierdo2008} and~\cite{Premnath2009a} presented preconditioned LB formulations based on multiple relaxation times. More recently,~\cite{Hajabdollahi2017} presented a preconditioned scheme for the central moment based
cascaded LB method~\cite{Geier2006} in the presence of forcing terms~\cite{Premnath2009b} and demonstrated significant convergence acceleration.

In general, such preconditioned LB schemes are intended to solve the preconditioned NS equations, which can be written as~(\cite{Choi1993},
~\cite{Turkel1999})
\begin{subequations}
\begin{equation}
\frac{\partial \rho}{\partial t} + \bm{\nabla}\cdot \left(\rho\bm{u}\right) = 0
\label{eq:1a},
\end{equation}
\begin{equation}
\frac{\partial \left(\rho\bm{u}\right)}{\partial t} +\bm{\nabla}\cdot \left(\frac{\rho\bm{u}\bm{u}}{\gamma}\right)=-\frac{1}{\gamma} \bm{\nabla} p^* +\frac{1}{\gamma}\bm{\nabla}\cdot\left(\rho\nu\tensr{S}\right)+\frac{\bm{F}}{\gamma},
\label{eq:1b}
\end{equation}
\end{subequations}
where $p^*$, $\tensr{S}$ and $\bm{F}$ are the pressure, strain rate tensor and the impressed force, respectively. Here, $\gamma$ is the preconditioning parameter, which can be used to tune the pseudo-sound speed, thereby alleviating the eigenvalue stiffness and improving convergence acceleration
(e.g.~\cite{Hajabdollahi2017}). However, the existing LB models for the preconditioned NS equations are not Galilean invariant and
are expected to involve both velocity- and parameter-dependent anisotropic form of the viscous stress tensor. Development of the Galilean invariant preconditioned central moment based LB method without cubic-velocity defects and parameter free truncation errors for steady flow simulations is the main focus of this study. It may be noted that the preconditioned NS equations may be considered as a specific example of what may be called as the generalized NS equations containing a free parameter. In the present case, such a parameter is imposed by numerics due to preconditioning. On the other hand, such generalized NS equations arise in other contexts such as in the simulation of the fluid saturated variable porous media flows represented by the Brinkman-Forchheimer-Darcy equation. In such cases, the free parameter appearing in the generalized NS equations is imposed by physics, viz., the porosity. Thus, our present investigation on the development of the Galilean invariant LB models for the preconditioned NS equations on standard lattices without cubic-velocity and parameter dependent errors also has wider implications in other contexts.

In order to first identify such truncation errors, we  perform a Chapman-Enskog analysis of the preconditioned central moment LB formulation and isolate various cubic-velocity and parameter dependent errors at various moment orders. It will be seen that the anisotropy of the stress tensor depends not just on the cubic-velocity terms (like in the previous studies), but also on the preconditioning parameter $\gamma$. Furthermore, we will also demonstrate that even to achieve partial corrections for the GI defects on the standard lattice, the cubic velocity terms appearing in the off-diagonal components of the third-order moment equilibria need to be appropriately scaled by $\gamma$ (e.g. $\widehat{\kappa}^{eq'}_{xxy}=c_s^2\rho u_y+\rho u_x^2u_y/\gamma^2$). In general, the various truncation error terms that arise due to the degeneracy of the third-order diagonal elements will be seen to have complex dependence on both the velocities and the preconditioning parameter. Once such GI defect terms are identified, new corrections are derived for the preconditioned central moment LB formulation based on the extended moment equilibria. This results in a GI central moment LB method for the preconditioned NS equations without cubic-velocity and parameter based defects on standard lattices. The present scheme is targeted towards efficient and accurate low Reynolds number steady state laminar flows by a preconditioned LB formulation without the discrete cubic velocity and parameter dependent effects via corrections to the moment equilibria. On the other hand, for high Reynolds number turbulent flow simulations, higher-order lattice based LB methods such
as that presented in~\cite{chikatamarla2010lattice} appears as one of the attractive approaches.

This paper is organized as follows. In the next section (Sec.~2), our previous central moment based preconditioned LBM with forcing terms on the D2Q9 lattice is summarized first. Section 3 performs a more refined analysis based on the Chapman-Enskog expansion and identifies various cubic-velocity and parameter dependent GI defect errors. Then, Sec.~4 derives new corrections based on the extended moment equilibria and Sec.~5 presents a GI preconditioned central moment LB method free of cubic-velocity and parameter dependent errors. Numerical results are presented in Sec.~6, which compares our numerical results for a variety of benchmark problems, including the lid-driven cavity flow, flow over a square cylinder, backward facing step flow, the Hartmann flow and the four-roll mills flow problem for the purpose of validation. In addition, convergence acceleration due to preconditioning and improvement in accuracy due to the GI corrected LB scheme are also illustrated. Finally, the main findings of our study are summarized in Sec.~7.

\section{\label{2}Preconditioned Cascaded Central Moment Lattice Boltzmann Method: Non-Galilean Invariant Formulation}
In our previous work, we presented a modified cascaded central moment lattice Boltzmann method (LBM) with forcing terms for the computation of preconditioned NS equations~\cite{Hajabdollahi2017}. However, this preconditioned LBM formulation is not Galilean invariant on standard lattices. This is because it results in grid-independent cubic-velocity errors that are sensitive to the preconditioning parameter. In fact, the derivation of the precise expression for the non-GI truncation errors will be derived in the next section. It may be noted that all other prior preconditioned LB schemes are also not Galilean invariant. However, the choice of central moments here partially corrects parts of the cubic velocity defects in the off-diagonal third order moments naturally (Sec.~\ref{3}) and simplifies derivation of the correction terms to completely restore GI free of cubic velocity errors on  standard lattice (Sec.~\ref{4}). Here, we summarize our previous preconditioned central moment LB model setting the stage for further development in the following.\\
The preconditioned cascaded central moment LBM with forcing terms may be written as~\cite{Hajabdollahi2017}
 \begin{subequations}
 \begin{align}
\widetilde{\bar{f}_{\alpha}}(\bm{x},t)=\bar{f}_{\alpha}(\bm{x},t)+ (\tensr{K}\cdot \hat {\mathbf {g}})_{\alpha}+S_{\alpha}(\bm{x},t),\label{eq:2a}\\
\bar{f}_{\alpha}(\bm{x}+\bm{e}_{\alpha},t+1)=\widetilde{{\bar{f}}_{\alpha}}(\bm{x},t),\label{eq:2b}
\end{align}
\end{subequations}
where a variable transformation $\bar{f}_{\alpha}=f_{\alpha}-\frac{1}{2}S_{\alpha}$ is introduced to maintain second order accuracy in the presence of forcing  terms. In the above, $\tensr{K}$ is the orthogonal transformation matrix and $\hat{\mathbf {g}}$ is the collision operator. In order to list the expressions for the collision kernel for the standard two-dimensional, nine particle velocity (\mbox{D2Q9}) lattice, we first define various sets of raw moments as follows on which it is based:
\begin{eqnarray}
\left( {\begin{array}{*{20}{l}}
{{{\hat \kappa }_{{x^m}{y^n}}}}^{'}\\
{\hat {\kappa} _{{x^m}{y^n}}^{eq'}}\\
{{{\hat \sigma }_{{x^m}{y^n}}}}^{'}\\
{{\hat{ \bar \kappa}^{'}_{{x^m}{y^n}}}}
\end{array}} \right) = \sum\limits_\alpha  {\left( {\begin{array}{*{20}{l}}
{{f_\alpha }}\\
{f_\alpha ^{eq}}\\
{{S_\alpha }}\\
{{{\bar f}_\alpha }}
\end{array}} \right)} {{e^m_{\alpha x}}}{{e^n_{\alpha y}}}.
\label{eq:3}
\end{eqnarray}
The preconditioned collision kernel set for the orthogonal moment basis using the preconditioning parameter $\gamma$ can be written as~\cite{Hajabdollahi2017}
\begin{eqnarray}
&\widehat{g}_0=\widehat{g}_1=\widehat{g}_2=0, \label{eq:4a}\nonumber\\
&\widehat{g}_3=\frac{\omega_3}{12}\left\{ \frac{2}{3}\rho+\frac{\rho(u_x^2+u_y^2)}{\gamma}
-(\widehat{\overline{\kappa}}_{xx}^{'}+\widehat{\overline{\kappa}}_{yy}^{'})
-\frac{1}{2}(\widehat{\sigma}_{xx}^{'}+\widehat{\sigma}_{yy}^{'})
\right\}, \label{eq:4b}\nonumber\\
&\widehat{g}_4=\frac{\omega_4}{4}\left\{\frac{\rho(u_x^2-u_y^2)}{\gamma}
-(\widehat{\overline{\kappa}}_{xx}^{'}-\widehat{\overline{\kappa}}_{yy}^{'})
-\frac{1}{2}(\widehat{\sigma}_{xx}^{'}-\widehat{\sigma}_{yy}^{'})
\right\}, \label{eq:4c}\nonumber\\
&\widehat{g}_5=\frac{\omega_5}{4}\left\{\frac{\rho u_x u_y}{\gamma}
-\widehat{\overline{\kappa}}_{xy}^{'}
-\frac{1}{2}\widehat{\sigma}_{xy}^{'}
\right\}, \label{eq:4d}\\
&\widehat{g}_6=\frac{\omega_6}{4}\left\{2\rho u_x^2 u_y+\widehat{\overline{\kappa}}_{xxy}^{'}
              -2u_x\widehat{\overline{\kappa}}_{xy}^{'}-u_y\widehat{\overline{\kappa}}_{xx}^{'}-\frac{1}{2}\widehat{\sigma}_{xxy}
              \right\}-\frac{1}{2}u_y(3\widehat{g}_3+\widehat{g}_4)-2u_x\widehat{g}_5, \label{eq:4c}\nonumber\\
&\widehat{g}_7=\frac{\omega_7}{4}\left\{2\rho u_x u_y^2+\widehat{\overline{\kappa}}_{xyy}^{'}
              -2u_y\widehat{\overline{\kappa}}_{xy}^{'}-u_x\widehat{\overline{\kappa}}_{yy}^{'}-\frac{1}{2}\widehat{\sigma}_{xyy}
              \right\}-\frac{1}{2}u_x(3\widehat{g}_3-\widehat{g}_4)-2u_y\widehat{g}_5, \label{eq:4d}\nonumber \allowdisplaybreaks\\
&\widehat{g}_8=\frac{\omega_8}{4}\left\{\frac{1}{9}\rho+3\rho u_x^2 u_y^2-\left[\widehat{\overline{\kappa}}_{xxyy}^{'}
                                 -2u_x\widehat{\overline{\kappa}}_{xyy}^{'}-2u_y\widehat{\overline{\kappa}}_{xxy}^{'}
                                 +u_x^2\widehat{\overline{\kappa}}_{yy}^{'}+u_y^2\widehat{\overline{\kappa}}_{xx}^{'}\right.\right.
                                 \nonumber \\
                                 &\left.\left.+4u_xu_y\widehat{\overline{\kappa}}_{xy}^{'}
                                 \right]-\frac{1}{2}\widehat{\sigma}_{xxyy}^{'}
                                  \right\}-2\widehat{g}_3-\frac{1}{2}u_y^2(3\widehat{g}_3+\widehat{g}_4)
                                  -\frac{1}{2}u_x^2(3\widehat{g}_3-\widehat{g}_4)\nonumber\\
                                  &-4u_xu_y\widehat{g}_5-2u_y\widehat{g}_6
                                  -2u_x\widehat{g}_7.\label{eq:4e}\nonumber
\end{eqnarray}
For further details, and including the choice of the collision matrix $\tensr{K}$ and source raw moments ${{{\hat \sigma }_{{x^m}{y^n}}}}^{'}$, see~\cite{Hajabdollahi2017}. This scheme results in a tunable pseudo-sound speed $c_s^*=\gamma c_s$, where $c_s=\frac{1}{\sqrt{3}}\delta_x/\delta_t$, and the emergent viscosity $\nu$ is given by $\nu=\frac{\gamma}{3}(\frac{1}{\omega_\beta}-\frac{1}{2})$, $\beta=4,5$. While this scheme is intended to simulate the preconditioned NS equations given in Eq.~(1), as will be shown via a consistency analysis based on the Chapman-Enskog expansion in the next section that it leads to velocity-and preconditioning parameter-dependent non-GI truncation errors. In particular, it will be seen that the components of the non-equilibrium parts of the second order moments, which contribute to the viscous stress tensor, depends on cubic velocity truncation errors and modulated by the preconditioning parameter $\gamma$.

\section{\label{3} Derivation of Non-Galilean Invariant Spurious Terms in the Preconditioned Cascaded Central Moment LB Method: Chapman-Enskog Analysis}
In order to facilitate the Chapman-Enskog analysis, the central moment LB formulation can be equivalently rewritten in terms of a collision process involving relaxation to a generalized equilibria in the lattice or rest frame of reference~\cite{Hajabdollahi2017}. This strategy is considered in this work to further investigate the structure of the cubic velocity non-GI truncation errors for our preconditioned LB method. In this regard, it is convenient to define the non-orthogonal transformation matrix $\tensr{T}$ which is the basis to obtain the orthogonal collision matrix $\tensr{K}$ used in the previous section and on which the subsequent analysis follows:
\begin{align}
\left.\tensr{T}=[{\left| {{{\mathbf{\buildrel{\lower3pt\hbox{$\scriptscriptstyle$}}
\over e} }_\alpha }} \right|^0},\ket{e_{\alpha {x}}},\ket{e_{\alpha {y}}},\ket{e_{\alpha {x}}^2+e_{\alpha {y}}^2},\ket{e_{\alpha {x}}^2-e_{\alpha {y}}^2},\ket{e_{\alpha {x}}e_{\alpha {y}}},\right.\nonumber \\\left.\ket{e_{\alpha {x}}^2e_{\alpha {y}}},\ket{e_{\alpha {x}}e_{\alpha {y}}^2},\ket{e_{\alpha {x}}^2e_{\alpha {y}}^2}]\right.,
\label{eq:5}
\end{align}
where the usual bra-ket notation is used to represent the raw and column vectors in the q-dimensional space ($ {q}=9$) for the \mbox {D2Q9} lattice.
Then, the relation between the various sets of the raw moments and their corresponding states in the velocity space can be defined via this nominal,  non-orthogonal transformation matrix $\tensr{T}$ as
\begin{equation}
\mathbf{\widehat{\bar{m}}}=\tensr{T}\mathbf{\overline{f}}, \quad
\mathbf{\widehat{m}}=\tensr{T}\mathbf{f},\quad
\mathbf{\widehat{m}}^{eq}=\tensr{T}\mathbf{f}^{eq}, \quad
\mathbf{\widehat{S}}=\tensr{T}\mathbf{S},
\label{eq:6}
\end{equation}
where
\begin{eqnarray*}
&\mathbf{{\bar{f}}}=\left({\bar{{f}}_{0}},{\bar{f}}_{1},{\bar{{f}}_{2}},\ldots,{\bar{{f}}_{8}}\right)^{\dag}, \quad
\mathbf{{f}}=\left({f}_{0},{f}_{1},{f}_{2},\ldots,{f}_{8}\right)^{\dag}, \\ &\mathbf{{f}}^{eq}=\left({f}_{0}^{eq},{f}_{1}^{eq},{f}_{2}^{eq},\ldots,{f}_{8}^{eq}\right)^{\dag}, \quad \mathbf{{S}}=\left({S}_{0},{S}_{1},{S}_{2},\ldots,{S}_{8}\right)^{\dag}
\end{eqnarray*}
 are the various quantities in the velocity space, and
\begin{subequations}
\begin{eqnarray}
&\mathbf{\widehat{\bar{m}}}=\left(\widehat{\bar{{m}}}_{0},\widehat{\bar{{m}}}_{1},\widehat{\bar{{m}}}_{2},\ldots,\widehat{\bar{{m}}}_{8}\right)^{\dag}
=\left(\widehat{\overline{\kappa}}_{0}^{'},\widehat{\overline{\kappa}}_{x}^{'},\widehat{\overline{\kappa}}_{y}^{'},
\widehat{\overline{\kappa}}_{xx}^{'}+\widehat{\overline{\kappa}}_{yy}^{'},\widehat{\overline{\kappa}}_{xx}^{'}-\widehat{\overline{\kappa}}_{yy}^{'},
\widehat{\overline{\kappa}}_{xy}^{'},\widehat{\overline{\kappa}}_{xxy}^{'},\widehat{\overline{\kappa}}_{xyy}^{'},
\widehat{\overline{\kappa}}_{xxyy}^{'}\right)^{\dag},\\
&\mathbf{\widehat{m}}=\left(\widehat{m}_{0},\widehat{m}_{1},\widehat{m}_{2},\ldots,\widehat{m}_{8}\right)^{\dag}
=\left(\widehat{\kappa}_{0}^{'},\widehat{\kappa}_{x}^{'},\widehat{\kappa}_{y}^{'},
\widehat{\kappa}_{xx}^{'}+\widehat{\kappa}_{yy}^{'},\widehat{\kappa}_{xx}^{'}-\widehat{\kappa}_{yy}^{'},
\widehat{\kappa}_{xy}^{'},\widehat{\kappa}_{xxy}^{'},\widehat{\kappa}_{xyy}^{'},
\widehat{\kappa}_{xxyy}^{'}\right)^{\dag},\\
&\mathbf{\widehat{m}}^{eq}=\left(\widehat{m}_{0}^{eq},\widehat{m}_{1}^{eq},\widehat{m}_{2}^{eq},\ldots,\widehat{m}_{8}^{eq}\right)^{\dag}
=\left(\widehat{\kappa}_{0}^{eq'},\widehat{\kappa}_{x}^{eq'},\widehat{\kappa}_{y}^{eq'},
\widehat{\kappa}_{xx}^{eq'}+\widehat{\kappa}_{yy}^{eq'},\widehat{\kappa}_{xx}^{eq'}-\widehat{\kappa}_{yy}^{eq'},
\widehat{\kappa}_{xy}^{eq'},\widehat{\kappa}_{xxy}^{eq'},\widehat{\kappa}_{xyy}^{eq'},\right.\nonumber\\&
    \left.
\widehat{\kappa}_{xxyy}^{eq'}\right)^{\dag},\\
&\mathbf{\widehat{S}}=\left(\widehat{S}_{0},\widehat{S}_{1},\widehat{S}_{2},\ldots,\widehat{S}_{8}\right)^{\dag}
=\left(\widehat{\sigma}_{0}^{'},\widehat{\sigma}_{x}^{'},\widehat{\sigma}_{y}^{'},
\widehat{\sigma}_{xx}^{'}+\widehat{\sigma}_{yy}^{'},\widehat{\sigma}_{xx}^{'}-\widehat{\sigma}_{yy}^{'},
\widehat{\sigma}_{xy}^{'},\widehat{\sigma}_{xxy}^{'},\widehat{\sigma}_{xyy}^{'}
\widehat{\sigma}_{xxyy}^{'}\right)^{\dag}
\end{eqnarray}
\end{subequations}
are the corresponding states in the moment space.

To facilitate the Chapman-Enskog analysis, we can rewrite the preconditioned LB model presented in Eq.~(\ref{eq:2a}) and Eq.~(\ref{eq:2b}) in terms of the raw moment space given in  Eq.~(\ref{eq:6}) as~(\cite{Premnath2009b},~\cite{Hajabdollahi2017})
\begin{equation}
\begin{aligned}
\mathbf f\left( {\bm{x} + {{ \bm e}_\alpha }{\delta _t},t + {\delta _t}} \right) - \mathbf f\left( {\bm x,t} \right) = {\tensr{T}^{ - 1}}\left[ { - \hat {\tensr\Lambda} \left( {\hat {\mathbf m} - {{\hat {\mathbf m}}^{eq}}} \right)} \right] + {\tensr{T}^{ - 1}}\left[ {\left( {\tensr{I} - \frac{1}{2}\hat { \tensr\Lambda} } \right)\hat {\mathbf S}} \right]{\delta _t},
\label{eq:7}
\end{aligned}
\end{equation}
where the diagonal relaxation time matrix $\hat {\tensr \Lambda}$ is defined as
\begin{equation}
\hat {\tensr \Lambda}=\mbox{diag}(0,0,0,\omega_3,\omega_4,\omega_5,\omega_6,\omega_7,\omega_8).
\label{eq:8}
\end{equation}
The preconditioned raw moments of the equilibrium distribution and source terms can be represented as
\begin{eqnarray}
&\widehat{\kappa}^{eq'}_{0}=\rho,\
\widehat{\kappa}^{eq'}_{x}=\rho u_x, \
\widehat{\kappa}^{eq'}_{y}=\rho u_y, \nonumber\\
&\widehat{\kappa}^{eq'}_{xx}=\frac{1}{3}\rho+\frac{\rho u_x^2}{\gamma},\
\widehat{\kappa}^{eq'}_{yy}=\frac{1}{3}\rho+\frac{\rho u_y^2}{\gamma},\
\widehat{\kappa}^{eq'}_{xy}=\frac{\rho u_x u_y}{\gamma}, \nonumber\\
&\widehat{\kappa}^{eq'}_{xxy}=\frac{1}{3}\rho u_y+\boxed{\frac{\rho u_x^2u_y}{\gamma^2}}, \
\widehat{\kappa}^{eq'}_{xyy}=\frac{1}{3}\rho u_x+\boxed{\frac{\rho u_xu_y^2}{\gamma^2}},\nonumber\\&
\widehat{\kappa}^{eq'}_{xxyy}=\frac{1}{9}\rho+\frac{1}{3}\rho (u_x^2+u_y^2)+\rho u_x^2u_y^2. \label{eq:eqmrawmoment}
\end{eqnarray}
 and
\begin{eqnarray}
&\widehat{\sigma}^{'}_{0}=0,\
\widehat{\sigma}^{'}_{x}=\frac{F_x}{\gamma},\
\widehat{\sigma}^{'}_{y}=\frac{F_y}{\gamma}, \nonumber\\
&\widehat{\sigma}^{'}_{xx}=\frac{2F_xu_x}{\gamma^2},\
\widehat{\sigma}^{'}_{yy}=\frac{2F_yu_y}{\gamma^2},\
\widehat{\sigma}^{'}_{xy}=\frac{F_xu_y+F_yu_x}{\gamma^2},\nonumber\\
&\widehat{\sigma}^{'}_{xxy}=F_yu_x^2+2F_xu_xu_y,\
\widehat{\sigma}^{'}_{xyy}=F_xu_y^2+2F_yu_yu_x,\nonumber\\&
\widehat{\sigma}^{'}_{xxyy}=2(F_xu_xu_y^2+F_yu_yu_x^2).
\end{eqnarray}
The following comments are in order here. Up to the second order moments, the above expressions coincide with those presented in our previous work ~\cite{Hajabdollahi2017}). In other words, $u_iu_j$ terms in the moment equilibria are preconditioned by $\gamma$, while the first and second order moment terms, i.e. $F_i$ and $F_iu_j$ are preconditioned by $\gamma$ and $\gamma^2$, respectively. As a first new element towards a LB scheme with an improved GI, we precondition the third-order moment equilibria terms $u_iu^2_j$ terms by $\gamma^2$ (see the terms inside boxes in Eq.~(\ref{eq:eqmrawmoment})). This partially restores GI without cubic velocity defects for the preconditioned LB model for the off-diagonal components of the third-order moments. In fact, as will be shown later in this section, in order to remove the spurious cross-velocity derivative terms appearing in the equivalent macroscopic equations of our preconditioned LB scheme (e.g. $u_xu_y\partial_ xu_y$ and $u_yu_x\partial_ yu_x$), such a scaling of the cubic velocity terms in the third order moment equilibria is essential. Then, applying the standard Chapman-Enskog multiscale expansion to Eq.~(\ref{eq:7}), i.e.
\begin{eqnarray}
\mathbf{\widehat{m}}&=&\sum_{n=0}^{\infty}\epsilon^n \mathbf{\widehat{m}}^{(n)}, \label{eq:9}\\
\partial_t&=&\sum_{n=0}^{\infty}\epsilon^n \partial_{t_n}.\label{eq:10}
\end{eqnarray}
where $\epsilon$ is ${a}$ small bookkeeping perturbation parameter, and also using a Taylor expansion to simplify the streaming operator in Eq.~(\ref{eq:7}), i.e.
\begin{equation}
\mathbf{f}(\bm{x}+\bm{e}_{\alpha}\epsilon,t+\epsilon)=\sum_{n=0}^{n}\frac{\epsilon^n}{n!}(\partial_t+\bm{e}_{\alpha}\cdot\bm{\nabla})\mathbf{f}(\bm{x},t).
\label{eq:11}
\end{equation}
After converting all the resulting terms into the moment space using Eq.~(\ref{eq:6}), we get the following moment equations at consecutive order in $\epsilon$:
\begin{subequations}
\begin{eqnarray}
&O(\epsilon^0):\quad \mathbf{\widehat{m}}^{(0)}=\mathbf{\widehat{m}}^{eq},\label{eq:12a}\\
&O(\epsilon^1):\quad (\partial_{t_0}+\widehat{\tensr E}_i \partial_i)\mathbf{\widehat{m}}^{(0)}=-\widehat{\tensr\Lambda}\mathbf{\widehat{m}}^{(1)}+\mathbf{\widehat{S}},\label{eq:12b}\\
&O(\epsilon^2):\quad \partial_{t_1}\mathbf{\widehat{m}}^{(0)}+(\partial_{t_0}+\widehat{\tensr E}_i \partial_i)\left[ \tensr{I}-\frac{1}{2}\widehat{\tensr\Lambda}\right]\mathbf{\widehat{m}}^{(1)}=-\widehat{\tensr\Lambda}\mathbf{\widehat{m}}^{(2)},\label{eq:12c}
\end{eqnarray}
\end{subequations}
where $\widehat{\tensr E}_i=\tensr{T}(\bm e_{ i}\tensr{I})\tensr{T}^{-1}, i \in \{x,y\}$. The relevant components of the first-order $O(\epsilon)$ equations Eq.~(\ref{eq:12b}), i.e. up to the second order in moment space needed for deriving the preconditioned macroscopic hydrodynamics equations are given as
\begin{subequations}
 \begin{eqnarray}
&\partial_{t_0}\rho+\partial_x (\rho u_x)+\partial_y (\rho u_y) = 0,\label{eq:13a}\\&
\partial_{t_0}\left(\rho u_x\right)+\partial_x \left({\frac{1}{3}\rho}+\frac{\rho u_x^2}{\gamma}\right)+\partial_y \left(\frac{\rho u_xu_y}{\gamma}\right) = \frac{F_x}{\gamma},
\label{eq:13b}\\&
\partial_{t_0}\left(\rho u_y\right)+\partial_x \left(\frac{\rho u_xu_y}{\gamma}\right)+\partial_y \left(\frac{1}{3}\rho+\frac{\rho u_y^2}{\gamma}\right) = \frac{F_y}{\gamma},
\label{eq:13c}\\&
\partial_{t_0}\left(\frac{2}{3}\rho+\frac{\rho(u_x^2+u_y^2)}{\gamma}\right)+\partial_x \left(\frac{4}{3}\rho u_x+\frac{\rho u_xu_y^2}{\gamma^2}\right)+\partial_y \left(\frac{4}{3}\rho u_y+\frac{\rho u_x^2u_y}{\gamma^2}\right)\nonumber\\& =-\omega_3\widehat{m}_3^{(1)}+\frac{2({F_xu_x}+{F_yu_y})}{\gamma^2},
\label{eq:13d}\\&
\partial_{t_0}\left(\frac{\rho(u_x^2-u_y^2)}{\gamma}\right)+\partial_x \left(\frac{2}{3}\rho u_x-\frac{\rho u_xu_y^2}{\gamma^2}\right)+\partial_y \left(-\frac{2}{3}\rho u_y+\frac{\rho u_x^2u_y}{\gamma}\right)\nonumber\\&
=-\omega_4\widehat{m}_4^{(1)}+\frac{2(F_xu_x-F_yu_y)}{\gamma^2},
\label{eq:13e}\\&
\partial_{t_0}\left(\frac{\rho u_x u_y}{\gamma}\right)+\partial_x \left(\frac{1}{3}\rho u_y+\frac{\rho u_x^2u_y}{\gamma^2}\right)+\partial_y \left(\frac{1}{3}\rho u_x+\frac{\rho u_xu_y^2}{\gamma^2}\right)\nonumber\\&
 =-\omega_5\widehat{m}_5^{(1)}+\frac{F_xu_y+F_yu_x}{\gamma^2}.
\label{eq:13f}
\end{eqnarray}
\end{subequations}
Similarly, the leading order moment equations at $O(\epsilon^2)$ can be obtained from Eq.~(\ref{eq:12c}) as
\begin{subequations}
\begin{eqnarray}
&\partial_{t_1}\rho=0,
\label{eq:14a}\\&
\partial_{t_1}\left(\rho u_x\right)+\partial_x \left[\frac{1}{2}\left(1-\frac{1}{2}\omega_3\right)\widehat{m}_3^{(1)}+\frac{1}{2}\left(1-\frac{1}{2}\omega_4\right)\widehat{m}_4^{(1)}\right]
+\partial_y \left[\left(1-\frac{1}{2}\omega_5\right)\widehat{m}_5^{(1)}\right]\nonumber \\&=0,
\label{eq:14b}\\&
\partial_{t_1}\left(\rho u_y\right)+\partial_x \left[\left(1-\frac{1}{2}\omega_5\right)\widehat{m}_5^{(1)}\right]+\partial_y \left[\frac{1}{2}\left(1-\frac{1}{2}\omega_3\right)\widehat{m}_3^{(1)}-\frac{1}{2}\left(1-\frac{1}{2}\omega_4\right)\widehat{m}_4^{(1)}\right]\nonumber \\
&=0.
\label{eq:14c}
\end{eqnarray}
\end{subequations}
In the above equations, the second-order, non-equilibrium moments $\widehat{m}_3^{(1)}$, $\widehat{m}_4^{(1)}$ and $\widehat{m}_5^{(1)}$ (corresponding to, $\widehat{\kappa}_{xx}^{\prime(1)}+\widehat{\kappa}_{yy}^{\prime(1)}$, $\widehat{\kappa}_{xx}^{\prime(1)}-\widehat{\kappa}_{yy}^{\prime(1)}$ and $\widehat{\kappa}_{xy}^{\prime(1)}$, respectively) are unknowns. Ideally, they should only be related to the strain rate tensor components to recover the correct physics related to the viscous stress. However, as will be should below, on the standard \mbox {D2Q9} lattice there will be non-GI contributions dependent on the preconditioning parameter $\gamma$. In what follows, $\widehat{m}_3^{(1)}$, $\widehat{m}_4^{(1)}$ and $\widehat{m}_5^{(1)}$ will be obtained from Eq.~(\ref{eq:13d}),  Eq.~(\ref{eq:13e}) and  Eq.~(\ref{eq:13f}), respectively.\\
Now, from Eq.~(\ref{eq:13d}), the non-equilibrium moment $\widehat{m}_3^{(1)}$ can be written as
\begin{eqnarray}
&\widehat{m}_3^{(1)} = \frac{1}{\omega_3}\left[ -\partial_{t_0}\left(\frac{2}{3}\rho+\frac{\rho (u_x^2+u_y^2)}{\gamma}\right)
-\partial_x\left(\frac{4}{3}\rho u_x+\frac{\rho u_xu_y^2}{\gamma^2}\right)
-\partial_y\left(\frac{4}{3}\rho u_y+\frac{\rho u_x^2u_y}{\gamma^2}\right)
 \right.\nonumber\\
 &\left.+\frac{2(F_xu_x+F_yu_y)}{\gamma^2}\right].
\label{eq:15}
\end{eqnarray}
In order to simplify Eq.~(\ref{eq:15}) further, one needs to obtain expressions, in particular, for $\partial_{t_0}\left(\frac{\rho u_x^2}{\gamma}\right)$, $\partial_{t_0}\left(\frac{\rho u_y^2}{\gamma}\right)$, $\partial_{x}\left(\frac{\rho u_x u_y^2}{\gamma^2}\right)$ and $\partial_{y}\left(\frac{\rho u_x^2 u_y}{\gamma^2}\right)$. It follows from Eq.~(\ref{eq:13b}) that
\begin{eqnarray}
\partial_{t_0}(\rho u_x)=-\frac{1}{3}\partial_x {\rho}-\partial_x\left(\frac{\rho u_x^2}{\gamma}\right)-\partial_y\left(\frac{\rho u_x u_y}{\gamma}\right)+\frac{F_x}{\gamma}.
\label{eq:16}
\end{eqnarray}
Rearranging $\partial_{t_0}\left(\frac{\rho u_x^2}{\gamma}\right)$ as
\begin{eqnarray*}
\partial_{t_0}\left(\frac{\rho u_x^2}{\gamma}\right)=\frac{2u_x}{\gamma}\partial_{t_0}\left(\rho u_x\right)+\frac{u_x^2}{\gamma}\partial_{t_0}\rho.
\end{eqnarray*}
Using Eq.~(\ref{eq:16}) and Eq.~(\ref{eq:13a}) to replace the time derivative in the first and second terms  respectively, on the right hand side of the above equation, we get.
\begin{eqnarray}
\partial_{t_0}\left(\frac{{\rho}{u_x^2}}{\gamma}\right)&=\frac{2u_x}{\gamma}\left[-\frac{1}{3}\partial_{x}\rho-\partial_{x}\left(\frac{\rho{u_x^2}}{\gamma}\right)-\partial_{y}\left(\frac{\rho{u_xu_y}}{\gamma}\right)\right.\nonumber \\
   &\left.+\frac{F_x}{\gamma}\right]+\frac{u_x^2}{\gamma}\left[\partial_{x}\left({\rho}{u_x}\right)
+\partial_y\left({\rho}u_y\right)
\right].
\label{eq:17}
\end{eqnarray}
Similarly, we may write
\begin{eqnarray}
\partial_{t_0}\left(\frac{{\rho}{u_y^2}}{\gamma}\right)&=\frac{2u_y}{\gamma}\left[-\frac{1}{3}\partial_{y}\rho-\partial_{y}\left(\frac{\rho{u_y^2}}{\gamma}\right)-\partial_{x}\left(\frac{\rho{u_xu_y}}{\gamma}\right)\right.\nonumber \\
   &\left.+\frac{F_y}{\gamma}\right]+\frac{u_y^2}{\gamma}\left[\partial_{x}\left({\rho}{u_x}\right)
+\partial_y\left({\rho}u_y\right)\right].
\label{eq:18}
\end{eqnarray}
Thus, the time derivative can be replaced with the spatial derivative. Also , it readily follows that
\begin{subequations}
\begin{eqnarray}
&-\partial_x\left(\frac{\rho{u_xu_y^2}}{\gamma^2}\right)=-\frac{u_y^2}{\gamma^2}\partial_{x}({\rho}u_x)-\frac{2{\rho}u_xu_y}{\gamma^2}\partial_{x}u_y, \label{eq:19a}\\
&-\partial_y\left(\frac{\rho{u_x^2u_y}}{\gamma^2}\right)=-\frac{u_x^2}{\gamma^2}\partial_{y}({\rho}u_y)-\frac{2{\rho}u_xu_y}{\gamma^2}\partial_{y}u_x.
\label{eq:19b}
\end{eqnarray}
\end{subequations}
Rearranging Eq.~(\ref{eq:17}) and simplifying it further by retaining all cubic velocity terms and neglecting all others higher order terms in velocity (e.g. fifth order and higher) we get
 \begin{eqnarray}
&-\partial_{t_0}\left(\frac{{\rho}u_x^2}{\gamma}\right)=\frac{2u_x}{3\gamma}\partial_x\rho+\frac{2u_x}{\gamma^2}\partial_x(\rho{u_x^2})+\frac{2\rho u_x^2}{\gamma^2}\partial_y u_y+\frac{2\rho u_x u_y}{\gamma^2}\partial_y u_x-\frac{2F_xu_x}{\gamma^2}\nonumber\\&-\frac{u_x^2}{\gamma}\partial_x(\rho{u_x})-\frac{u_x^2}{\gamma}\partial_y(\rho{u_y}).
\label{eq:20}
\end{eqnarray}
Similarly, it follows from Eq.~(\ref{eq:18}) that
\begin{eqnarray}
&-\partial_{t_0}\left(\frac{{\rho}u_y^2}{\gamma}\right)=\frac{2u_y}{3\gamma}\partial_y\rho+\frac{2u_y}{\gamma^2}\partial_y(\rho{u_y^2})+\frac{2\rho u_y^2}{\gamma^2}\partial_x u_x+\frac{2\rho u_x u_y}{\gamma^2}\partial_x u_y+\frac{2F_yu_y}{\gamma^2}\nonumber\\&-\frac{u_y^2}{\gamma}\partial_x(\rho{u_x})-\frac{u_y^2}{\gamma}\partial_y(\rho{u_y}).
\label{eq:21}
\end{eqnarray}
Now, to obtain an expression for $\widehat{m}_3^{(1)}$, we group all the higher order terms given in Eqs.~(\ref{eq:19a}),~(\ref{eq:19b}),~(\ref{eq:20}) and~(\ref{eq:21}). It follows that owing to the choice of the off-diagonal third-order equilibrium moments with the cubic velocity terms scaled by $\gamma^2$ (i.e.~$\widehat{\kappa}^{eq'}_{xxy}=\frac{1}{3}\rho u_y+\frac{\rho u_x^2u_y}{\gamma^2},\ \widehat{\kappa}^{eq'}_{xyy}=\frac{1}{3}\rho u_x+\frac{\rho u_xu_y^2}{\gamma^2})$  at the outset following Eq.~(\ref{eq:8}) earlier, all the cross-derivative spurious terms, i.e.$-2{\rho}u_xu_y\partial_xu_y$ and $-2{\rho}u_xu_y\partial_yu_x$ cancel. Then, simplifying the grouping of all the remaining higher order terms in Eq.~(\ref{eq:19a}), Eq.~(\ref{eq:19b}), Eq.~(\ref{eq:20}) and Eq.~(\ref{eq:21}) and retaining all cubic velocity terms and neglecting terms of negligible higher orders and after considerable rearrangement, we get
\begin{eqnarray}
&-\partial_{t_0}\left(\frac{\rho u_x^2}{\gamma}\right)-\partial_{t_0}\left(\frac{\rho u_y^2}{\gamma}\right)-\partial_{x}\left(\frac{\rho u_x u_y^2}{\gamma^2}\right)-\partial_{y}\left(\frac{\rho u_x^2 u_y}{\gamma^2}\right)\approx\nonumber\\&
\frac{2}{3\gamma}(u_x\partial_x\rho+u_y\partial_y\rho)-\frac{2}{\gamma^2}(F_xu_x+F_yu_y)+\rho\left[\left(\frac{4}{\gamma^2}-
\frac{1}{\gamma}\right)u_x^2+\left(\frac{1}{\gamma^2}-\frac{1}{\gamma}\right)u_y^2\right]\partial_xu_x\nonumber \\&+\rho\left[\left(\frac{4}{\gamma^2}-
\frac{1}{\gamma}\right)u_y^2+\left(\frac{1}{\gamma^2}-\frac{1}{\gamma}\right)u_x^2\right]\partial_yu_y.
\label{eq:22}
\end{eqnarray}
By substituting the above equation (Eq.~(\ref{eq:22})) in Eq.~(\ref{eq:15}) and using $\partial_{t_0} \rho=-\partial_x(\rho u_x)-\partial_y(\rho u_y)$ from Eq.~(\ref{eq:13a}) to further simplify the resulting expressions, we finally get the form of the non-equilibrium moment $\widehat{m}_3^{(1)}$ as
\begin{eqnarray}
&\widehat{m}_3^{(1)} =\underline{- \frac{2\rho}{3\omega_3 }(\partial_xu_x+\partial_yu_y)}+\frac{2}{3\omega_3}\left(\frac{1}{\gamma}-1\right)\left(u_x\partial_x \rho+u_y\partial_y \rho\right)+\nonumber \\&
\frac{\rho}{\omega_3}\left[\left(\frac{4}{\gamma^2}-
\frac{1}{\gamma}\right)u_x^2+\left(\frac{1}{\gamma^2}-\frac{1}{\gamma}\right)u_y^2\right]\partial_xu_x\nonumber \\&+\frac{\rho}{\omega_3}\left[\left(\frac{4}{\gamma^2}-
\frac{1}{\gamma}\right)u_y^2+\left(\frac{1}{\gamma^2}-\frac{1}{\gamma}\right)u_x^2\right]\partial_yu_y.
\label{eq:23}
\end{eqnarray}
Similarly, using Eq.~(\ref{eq:13e}) and following analogous procedure as above for $\widehat{m}_4^{(1)}$ and using Eq.~(\ref{eq:13f}) for $\widehat{m}_5^{(1)}$  after considerable algebraic manipulations and simplifications we get the expressions for the remaining non-equilibrium second-order moments as
\begin{eqnarray}
&\widehat{m}_4^{(1)} =\underline{- \frac{2\rho}{3\omega_4 }(\partial_xu_x-\partial_yu_y)}+\frac{2}{3\omega_4}\left(\frac{1}{\gamma}-1\right)\left(u_x\partial_x \rho-u_y\partial_y \rho\right)+\nonumber \\&
\frac{\rho}{\omega_4}\left[\left(\frac{4}{\gamma^2}-
\frac{1}{\gamma}\right)u_x^2-\left(\frac{1}{\gamma^2}-\frac{1}{\gamma}\right)u_y^2\right]\partial_xu_x\nonumber \\&+\frac{\rho}{\omega_4}\left[-\left(\frac{4}{\gamma^2}-
\frac{1}{\gamma}\right)u_y^2+\left(\frac{1}{\gamma^2}-\frac{1}{\gamma}\right)u_x^2\right]\partial_yu_y,
\label{eq:24}
\end{eqnarray}
 and
 \begin{eqnarray}
&\widehat{m}_5^{(1)} =\underline{- \frac{\rho}{3\omega_5 }(\partial_xu_y+\partial_yu_x)}+\frac{1}{3\omega_5}\left(\frac{1}{\gamma}-1\right)\left(u_x\partial_y \rho+u_y\partial_x \rho\right)+\nonumber \\&
\frac{1}{\omega_5}\left(\frac{1}{\gamma^2}-\frac{1}{\gamma}\right)\rho u_xu_y\left(\partial_xu_x+\partial_yu_y\right).
\label{eq:25}
\end{eqnarray}
The first terms, which are underlined, in the right hand sides of Eq.~(\ref{eq:23}), Eq.~(\ref{eq:24}) and Eq.~(\ref{eq:25}) are associated with the required flow physics related to the components of the viscous stress tensor. All the remaining terms in these equations are non-Galilean invariant terms for the preconditioned LB scheme. These spurious terms arise because the diagonal third-order moments $\widehat{\kappa}_{xxx}^{eq'}$ and $\widehat{\kappa}_{yyy}^{eq'}$ are not supported by the standard \mbox {D2Q9} lattice. However, such discrete effects are not observed in the C-E analysis of the continuous Boltzmann equation. In order to eliminate the non-GI error terms by other means in the next section on the standard lattice, we explicitly identify the various non-GI terms in the components of the second-order non-equilibrium moments as
\begin{subequations}
\begin{eqnarray}
E_{g{\rho}}^3&=&\frac{2}{3\omega_3}\left(\frac{1}{\gamma}-1\right)(u_x\partial_x\rho+u_y\partial_y\rho), \label{eq:26a}\\
E_{g{u}}^3&=&\frac{\rho}{\omega_3}\left[\left(\frac{4}{\gamma^2}-
\frac{1}{\gamma}\right)u_x^2+\left(\frac{1}{\gamma^2}-\frac{1}{\gamma}\right)u_y^2\right]\partial_xu_x\nonumber\\ &&+\frac{\rho}{\omega_3}\left[\left(\frac{4}{\gamma^2}-
\frac{1}{\gamma}\right)u_y^2+\left(\frac{1}{\gamma^2}-\frac{1}{\gamma}\right)u_x^2\right]\partial_yu_y\label{eq:26b}.
 \end{eqnarray}
 \end{subequations}
 \begin{subequations}
 \begin{eqnarray}
E_{g{\rho}}^4&=&\frac{2}{3\omega_4}\left(\frac{1}{\gamma}-1\right)(u_x\partial_x\rho-u_y\partial_y\rho)\label{eq:27a},\\
E_{g{u}}^4&=&\frac{\rho}{\omega_4}\left[\left(\frac{4}{\gamma^2}-
\frac{1}{\gamma}\right)u_x^2-\left(\frac{1}{\gamma^2}-\frac{1}{\gamma}\right)u_y^2\right]\partial_xu_x\nonumber\\&&+\frac{\rho}{\omega_4}\left[-\left(\frac{4}{\gamma^2}-
\frac{1}{\gamma}\right)u_y^2+\left(\frac{1}{\gamma^2}-\frac{1}{\gamma}\right)u_x^2\right]\partial_yu_y,\label{eq:27b}
  \end{eqnarray}
   \end{subequations}
and
\begin{subequations}
\begin{eqnarray}
E_{g{\rho}}^5&=&\frac{1}{3\omega_5}\left(\frac{1}{\gamma}-1\right)(u_x\partial_y\rho+u_y\partial_x\rho),\label{eq:28a}\\
E_{g{u}}^5&=&\frac{\rho}{\omega_5}\left(\frac{1}{\gamma^2}-\frac{1}{\gamma}\right)u_xu_y\partial_xu_x+\frac{\rho}{\omega_5}\left(\frac{1}{\gamma^2}-\frac{1}{\gamma}\right)u_xu_y\partial_yu_y.
\label{eq:28b}
\end{eqnarray}
\end{subequations}
Then, we can rewrite the non-equilibrium second-order moments
\begin{eqnarray}
\widehat{m}_3^{(1)}=\widehat{\kappa}_{xx}^{(1)'}+\widehat{\kappa}_{yy}^{(1)'}&=&-\frac{2\rho}{3\omega_3 }(\partial_xu_x+\partial_yu_y)+E_{g{\rho}}^3+E_{g{u}}^3, \label{eq:29}\\
\widehat{m}_4^{(1)}=\widehat{\kappa}_{xx}^{(1)'}-\widehat{\kappa}_{yy}^{(1)'}&=&-\frac{2\rho}{3\omega_4 }(\partial_xu_x-\partial_yu_y)+E_{g{\rho}}^4+E_{g{u}}^4,\label{eq:30}\\
\widehat{m}_5^{(1)}=\widehat{\kappa}_{xy}^{(1)'}&=&-\frac{\rho}{3\omega_5 }(\partial_xu_y+\partial_yu_x)+E_{g{\rho}}^5+E_{g{u}}^5. \label{eq:31}
\end{eqnarray}
Some interesting observations can be made from the above analysis: $(i)$ when the LB scheme is preconditioned, i.e. $\gamma \neq 1$, non-GI terms persist in terms of velocity and density gradients for all the second-order non-equilibrium moments, including the off-diagonal moment ($\widehat{m}_5^{(1)}=\widehat{\kappa}_{xy}^{(1)'}$), unlike that for the simulation of the standard NS equations (i.e. with $\gamma=1$). However, the non-GI cubic velocity contributions in $\widehat{m}_5^{(1)}$ vanish for incompressible flow ($\mathbf{\nabla}\cdot\mathbf{u}=0$), i.e. $E_{gu}^{5}=0$ . $(ii)$. In general the prefactors appearing in the non-GI terms for the diagonal components, i.e. in $\widehat{m}_3^{(1)}$ and $\widehat{m}_4^{(1)}$ exhibit dramatically different behaviour for the asymptotic limit cases: No preconditioning case ($\gamma\rightarrow 1$): $\left(\frac{4}{\gamma^2}-\frac{1}{\gamma}\right)\sim 3$,$\left(\frac{1}{\gamma^2}-\frac{1}{\gamma}\right)\sim 0$; strong preconditioning case ($\gamma\rightarrow 0$): $\left(\frac{4}{\gamma^2}-\frac{1}{\gamma}\right)\sim \frac{4}{\gamma^2}$, $\left(\frac{1}{\gamma^2}-\frac{1}{\gamma}\right)\sim \frac{1}{\gamma^2}$. Thus, due to the complicated structure of the truncation errors and their dependence on $\gamma$, the non-GI terms in the diagonal moment components modify significantly as $\gamma$ varies due to preconditioning. $(iii)$ when $\gamma=1$, i.e.~when our preconditioned LB scheme reverts  to the solution of the standard NS equations, $E_{g\rho}^3=E_{g\rho}^4=E_{g\rho}^5=0$, $E_{gu}^3=\frac{3\rho}{\omega_3}\left(u_x^2\partial_xu_x+u_y^2\partial_yu_y\right)$, $E_{gu}^4=\frac{3\rho}{\omega_4}\left(u_x^2\partial_xu_x-u_y^2\partial_yu_y\right)$, and $E_{gu}^5=0$. That is, the non-GI terms become identical to the results reported by~\cite{Dellar2014} and~\cite{Geier2015}.

\section{\label{4} Derivation of Corrections via Extended Moment Equilibria for Elimination of Cubic Velocity errors in Preconditioned Macroscopic Equations}
In order to effectively eliminate the non-GI error terms given in Eq.~(\ref{eq:26a})-(\ref{eq:28b}) that appear in the non-equilibrium moments $\widehat{m}_3^{(1)}$, $\widehat{m}_4^{(1)}$ and $\widehat{m}_5^{(1)}$ in the previous section (see Eqs.~(\ref{eq:29})-(\ref{eq:31})) arising  due to the third-order diagonal equilibrium moments $(\widehat{\kappa}_{xxx}^{eq'}$ and $\widehat{\kappa}_{yyy}^{eq'}) $ not being independently supported by the \mbox {D2Q9} lattice, we consider an approach based on the extended moment equilibria. In other words, we extended the second-order moment equilibria by including extra gradient terms with unknown coefficients as follows:
\begin{eqnarray}
\mathbf{\widehat f}^{eq}=\left[ \begin{array}{l}
\widehat{m}^{eq(0)}_{0}\\\widehat{m}^{eq(0)}_{1}\\\widehat{m}^{eq(0)}_{2}\\\widehat{m}^{eq(0)}_{3}\\
\widehat{m}^{eq(0)}_{4}\\\widehat{f}^{eq(0)}_{5}\\\widehat{m}^{eq(0)}_{6}\\\widehat{m}^{eq(0)}_{7}\\\widehat{m}^{eq(0)}_{8}
\end{array} \right]+
\delta_t{\left[ \begin{array}{l}
0\\0\\0\\\widehat{m}^{eq(1)}_{3}\\
\widehat{m}^{eq(1)}_{4}\\\widehat{m}^{eq(1)}_{5}\\0\\0\\0
\end{array} \right]}=
\left[ \begin{array}{l}
\widehat{\kappa}^{eq'}_{0}\\\widehat{\kappa}^{eq'}_{x}\\\widehat{\kappa}^{eq'}_{y}\\\widehat{\kappa}^{eq'}_{xx}+\widehat{\kappa}^{eq'}_{yy}\\
\widehat{\kappa}^{eq'}_{xx}-\widehat{\kappa}^{eq'}_{yy}\\\widehat{\kappa}^{eq'}_{xy}\\\widehat{\kappa}^{eq'}_{xxy}\\\widehat{\kappa}^{eq'}_{xyy}\\\widehat{\kappa}^{eq'}_{xxyy}
\end{array} \right]+
\delta_t{\left[ \begin{array}{l}
0\\0\\0\\\theta_x^3\partial_xu_x+\theta_y^3\partial_yu_y+\lambda_x^3\partial_x\rho+\lambda_y^3\partial_y\rho\\\theta_x^4\partial_xu_x-\theta_y^4\partial_yu_y+\lambda_x^4\partial_x\rho-\lambda_y^4\partial_y\rho
\\\theta_x^5\partial_xu_x+\theta_y^5\partial_yu_y+\lambda_x^5\partial_x\rho+\lambda_y^5\partial_y\rho\\0\\0\\0
\end{array} \right]}.
\label{eq:32}
\end{eqnarray}
In other words, the corrections to the second-order moments are given by
\begin{subequations}
\begin{eqnarray}
 &\widehat{m}_3^{eq(1)}=(\theta_x^3\partial_xu_x+\theta_y^3\partial_yu_y)+(\lambda_x^3\partial_x{\rho}+\lambda_y^3\partial_y\rho), \label{eq:33a}\\
   &\widehat{m}_4^{eq(1)}=(\theta_x^4\partial_xu_x-\theta_y^4\partial_yu_y)+(\lambda_x^4\partial_x{\rho}-\lambda_y^4\partial_y\rho), \label{eq:33b}\\
&\widehat{m}_5^{eq(1)}=(\theta_x^5\partial_xu_x+\theta_y^5\partial_yu_y)+(\lambda_x^5\partial_x{\rho}+\lambda_y^5\partial_y\rho),
\label{eq:33c}
 \end{eqnarray}
\end{subequations}
where the coefficients $\theta_x^j$, $\theta_y^j$, $\lambda_x^j$ and $\lambda_y^j$, where $j=3,4,5$ are to be determined  from a modified Chapman-Enskog analysis so that the non-GI cubic velocity terms are effectively removed from the emergent preconditioned macroscopic moment equations.\\
We now apply a Chapman-Enskog (C-E) expansion by taking into account the modified equilibria which is now given as  $\mathbf{\widehat m}^{eq}=\mathbf{\widehat m}^{eq(0)}+\delta_t{\mathbf{\widehat m}^{eq(1)}}$, where $\mathbf{\widehat m}^{eq(0)}$ is the moment equilibria presented in the previous section and $\mathbf{\widehat m}^{eq(1)}$ is the correction to this equilibria. As a result, the C-E expansion given as Eq.~(\ref{eq:9}) and Eq.~(\ref{eq:10}) are now replaced with
\begin{equation}
  \begin{aligned}
  \widehat {\mathbf{m}}=\mathbf{\widehat m}^{eq(0)}+\underline{\epsilon \mathbf{\widehat m}^{eq(1)}}+\epsilon \mathbf{\widehat m}^{(1)}+\epsilon^2 \mathbf{\widehat m}^{(2)}+\cdots ,\quad
 \partial_t=\partial_{t_0}+\epsilon{\partial_{t_1}}+\epsilon^2\partial_{t_2}+\cdots .
 \label{eq:34}
  \end{aligned}
\end{equation}
Then, by using a Taylor expansion given in Eq.~(\ref{eq:11}) for the streaming operator in Eq.~(2b) along with above modified C-E expansion Eq.~(\ref{eq:34}), we get the following hierarchy of moment equations at different orders in $\epsilon$:
\begin{subequations}
\begin{eqnarray}
&O(\epsilon^0):\ \mathbf{\widehat{m}}^{(0)}=\mathbf{\widehat{m}}^{eq},\label{eq:35a}\\
&O(\epsilon^1):\ (\partial_{t_0}+\widehat{ \tensr E}_i \partial_i)\mathbf{\widehat{m}}^{(0)}=-\widehat{\tensr{\Lambda}}\left[\mathbf{\widehat{m}}^{(1)}-\underline{\mathbf{\widehat{m}}^{eq(1)}}\right]+\mathbf{\widehat{S}},\label{eq:35b}\\
&O(\epsilon^2):\ \partial_{t_1}\mathbf{\widehat{m}}^{(0)}+(\partial_{t_0}+\widehat{\tensr E}_i \partial_i)\left[ \mathcal{\tensr I}-\frac{1}{2}\widehat{\tensr\Lambda}\right]\mathbf{\widehat{m}}^{(1)}+\underline{(\partial_{t_0}+\widehat{\tensr E}_i \partial_i)\left[ \frac{1}{2}\widehat{\tensr\Lambda}\mathbf{\widehat{m}}^{eq(1)}\right]}=-\widehat{\tensr\Lambda}\mathbf{\widehat{m}}^{(2)}, \label{eq:35c}
\end{eqnarray}
\end{subequations}
where $\widehat{\tensr E}_i=\tensr{T}(\mathbf e_{ i}\tensr{I})\tensr{T}^{-1}$ and  $i \in \{x,y\}$ . The relevant $O(\epsilon)$ equations for the first order moments are given in Eqs.~(\ref{eq:13a})-(\ref{eq:13c}). However, the equations of the second order moments are now modified due to the presence of the extended moment equilibria $\widehat{\mathbf{m}}^{eq(1)}$ in Eq.~(\ref{eq:35b}) which are now given by (instead of Eqs.~(\ref{eq:13d})-(\ref{eq:13f}))

\begin{subequations}
\begin{eqnarray}
&\partial_{t_0}\left(\frac{2}{3}\rho+\frac{\rho(u_x^2+u_y^2)}{\gamma}\right)+\partial_x \left(\frac{4}{3}\rho u_x+\frac{\rho u_xu_y^2}{\gamma^2}\right)+\partial_y \left(\frac{4}{3}\rho u_y+\frac{\rho u_x^2u_y}{\gamma^2}\right)\nonumber\\& =-\omega_3\widehat{m}_3^{(1)}+\omega_3\widehat{m}_3^{eq(1)}+\frac{2(F_xu_x+F_yu_y)}{\gamma^2},
\label{eq:36a}\\&
\partial_{t_0}\left(\frac{\rho(u_x^2-u_y^2)}{\gamma}\right)+\partial_x \left(\frac{2}{3}\rho u_x-\frac{\rho u_xu_y^2}{\gamma^2}\right)+\partial_y \left(-\frac{2}{3}\rho u_y+\frac{\rho u_x^2u_y}{\gamma}\right)\nonumber\\&
=-\omega_4\widehat{m}_4^{(1)}+\omega_4\widehat{m}_4^{eq(1)}+\frac{2(F_xu_x-F_yu_y)}{\gamma^2},
\label{eq:36b}\\&
\partial_{t_0}\left(\frac{\rho u_x u_y}{\gamma}\right)+\partial_x \left(\frac{1}{3}\rho u_y+\frac{\rho u_x^2u_y}{\gamma^2}\right)+\partial_y \left(\frac{1}{3}\rho u_x+\frac{\rho u_xu_y^2}{\gamma^2}\right)\nonumber\\&
 =-\omega_5\widehat{m}_5^{(1)}+\omega_5\widehat{m}_5^{eq(1)}+\frac{F_xu_y+F_yu_x}{\gamma^2}.
\label{eq:36c}
\end{eqnarray}
\end{subequations}
Similarly, the leading order moment equations of $O(\epsilon^2)$ which are modified by $\mathbf{\widehat m}^{eq(1)}$ as shown in Eq.~(\ref{eq:35c}) are obtained as (instead of \mbox {Eqs}.~(\ref{eq:14a})-(\ref{eq:14c}))
\begin{subequations}
\begin{eqnarray}
&\partial_{t_1}\rho=0,
\label{eq:37a}\\&
\partial_{t_1}\left(\rho u_x\right)+\partial_x \left[\frac{1}{2}\left(1-\frac{1}{2}\omega_3\right)\widehat{m}_3^{(1)}+\frac{1}{2}\left(1-\frac{1}{2}\omega_4\right)\widehat{m}_4^{(1)}\right]
+\partial_y \left[\left(1-\frac{1}{2}\omega_5\right)\widehat{m}_5^{(1)}\right]+\nonumber \\&
\partial_x\left[\frac{1}{4}\omega_3\widehat{m}_3^{eq(1)}+\frac{1}{4}\omega_4\widehat{m}_4^{eq(1)} \right]+
\partial_y\left[\frac{1}{2}\omega_5\widehat{m}_5^{eq(1)}\right]=0,
\label{eq:37b}\\&
\partial_{t_1}\left(\rho u_y\right)+\partial_x \left[\left(1-\frac{1}{2}\omega_5\right)\widehat{m}_5^{(1)}\right]+\partial_y \left[\frac{1}{2}\left(1-\frac{1}{2}\omega_3\right)\widehat{m}_3^{(1)}-\frac{1}{2}\left(1-\frac{1}{2}\omega_4\right)\widehat{m}_4^{(1)}\right]
+ \nonumber \\& \partial_x\left[\frac{1}{2}\omega_5\widehat{m}_5^{eq(1)}\right]+\partial_y\left[\frac{1}{4}\omega_3\widehat{m}_3^{eq(1)}-\frac{1}{4}\omega_4\widehat{m}_4^{eq(1)}\right]=0.
\label{eq:37c}
\end{eqnarray}
\end{subequations}
The non-equilibrium moment $\widehat{m}_3^{(1)}$ is now obtained from Eq.~(\ref{eq:36a}) as
\begin{eqnarray}
&\widehat{m}_3^{(1)} \cong \frac{1}{\omega_3}\left[ -\partial_{t_0}\left(\frac{2}{3}\rho+\frac{\rho (u_x^2+u_y^2)}{\gamma}\right)
-\partial_x\left(\frac{4}{3}\rho u_x+\frac{\rho u_xu_y^2}{\gamma^2}\right)
-\partial_y\left(\frac{4}{3}\rho u_y+\frac{\rho u_x^2u_y}{\gamma^2}\right)
 \right.\nonumber\\
 &\left.+\frac{2(F_xu_x+F_yu_y)}{\gamma^2}\right]+\widehat{m}_3^{eq(1)}.
\label{eq:38}
\end{eqnarray}
All the terms within the square brackets in the above equation exactly corresponds to Eq.~(\ref{eq:29}). Hence, Eq.~(\ref{eq:38}) reduces to
\begin{eqnarray}
\widehat{m}_3^{(1)}=- \frac{2\rho}{3\omega_3 }(\partial_xu_x+\partial_yu_y)+E_{g{\rho}}^3+E_{g{u}}^3+\widehat{m}_3^{eq(1)},
\label{eq:39}
 \end{eqnarray}
where the non-GI error terms $E_{g{\rho}}^3$ and $E_{g{u}}^3$ are given in Eqs.~(\ref{eq:26a}) and (\ref{eq:26b}), respectively, and the extended moment equilibrium $\widehat{m}_3^{eq(1)}$ in Eq.~(\ref{eq:33a}). Similarly, the non-equilibrium moment $\widehat{m}_4^{(1)}$ is obtained from Eq.~(\ref{eq:36b}) and using Eq.~(\ref{eq:30}) for simplification, and for $\widehat{m}_5^{(1)}$ using Eqs.~(\ref{eq:36c}) and (\ref{eq:31}), we finally get
\begin{eqnarray}
&\widehat{m}_4^{(1)}=- \frac{2\rho}{3\omega_4 }(\partial_xu_x-\partial_yu_y)+E_{g{\rho}}^4+E_{g{u}}^4+\widehat{m}_4^{eq(1)},
\label{eq:40}\\
&\widehat{m}_5^{(1)}=- \frac{\rho}{3\omega_5 }(\partial_xu_y+\partial_yu_x)+E_{g{\rho}}^5+E_{g{u}}^5+\widehat{m}_5^{eq(1)}.
\label{eq:41}
 \end{eqnarray}
Here, the non-GI error terms $E_{g{\rho}}^4$ and $E_{g{u}}^4$ are given in Eq.~(\ref{eq:27a}) and Eq.~(\ref{eq:27b}), respectively, and the correction equilibrium moment $\widehat{m}_4^{eq(1)}$ in Eq.~(\ref{eq:33b}). Likewise, $E_{g{\rho}}^5$ and $E_{g{u}}^5$ are obtained from Eqs.~(\ref{eq:28a}) and (\ref{eq:28b}) respectively and $\widehat{m}_5^{eq(1)}$ is presented in Eq.~(\ref{eq:33c}).

Now, in order to obtain the preconditioned moment system for the conserved moments, we combine $O(\epsilon)$ equations Eqs.~(\ref{eq:13a})-(\ref{eq:13c}) with $\epsilon \times $ Eq.~(\ref{eq:37a})-(\ref{eq:37c}) for the corresponding equations at $O(\epsilon^2)$, and using $\partial_t=\partial_{t_0}+\epsilon\partial_{t_1}$, we get
\begin{subequations}
\begin{eqnarray}
&\partial_{t}\rho+\partial_x (\rho u_x)+\partial_y (\rho u_y) = 0,\label{eq:42a}\\
&\partial_{t}\left(\rho u_x\right)+\partial_x \left(\frac{1}{3}\rho+\frac{\rho u_x^2}{\gamma}\right)+\partial_y \left(\frac{\rho u_xu_y}{\gamma}\right) = \nonumber \\& \frac{F_x}{\gamma}-\epsilon\partial_x \left[\frac{1}{2}\left(1-\frac{\omega_3}{2}\right)\widehat{m}_3^{(1)}+\frac{1}{2}\left(1-\frac{\omega_4}{2}\right)\widehat{m}_4^{(1)}\right]-\epsilon\partial_y\left[\frac{1}{2}\left(1-\frac{\omega_5}{2}\right)\widehat{m}_5^{(1)}\right]\nonumber \\&
-\epsilon\partial_x\left[\frac{\omega_3}{4}\widehat{m}_3^{eq(1)}+\frac{\omega_4}{4}\widehat{m}_4^{eq(1)}\right]-\epsilon\partial_y\left[\frac{\omega_5}{2}\widehat{m}_5^{eq(1)}\right],\label{eq:42b}\\&
\partial_{t}\left(\rho u_y\right)+\partial_x \left(\frac{\rho u_xu_y}{\gamma}\right)+\partial_y \left(\frac{1}{3}\rho+\frac{\rho u_y^2}{\gamma}\right) = \frac{F_y}{\gamma}-\epsilon\partial_x\left[\left(1-\frac{\omega_5}{2}\right)\widehat{m}_5^{eq(1)}\right]\nonumber \\&
-\epsilon\partial_y\left[\frac{1}{2}\left(1-\frac{\omega_3}{2}\right)\widehat{m}_3^{(1)}-\frac{1}{2}\left(1-\frac{\omega_4}{2}\right)\widehat{m}_4^{(1)}\right]\nonumber \\&
-\epsilon\partial_x\left[\frac{\omega_5}{2}\widehat{m}_5^{eq(1)}\right]-\epsilon\partial_y\left[\frac{\omega_3}{4}\widehat{m}_3^{eq(1)}-\frac{\omega_3}{4}\widehat{m}_4^{eq(1)}\right].
\label{eq:42c}
\end{eqnarray}
\end{subequations}
Our goal is to show that the above equations (Eq.~(\ref{eq:42a})-(\ref{eq:42c})) is consistent with the preconditioned NS equations (Eq.~(1)) presented in Sec.~\ref{1} without the identified truncation errors, i.e. without involving the non-GI cubic velocity defects. Now, in order to relate the moment corrections $\widehat{m}_3^{eq(1)}$, $\widehat{m}_4^{eq(1)}$ and $\widehat{m}_5^{eq(1)}$ appearing in the equilibria with the non-GI error terms, with a view to eliminate them, consider the right hand side of Eq.~(\ref{eq:42b}) (i.e.the $x$-momentum equation) and substitute  for $\widehat{m}_3^{(1)}$, $\widehat{m}_4^{(1)}$ and $\widehat{m}_5^{(1)}$ from Eq.~(\ref{eq:39}), Eq.~(\ref{eq:40}) and Eq.~(\ref{eq:41}), respectively, which becomes
\begin{eqnarray}
&=\frac{F_x}{\gamma}+\epsilon\partial_x \left[+\frac{1}{3}\left(\frac{1}{\omega_3}-\frac{1}{2}\right)\rho (\partial_xu_x+\partial_yu_y)+\frac{1}{3}\left(\frac{1}{\omega_4}-\frac{1}{2}\right)\rho(\partial_xu_x-\partial_yu_y)\right]\nonumber \\&
+\epsilon\partial_y\left[\frac{1}{3}\left(\frac{1}{\omega_5}-\frac{1}{2}\right)\rho(\partial_xu_y+\partial_yu_x)\right]
\nonumber \\&-\epsilon\partial_x\left[\frac{1}{2}\left(1-\frac{\omega_3}{2}\right)\left\{E_{g\rho}^3+E_{gu}^3\right\}+\frac{1}{2}\left(1-\frac{\omega_4}{2}\right)\left\{E_{g\rho}^4+E_{gu}^4\right\}
 \right]-\epsilon\partial_y\left[\left(1-\frac{\omega_5}{2}\right)\left\{E_{g\rho}^5+E_{gu}^5\right\}\right]\nonumber \\&
 -\epsilon\partial_x\left[\frac{1}{2}\widehat{m}_3^{eq(1)}+\frac{1}{2}\widehat{m}_4^{eq(1)}\right]-\epsilon\partial_y\left[\widehat{m}_5^{eq(1)}\right].
 \label{eq:43}
\end{eqnarray}
The first two lines in the above equations correspond to the physics, while the third line corresponds to the spurious non-GI terms arising from discrete lattice effects and the fourth line are related to equilibrium corrections.

In order to eliminate the cubic velocity truncation errors, it follows that the third and fourth lines in the above equation (Eq.~(\ref{eq:43})) sum to zero. This yields
\begin{subequations}
  \begin{eqnarray}
  \left(1-\frac{\omega_3}{2}\right)\left\{E_{g\rho}^3+E_{gu}^3\right\}+\widehat{m}_3^{eq(1)}=0,
   \label{eq:44a}\\
   \left(1-\frac{\omega_4}{2}\right)\left\{E_{g\rho}^4+E_{gu}^4\right\}+\widehat{m}_4^{eq(1)}=0,
    \label{eq:44b}\\
      \left(1-\frac{\omega_5}{2}\right)\left\{E_{g\rho}^5+E_{gu}^5\right\}+\widehat{m}_5^{eq(1)}=0.
       \label{eq:44c}
   \end{eqnarray}
\end{subequations}
The above equations Eqs.~(\ref{eq:44a})-(\ref{eq:44c}), represent the key constraint relations between the non-GI error terms and the moment equilibria correction terms to obtain a preconditioned cascaded central moment LB model without cubic velocity defects.

Further analysis shows that these constraints hold identically for the $y$-momentum as well (Eq.~\ref{eq:42c})). Now considering Eq.~(\ref{eq:44a}) and using Eq.~(\ref{eq:26a}) and (\ref{eq:26b}) for $E_{g \rho}^3$ and $E_{gu}^3$, respectively, the extend moment equilibrium $\widehat{m}_3^{eq(1)}$ is given as
\begin{eqnarray*}
\widehat{m}_3^{eq(1)}=(\theta_x^3\partial_xu_x+\theta_y^3\partial_yu_y)+(\lambda_x^3\partial_x\rho+\lambda_y^3\partial_y\rho),
\end{eqnarray*}
where the coefficients obtained after matching are given by
\begin{subequations}
 \begin{eqnarray}
&\theta_x^3=-\left(\frac{1}{\omega_3}-\frac{1}{2}\right)\rho\left[\left(\frac{4}{\gamma^2}-\frac{1}{\gamma}\right)u_x^2+\left(\frac{1}{\gamma^2}-\frac{1}{\gamma}\right)u_y^2\right],\\&\label{eq:45a}
\theta_y^3=-\left(\frac{1}{\omega_3}-\frac{1}{2}\right)\rho\left[\left(\frac{4}{\gamma^2}-\frac{1}{\gamma}\right)u_y^2+\left(\frac{1}{\gamma^2}-\frac{1}{\gamma}\right)u_x^2\right],\\&\label{eq:45b}
\lambda_x^3=-\frac{2}{3}\left(\frac{1}{\omega_3}-\frac{1}{2}\right)\left(\frac{1}{\gamma}-1\right)u_x,\\&\label{eq:45c}
\lambda_y^3=-\frac{2}{3}\left(\frac{1}{\omega_3}-\frac{1}{2}\right)\left(\frac{1}{\gamma}-1\right)u_y. \label{eq:45d}
  \end{eqnarray}
\end{subequations}
Similarly, from Eq.~(\ref{eq:27a}),(\ref{eq:27b}), (\ref{eq:33b}) and (\ref{eq:44b}), we can obtain the coefficient of $\widehat{m}_4^{eq(1)}$, and from Eq.~(\ref{eq:28a}), Eq.~(\ref{eq:28b}), Eq.~(\ref{eq:33c}) and Eq.~(\ref{eq:44c}), those for $\widehat{m}_5^{eq(1)}$ can be determined. The results read as follows:
\begin{eqnarray*}
\widehat{m}_4^{eq(1)}=(\theta_x^4\partial_xu_x-\theta_y^4\partial_yu_y)+(\lambda_x^4\partial_x\rho-\lambda_y^4\partial_y\rho),
\end{eqnarray*}
where
\begin{subequations}
\begin{eqnarray}
&\theta_x^4=-\left(\frac{1}{\omega_4}-\frac{1}{2}\right)\rho\left[\left(\frac{4}{\gamma^2}-\frac{1}{\gamma}\right)u_x^2-\left(\frac{1}{\gamma^2}-\frac{1}{\gamma}\right)u_y^2\right], \\ \label{eq:46a}& \theta_y^4=\left(\frac{1}{\omega_4}-\frac{1}{2}\right)\rho\left[-\left(\frac{4}{\gamma^2}-\frac{1}{\gamma}\right)u_y^2+\left(\frac{1}{\gamma^2}-\frac{1}{\gamma}\right)u_x^2\right],\\\label{eq:46b}&
\lambda_x^4=-\frac{2}{3}\left(\frac{1}{\omega_4}-\frac{1}{2}\right)\left(\frac{1}{\gamma}-1\right)u_x,\\\label{eq:46c}&
\lambda_y^4=-\frac{2}{3}\left(\frac{1}{\omega_4}-\frac{1}{2}\right)\left(\frac{1}{\gamma}-1\right)u_y, \label{eq:46d}
\end{eqnarray}
\end{subequations}
and
\begin{eqnarray*}
 \widehat{m}_5^{eq(1)}=(\theta_x^5\partial_xu_x+\theta_y^5\partial_yu_y)+(\lambda_x^5\partial_x{\rho}+\lambda_y^5\partial_y\rho),
 \label{eq:sh3}
 \end{eqnarray*}
where
\begin{subequations}
\begin{eqnarray}
    &\theta_x^5=-\left(\frac{1}{\omega_5}-\frac{1}{2}\right)\rho\left(\frac{1}{\gamma^2}-\frac{1}{\gamma}\right)u_xu_y,\label{eq:47a} \\
    &\theta_y^5=-\left(\frac{1}{\omega_5}-\frac{1}{2}\right)\rho\left(\frac{1}{\gamma^2}-\frac{1}{\gamma}\right)u_xu_y,\\\label{eq:47b}&
    \lambda_x^5=-\frac{1}{3}\left(\frac{1}{\omega_5}-\frac{1}{2}\right)\rho\left(\frac{1}{\gamma}-1\right)u_y,\label{eq:47c} \\
        &\lambda_y^5=-\frac{1}{3}\left(\frac{1}{\omega_5}-\frac{1}{2}\right)\rho\left(\frac{1}{\gamma}-1\right)u_x.
         \label{eq:47d}
   \end{eqnarray}
\end{subequations}
Note that, as a special case, when $\gamma=1$, i.e. the LB model is used to solve the standard NS equations without preconditioning, then $\theta_x^3=-3\rho (\frac{1}{\omega_3}-\frac{1}{2})u_x^2$, $\theta_y^3=-3\rho (\frac{1}{\omega_3}-\frac{1}{2})u_y^2$, $\theta_x^4=-3\rho (\frac{1}{\omega_4}-\frac{1}{2})u_x^2$, $\theta_x^4=-3\rho (\frac{1}{\omega_4}-\frac{1}{2})u_y^2$, and all the remaining coefficient go to zero. In such a case, these moment corrections to the equilibria become identical to the GI corrections presented by~\cite{Geier2015} and equivalent to the alternative GI formulation without cubic velocity errors introduced by~\cite{Dellar2014}.

Finally, using the above extended moment equilibria ($\widehat{m}_3^{eq(1)}$, $\widehat{m}_4^{eq(1)}$ and $\widehat{m}_5^{eq(1)}$) and the expression for the non-equilibrium moments ($\widehat{m}_3^{(1)}$, $\widehat{m}_4^{(1)}$ and $\widehat{m}_5^{(1)}$) from Eq.~(\ref{eq:39})-Eq.~(\ref{eq:41}) along with the constraint relations, i.e. Eqs.~(\ref{eq:44a})-(\ref{eq:44c}) in Eqs.~(\ref{eq:42a})-(\ref{eq:42c}), we get
\begin{equation}
\partial_t \rho + \bm{\nabla}\cdot \bm{j} = 0,
\label{eq:48a},
\end{equation}
\begin{eqnarray}
\partial_t j_x+\bm{\nabla}\cdot \left(\frac{\bm{j}u_x}{\gamma}\right)&=&-\partial_x \frac{ p^*}{\gamma}+\partial_x\left[\frac{\vartheta_4}{\gamma}(2\partial_x j_x-\bm{\nabla}\cdot\bm{j})+\frac{\vartheta_3}{\gamma}\bm{\nabla}\cdot\bm{j}\right]\nonumber \\
&&+\partial_y\left[\frac{\vartheta_5}{\gamma}(\partial_x j_y+\partial_y j_x) \right]+\frac{F_x}{\gamma},
\label{eq:48b}
\end{eqnarray}
\begin{align}
&\partial_t j_y+\bm{\nabla}\cdot \left(\frac{\bm{j}u_y}{\gamma}\right)=-\partial_y \frac{p^*}{\gamma}+\partial_x\left[\frac{\vartheta_5}{\gamma}(\partial_x j_y+\partial_y j_x)\right]\nonumber \\
&+\partial_y\left[\frac{\vartheta_4}{\gamma}(2\partial_y j_y-\bm{\nabla}\cdot\bm{j})+\frac{\vartheta_3}{\gamma}\bm{\nabla}\cdot\bm{j} \right]+\frac{F_y}{\gamma},
\label{eq:48c}
\end{align}
where $p^*=\frac{\gamma}{3}\rho$ is the pressure, $\bm{j}=\rho \bm{u}$, and the bulk and shear viscosities are, respectively given by
\begin{equation}
\vartheta_3=\frac{\gamma}{3}\left(\frac{1}{\omega_3}-\frac{1}{2}\right),\quad
\vartheta_4=\frac{\gamma}{3}\left(\frac{1}{\omega_4}-\frac{1}{2}\right),\quad
\vartheta_5=\frac{\gamma}{3}\left(\frac{1}{\omega_5}-\frac{1}{2}\right). \label{eq:49}
\end{equation}
Thus, Eqs.~(\ref{eq:48a})-(\ref{eq:48c}) are consistent with the preconditioned NS equations given in Eqs.~(\ref{eq:1a})-(\ref{eq:1b}) without cubic velocity defects in GI due to the use of the extended moment equilibria presented earlier.

\section{\label{sec:5} Galilean Invariant Preconditioned Cascaded Central Moment LBM without Cubic Velocity Errors on a Standard Lattice}
The cascaded central moment LBM with forcing term presented in Eqs.~(\ref{eq:2a}), (\ref{eq:2b}), (\ref{eq:3}) and (4) modify to enforce GI without cubic velocity errors as follows. Equations Eq.~(\ref{eq:2a}), Eq.~(\ref{eq:2b}) and Eq.~(\ref{eq:3}) remains the same as before and the collision kernel given in Eq.~(4) is modified to account for the extended moment equilibria in the second order moments as well as corrections to the third-order equilibrium moments. The change of moments $\widehat{g}_3$, $\widehat{g}_4$ and $\widehat{g}_5$ for the second order components follow by
augmenting the corresponding moment equilibria with the extended moment equilibria incorporating the GI corrections identified in the previous section.
On the other hand, owing to the cascaded structure of the collision kernel, the GI corrections to the third order moment changes $\widehat{g}_6$ and $\widehat{g}_7$, which depend on the lower order moment changes, for the preconditioned central moment LB scheme need to be constructed carefully. They are obtained by prescribing the relaxation of the third order central moment components to their corresponding central moment equilibria. Following the derivation given in~\cite{Premnath2009b}, they can then be represented as $-6u_y\widehat{g}_3-2u_y\widehat{g}_4-8u_x\widehat{g}_5-4\widehat{g}_6=\omega_6[\widehat{\kappa}_{xxy}^{eq}-\widehat{\kappa}_{xxy}]$ and $-6u_x\widehat{g}_3+2u_x\widehat{g}_4-8u_y\widehat{g}_5-4\widehat{g}_7=\omega_7[\widehat{\kappa}_{xyy}^{eq}-\widehat{\kappa}_{xyy}]$, where
$\widehat{\kappa}_{xxy}$ and $\widehat{\kappa}_{xyy}$ are the third order central moment components, and $\widehat{\kappa}_{xxy}^{eq}$ and $\widehat{\kappa}_{xyy}^{eq}$, respectively, are their equilibria. Rewriting these central moment relaxations in terms of the relaxations of the raw moment components of the third and lower orders via the binomial theorem, it follows that
\begin{eqnarray*}
\widehat{g}_6&=& \frac{\omega_6}{4}\left[(\widehat{\kappa}_{xxy}^{'}-\widehat{\kappa}_{xxy}^{eq'})-2u_x(\widehat{\kappa}_{xy}^{'}-\widehat{\kappa}_{xy}^{eq'})
-u_y(\widehat{\kappa}_{xx}^{'}-\widehat{\kappa}_{xx}^{eq'})\right]-u_y\left(\frac{3}{2}\widehat{g}_3+\frac{1}{2}\widehat{g}_4\right)-2u_x\widehat{g}_5,\\
\widehat{g}_7&=&
\frac{\omega_7}{4}\left[(\widehat{\kappa}_{xyy}^{'}-\widehat{\kappa}_{xyy}^{eq'})-2u_y(\widehat{\kappa}_{xy}^{'}-\widehat{\kappa}_{xy}^{eq'})
-u_x(\widehat{\kappa}_{yy}^{'}-\widehat{\kappa}_{yy}^{eq'})\right]-u_x\left(\frac{3}{2}\widehat{g}_3-\frac{1}{2}\widehat{g}_4\right)-2u_y\widehat{g}_5.
\end{eqnarray*}
Now, using the components of the preconditioned raw moment equilibria, including those for the third order equilibrium moments with the GI corrections
from Eq.~(\ref{eq:eqmrawmoment}), the final expressions for the change in moments for the collision kernel $\widehat{g}_6$ and $\widehat{g}_7$ can be
derived. Thus, the modified preconditioned collision kernel with the GI corrections reads
\begin{eqnarray*}
&\widehat{g}_0=0,\quad \widehat{g}_1=0,\quad\widehat{g}_2=0,\\
&\widehat{g}_3= \frac{\omega_3}{12}\left\{ \frac{2}{3}\rho+\rho(u_x^2+u_y^2)/\gamma
-(\widehat{\overline{\kappa}}_{xx}^{'}+\widehat{\overline{\kappa}}_{yy}^{'})
-\frac{1}{2}(\widehat{\sigma}_{xx}^{'}+\widehat{\sigma}_{yy}^{'})+ \right.\nonumber \\
&\left.\underline{(\theta_x^3\partial_xu_x+\theta_y^3\partial_yu_y)\delta_t} +\underline{(\lambda_x^3\partial_x{\rho}+\lambda_y^3\partial_y\rho)\delta_t}
\right\}, \label{eq:collisionkernelcompacta2}\\
&\widehat{g}_4=\frac{\omega_4}{4}\left\{\rho(u_x^2-u_y^2)/\gamma
-(\widehat{\overline{\kappa}}_{xx}^{'}-\widehat{\overline{\kappa}}_{yy}^{'})
-\frac{1}{2}(\widehat{\sigma}_{xx}^{'}-\widehat{\sigma}_{yy}^{'})+\right.\nonumber \\
&\left.\underline{(\theta_x^4\partial_xu_x-\theta_y^4\partial_yu_y)\delta_t}+\underline{(\lambda_x^4\partial_x{\rho}-\lambda_y^4\partial_y\rho)\delta_t}
\right\}, \label{eq:collisionkernelcompacta3}\\
&\widehat{g}_5=\frac{\omega_5}{4}\left\{\rho u_x u_y/\gamma
-\widehat{\overline{\kappa}}_{xy}^{'}
-\frac{1}{2}\widehat{\sigma}_{xy}^{'}+\underline{(\theta_x^5\partial_xu_x+\theta_y^5\partial_yu_y)\delta_t}+\underline{(\lambda_x^5\partial_x{\rho}+\lambda_y^5\partial_y\rho)\delta_t}
\right\}, \label{eq:collisionkernelcompacta4}\\
&\widehat{g}_6=\frac{\omega_6}{4}\left\{\underline{\left(\frac{3}{\gamma}-\frac{1}{\gamma^2}\right)\rho u_x^2 u_y}+\widehat{\kappa}_{xxy}^{'}
              -2u_x\widehat{\kappa}_{xy}^{'}-u_y\widehat{\kappa}_{xx}^{'}
              \right\}-\frac{1}{2}u_y(3\widehat{g}_3+\widehat{g}_4)-2u_x\widehat{g}_5, \label{eq:collisionkernelcompacta5}\\
&\widehat{g}_7=\frac{\omega_7}{4}\left\{\underline{\left(\frac{3}{\gamma}-\frac{1}{\gamma^2}\right)\rho u_x u_y^2}+\widehat{\kappa}_{xyy}^{'}
              -2u_y\widehat{\kappa}_{xy}^{'}-u_x\widehat{\kappa}_{yy}^{'}
              \right\}-\frac{1}{2}u_x(3\widehat{g}_3-\widehat{g}_4)-2u_y\widehat{g}_5, \label{eq:collisionkernelcompacta6}\\
&\widehat{g}_8=\frac{\omega_8}{4}\left\{\frac{1}{9}\rho+3\rho u_x^2 u_y^2-\left[\widehat{\kappa}_{xxyy}^{'}
                                 -2u_x\widehat{\kappa}_{xyy}^{'}-2u_y\widehat{\kappa}_{xxy}^{'}
                                 +u_x^2\widehat{\kappa}_{yy}^{'}+u_y^2\widehat{\kappa}_{xx}^{'}\right.\right.
                                 \nonumber \\
                                  &\qquad\left.\left.+4u_xu_y\widehat{\kappa}_{xy}^{'}
                                 \right]
                                  \right\}-2\widehat{g}_3-\frac{1}{2}u_y^2(3\widehat{g}_3+\widehat{g}_4)
                                  -\frac{1}{2}u_x^2(3\widehat{g}_3-\widehat{g}_4)\nonumber\\
                                   &\qquad-4u_xu_y\widehat{g}_5-2u_y\widehat{g}_6
                                  -2u_x\widehat{g}_7.\label{eq:collisionkernelcompacta7}
\end{eqnarray*}
where the various coefficients $\theta_x^j$, $\theta_y^j$, $\lambda_x^j$ and $\lambda_y^j$ where $j=3,4$  and $5$ are given in Eqs.~(\ref{eq:45a})-(\ref{eq:45d}), and (\ref{eq:46a})-(\ref{eq:46d}) and (\ref{eq:47a})-(\ref{eq:47d}). The GI corrections are
identified by means of the underlined terms in the cascaded collision kernel terms in the above equation.

It may be noted that other GI preconditioned LB schemes without cubic velocity errors can be constructed from our results in the
previous section. For example, a non-orthogonal moment based multiple relaxation time LB method readily follows from the analysis
presented before. The spatial gradients for the velocity components and the density appearing in the extended moment equilibria
can be calculated using isotropic finite difference schemes. Alternatively, the diagonal strain rate components
$\partial_xu_x$ and $\partial_yu_y$ can be locally obtained from non-equilibrium moments as follows, which is used in our simulation
studies presented in the next section. From Eqs.~(\ref{eq:39}) and (\ref{eq:44a}) and rearranging, one may write the resulting
expression as follows:
\begin{eqnarray}
-c_1\partial_x u_x -c_2\partial_y u_y=\widehat{m}_3^{(1)}-{e_{\rho}}. \label{eq:51}
\end{eqnarray}
Similarly, from Eq.~(\ref{eq:40}) and Eq.~(\ref{eq:44b}), it follows that
\begin{eqnarray}
-\widetilde c_1\partial_x u_x +\widetilde c_2\partial_y u_y=\widehat{m}_4^{(1)}-\widetilde{e_{\rho}}, \label{eq:52}
\end{eqnarray}
where the coefficients $c_1$, $c_2$, $\widetilde c_1$ and $\widetilde c_1$ and the parameters ${e_{\rho}}$ and $\widetilde{e_{\rho}}$ are defined as
\begin{eqnarray}
&c_1=\left[\frac{2}{3\omega_3}+P_{\gamma}\right]\rho,\quad \widetilde c_1=\left[\frac{2}{3\omega_4}+\widetilde{P}_{\gamma}\right]\rho, \label{eq:53a}\\
&c_2=\left[\frac{2}{3\omega_3}+Q_{\gamma}\right]\rho,\quad \widetilde c_2=\left[\frac{2}{3\omega_4}+\widetilde{Q}_{\gamma}\right]\rho.\label{eq:53b}
\end{eqnarray}
Here,
\begin{eqnarray*}
&P_{\gamma}=-\frac{1}{2}\left(A_{\gamma}u_x^2+B_{\gamma}u_y^2\right),\quad Q_{\gamma}=-\frac{1}{2}\left(A_{\gamma}u_y^2+B_{\gamma}u_x^2\right),\\
&\widetilde{P}_{\gamma}=-\frac{1}{2}\left(A_{\gamma}u_x^2-B_{\gamma}u_y^2\right),\quad \widetilde{Q}_{\gamma}=-\frac{1}{2}\left(A_{\gamma}u_y^2-B_{\gamma}u_x^2\right)
\end{eqnarray*}
where
$A_{\gamma}=\left(\frac{4}{\gamma^2}-\frac{1}{\gamma}\right)$, $B_{\gamma}=\left(\frac{1}{\gamma^2}-\frac{1}{\gamma}\right)$, $C_{\gamma}=\left(\frac{1}{\gamma}-1\right)$
and
\begin{eqnarray}
{e_{\rho}}=\frac{1}{3}C_{\gamma}\left(u_x\partial_x \rho+u_y\partial_y \rho\right),\quad {\widetilde e_{\rho}}=\frac{1}{3}C_{\gamma}\left(u_x\partial_x \rho-u_y\partial_y \rho\right).
\label{eq:54}
\end{eqnarray}
Solving Eqs.~(\ref{eq:51}) and (\ref{eq:52}) for $\partial_xu_x$ and $\partial_yu_y$ , we get
\begin{subequations}
\begin{eqnarray}
\partial_xu_x&=&\quad\left[\widetilde c_2 (\widehat{m}_3^{(1)}-{e_{\rho}})+ c_2(\widehat{m}_4^{(1)}-\widetilde{e_{\rho}})\right]/\left[- c_1 \widetilde c_2-\widetilde c_1 c_2\right],\label{eq:55a}\\
\partial_yu_y&=&-\left[c_1 (\widehat{m}_4^{(1)}-{\widetilde e_{\rho}})+ \widetilde c_1(\widehat{m}_3^{(1)}-{e_{\rho}})\right]/\left[- c_1 \widetilde c_2-\widetilde c_1 c_2\right].
\label{eq:55b}
\end{eqnarray}
\end{subequations}
Here, the density gradients appearing in  $e_{\rho}$ and $\widetilde e_{\rho}$ Eqs.~(\ref{eq:54}) may be computed using a isotropic finite difference scheme. In Eqs.~(\ref{eq:55a}) and (\ref{eq:55b}), all the coefficients involving $\gamma$ need to be computed only once before the start of computations for efficient implementation; quantities such as $u_x^2$  and $u_y^2$ appearing in the factors $P$, $Q$, $\widetilde{P}$ and $\widetilde{Q}$ above need to be reused rather than perform the product calculations for every occurrence. A comparison of the computational costs for the uncorrected preconditioned LB scheme and the GI corrected preconditioned formulation is presented for a benchmark case study on the four-rolls mill flow problem at the end of the numerical results section (see Sec.~\ref{sec:fourrollsmill}), which also demonstrates a quantitative improvement in accuracy achieved with correction. The non-equilibrium moments $\widehat{m}_3^{(1)}$ and $\widehat{m}_4^{(1)}$ required in Eqs.~(\ref{eq:55a}) and Eqs.~(\ref{eq:55b}) are obtained as
\begin{subequations}
 \begin{eqnarray}
\widehat{m}_3^{(1)}&=&\sum_{\alpha}(e_{\alpha x}^2+e_{\alpha y}^2)f_{\alpha}-\left[\frac{2}{3}\rho+\frac{\rho(u_x^2+u_y^2)}{\gamma}\right],\\
\widehat{m}_4^{(1)}&=&\sum_{\alpha}(e_{\alpha x}^2-e_{\alpha y}^2)f_{\alpha}-\frac{\rho(u_x^2-u_y^2)}{\gamma}.
\end{eqnarray}
\end{subequations}

\section{\label{sec:6}Numerical Results}

We will now present the validation of our new Galilean invariant preconditioned cascaded central moment LBM by making comparisons against prior numerical solutions for various complex flow benchmark problems. These include the lid-driven cavity flow, flow over a square cylinder, backward-facing step flow, the Hartmann flow and the four-roll mills flow problem. In addition, we will also demonstrate the convergence acceleration achieved using our preconditioning LB model for some of the benchmark flow problems. \par
\subsection{Lid-driven Cavity Flow}

\begin{figure}[htbp]
\centering
\advance\leftskip2cm
\advance\rightskip3cm
    \subfloat[Re=3200\label{subfig-2:dummy}]{
        \includegraphics[scale=0.5] {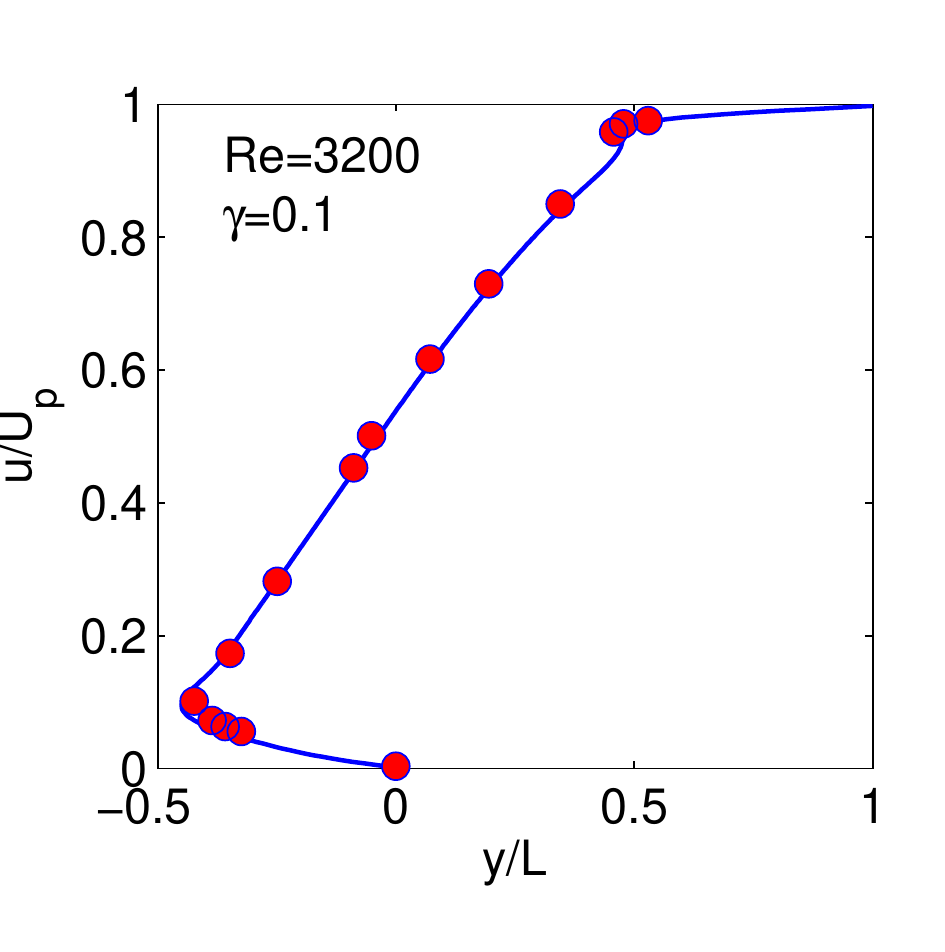}
         }
    \hfill
    \subfloat[Re=3200\label{subfig-2:dummy}]{
       \includegraphics[scale=0.5] {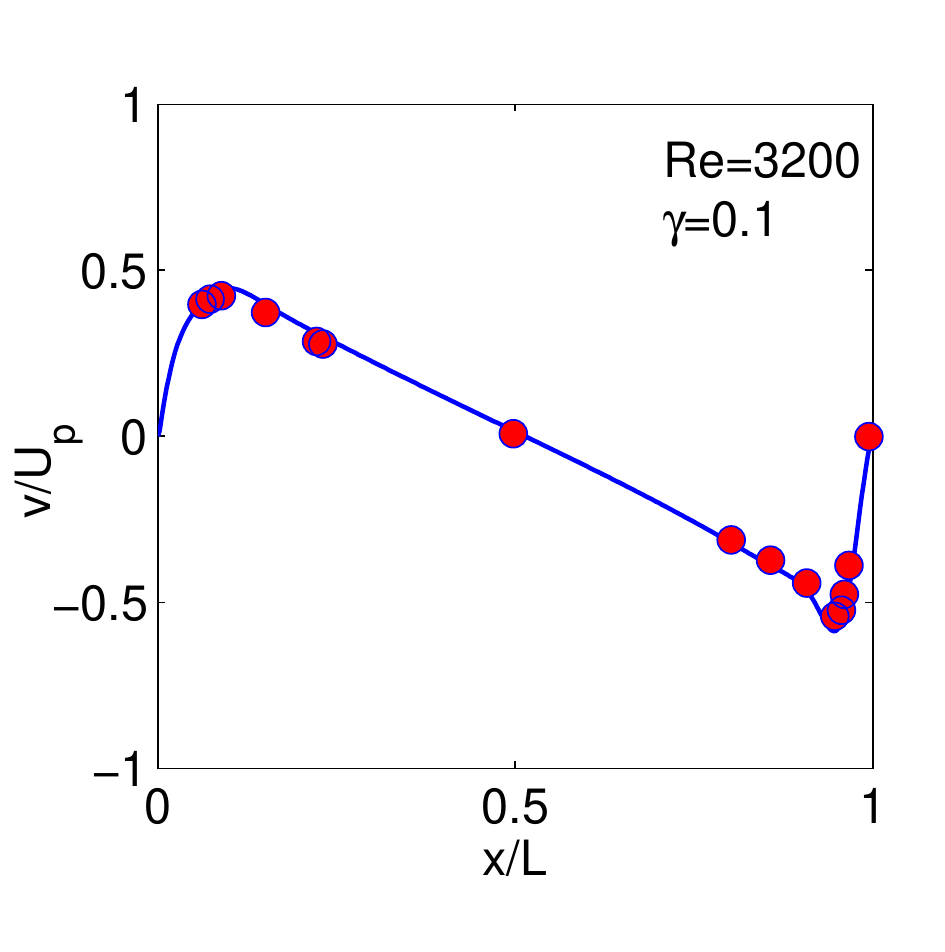}
        } \\
    \subfloat[Re=5000\label{subfig-2:dummy}]{
        \includegraphics[scale=0.5] {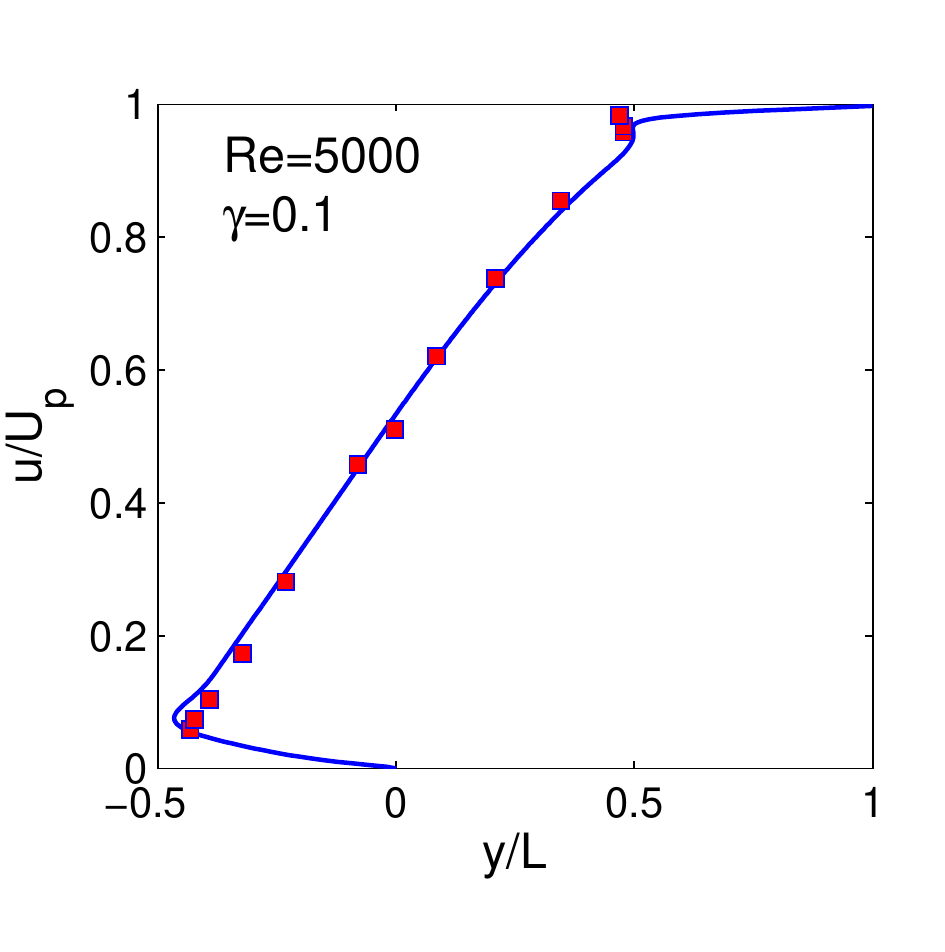}
         }
     \hfill
    \subfloat[Re=5000\label{subfig-2:dummy}]{
        \includegraphics[scale=0.5] {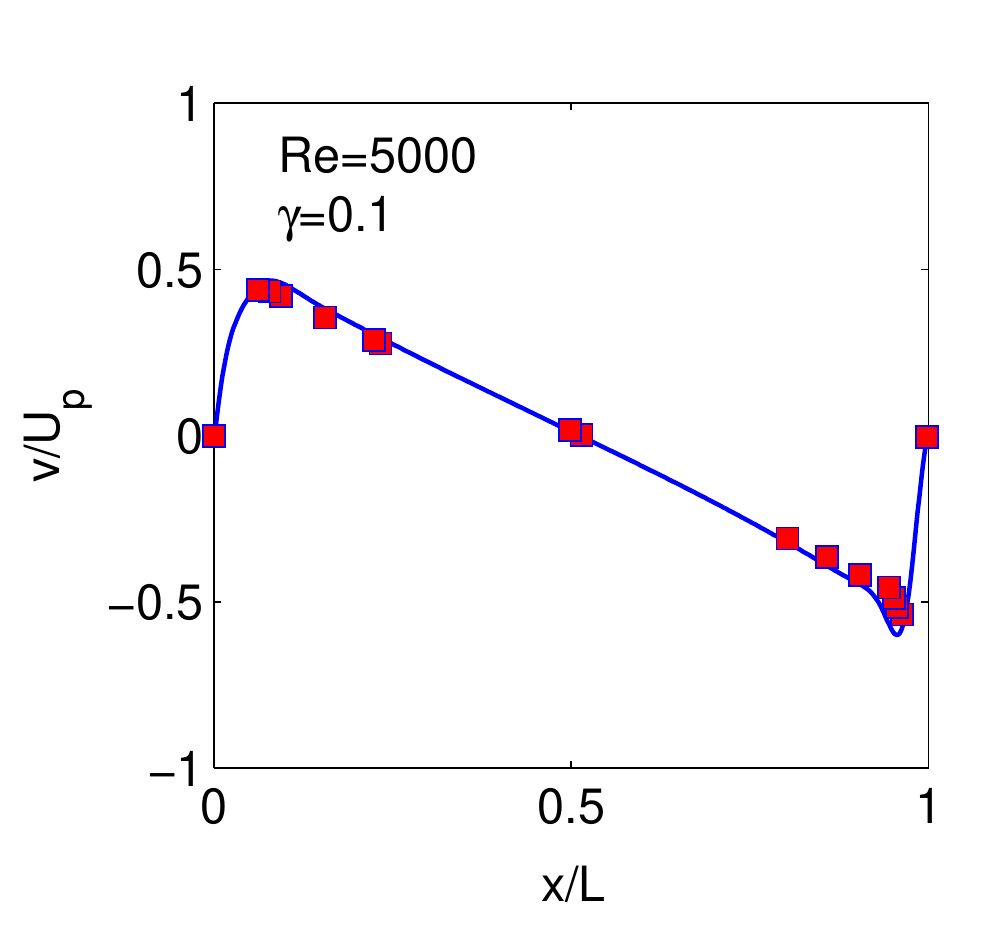}
         }
    \caption{Comparison of the computed horizontal   velocity $u/U_p$ and vertical  velocity $v/U_p$ profiles along the geometric centerlines of the cavity using the Galilean invariant preconditioned cascaded central moment LBM with the benchmark results of~\cite{Ghia1982} (symbols) for Re=3200 and 5000  and $\gamma=0.1$.}
    \label{fig:velcavity}
\end{figure}

\begin{figure}[htbp]
\centering
\advance\leftskip-4cm
\advance\rightskip-4cm
        \includegraphics[scale=0.6] {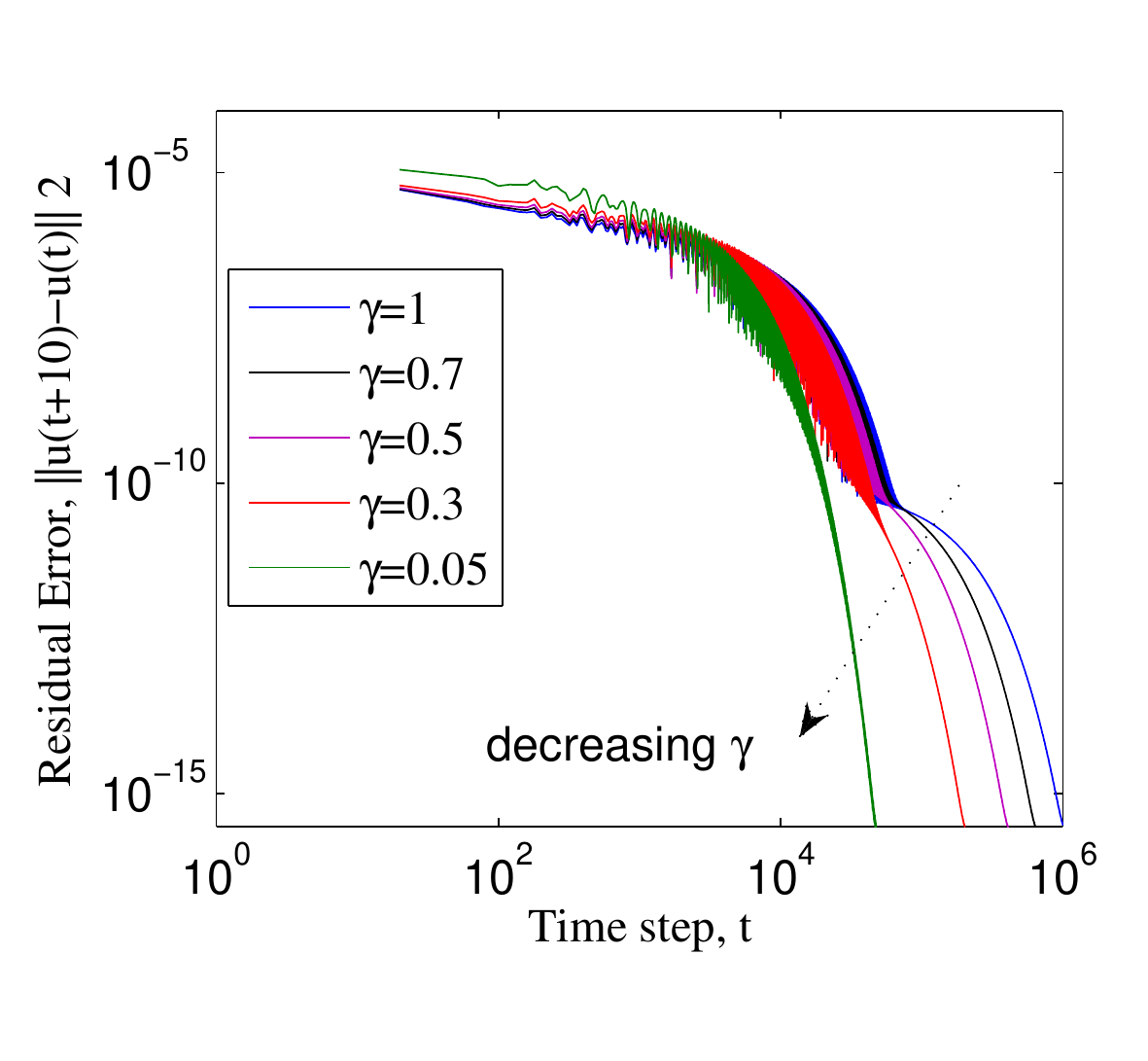}
        \caption{Convergence histories of  the GI preconditioned cascaded central moment LBM and the standard cascaded LBM ($\gamma=1$) for lid-driven cavity flow for Re=3200.}
        \label{fig:hiscavity}
    \end{figure}

As the first test problem, the GI preconditioned central moment LB model is applied for the simulation of steady, two-dimensional flow within a square cavity driven by the motion of the top lid. This is one of the classical internal flow benchmark problems with complex flow structures. The numerical simulations are computed at two different Reynolds numbers of 3200 and 5000, which are resolved by computational meshes with a resolution of $400\times 400$. To implement the moving top wall at a velocity $U_p$, the standard momentum augmented half-way bounce back scheme is considered. In order to validate the numerical simulation results obtained with our GI preconditioned LB scheme, the computed dimensionless horizontal and vertical velocity profiles along the vertical and horizontal centerlines, respectively, for Reynolds number $Re=3200$  and 5000 and preconditioning parameter $\gamma=0.1$, are presented with benchmark solutions of~\cite{Ghia1982} in Fig.~\ref{fig:velcavity}. The Mach number \mbox{Ma} considered in the simulations is $0.05$. It is clear that the velocity profiles for all the cases agree very well with the prior numerical data.
Next, we investigate how the steady state convergence histories are influenced by the use of our new preconditioned formulation for this benchmark problem. Figure~\ref{fig:hiscavity} presents the convergence histories for $Re=3200$ obtained by varying the  preconditioning parameter $\gamma$. Here $\gamma=1$ corresponds to results without preconditioning. Obviously, the use of preconditioning  accelerates the steady state convergence by at least one order of magnitude. For example, it can be seen that when compared to the case without preconditioning ($\gamma=1$), the preconditioned GI cascaded LBM with $\gamma=0.05$, is at least 15 times faster.

\subsection{ Laminar Flow over a Square Cylinder}
\begin{figure}[htbp]
\centering
\advance\leftskip-4cm
\advance\rightskip-4cm
\includegraphics[clip,trim=.0cm 8cm 0cm 0cm,width=0.6 \textwidth] {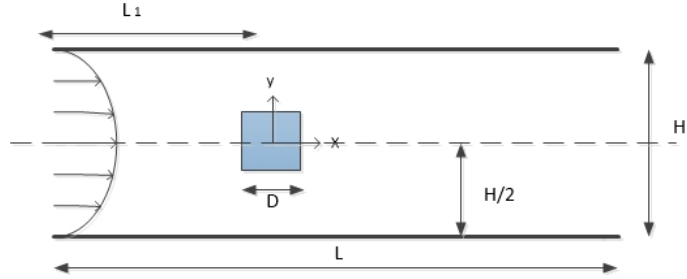}
\caption{Schematic representation of the flow over a square cylinder in a 2D channel.}
\label{fig:schsquare}
\end{figure}

\begin{figure}[htbp]
\centering
\advance\leftskip-2cm
\advance\rightskip2cm
    \subfloat[Re=1\label{subfig-2:dummy}]{
        \includegraphics[width=.7\textwidth,height=0.15\textheight] {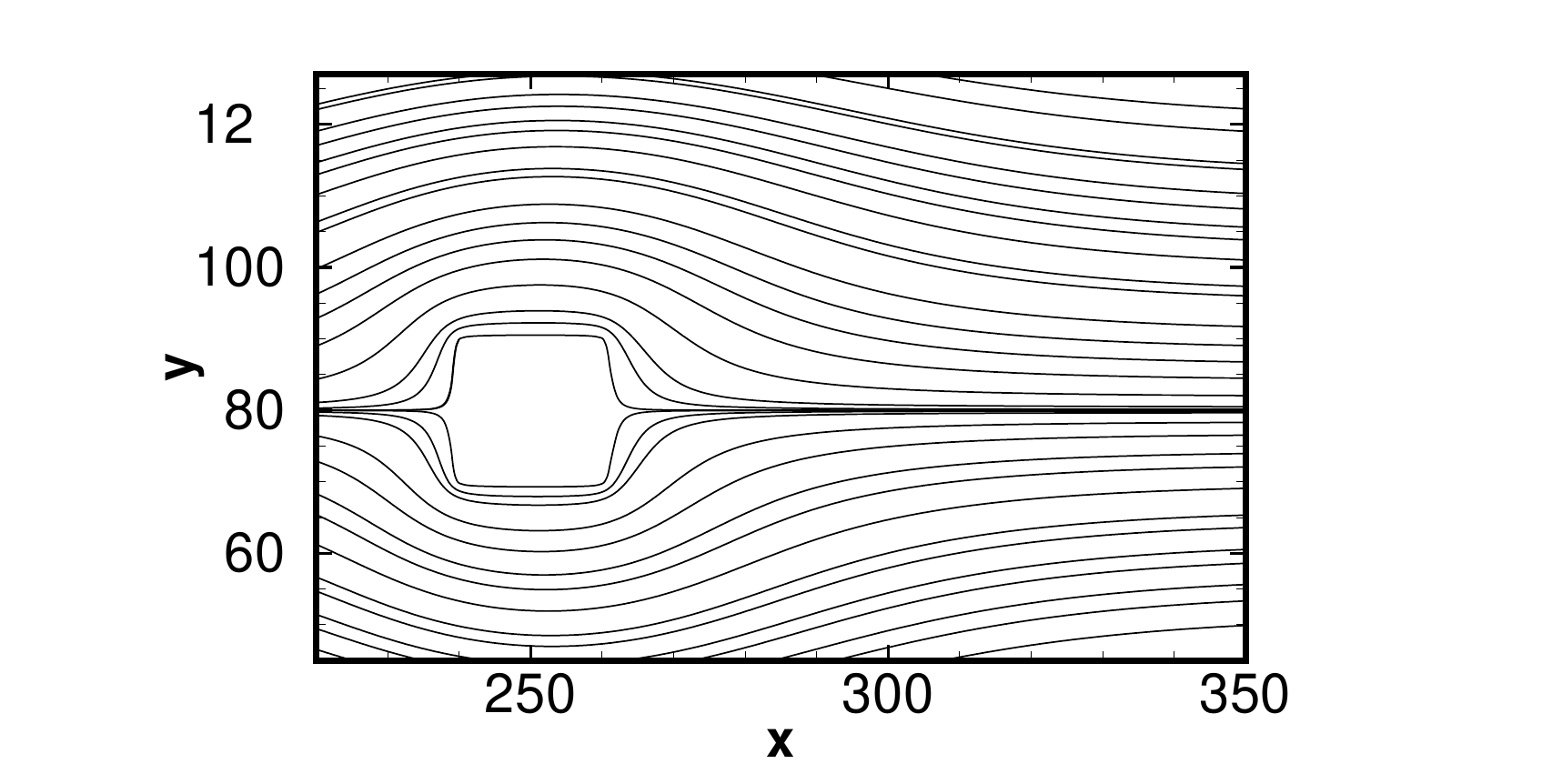}
         } \hspace*{-8em}
    \hfill
    \subfloat[Re=15\label{subfig-2:dummy}]{
        \includegraphics[width=.7\textwidth,height=0.15\textheight] {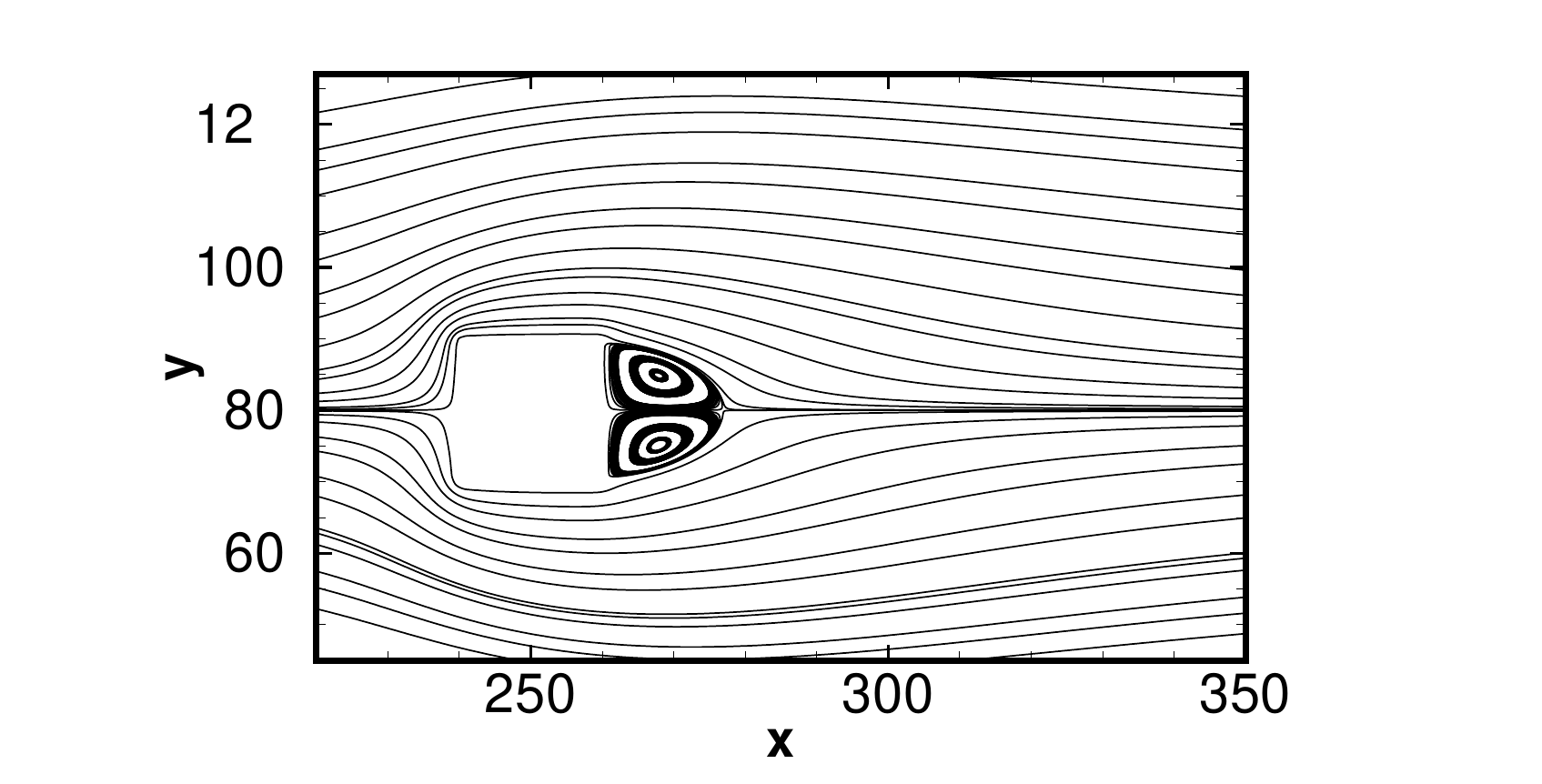}
        } \\
        \advance\leftskip0cm
    \subfloat[Re=30\label{subfig-2:dummy}]{
        \includegraphics[width=.7\textwidth,height=0.15\textheight] {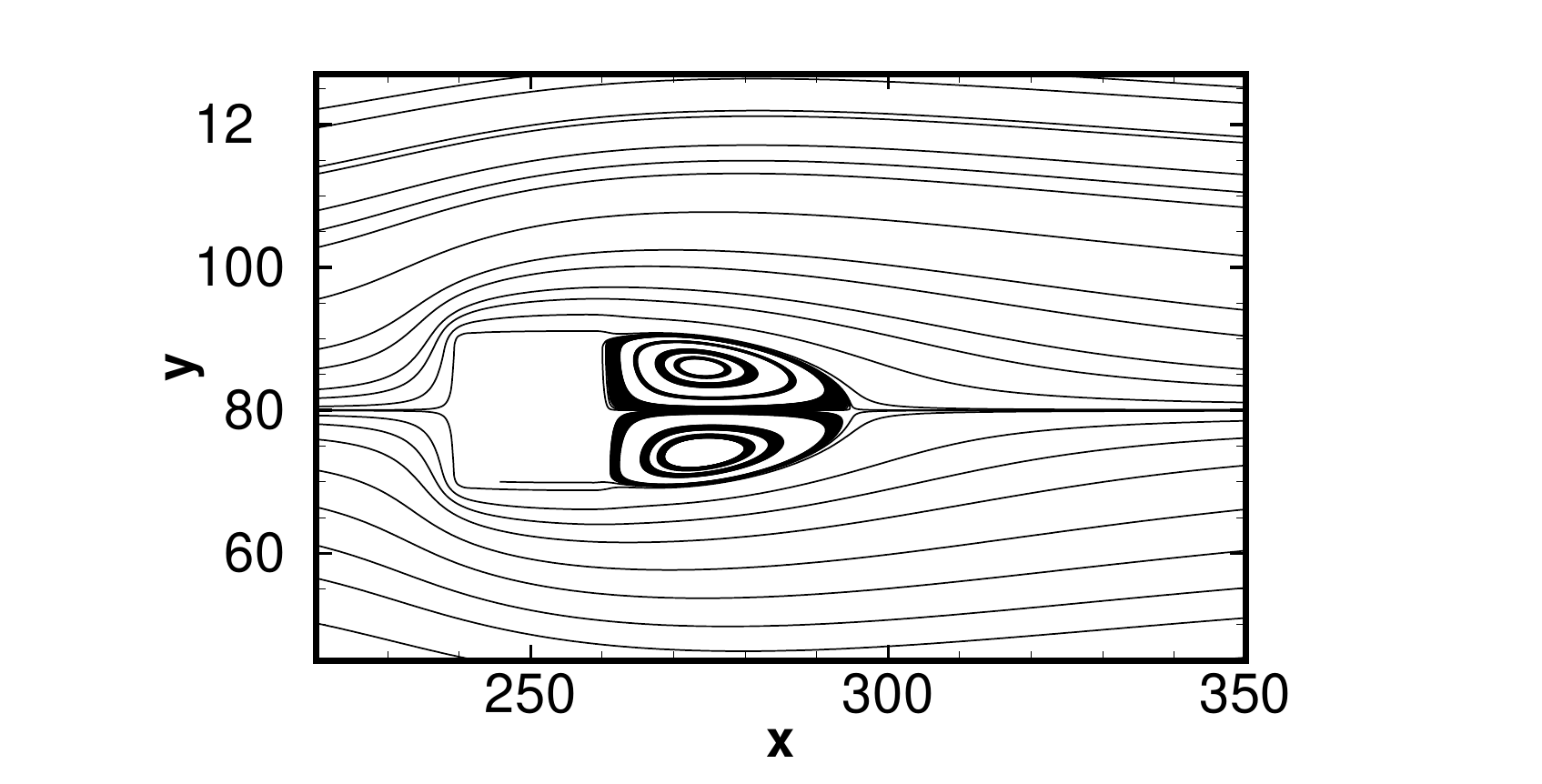}
         } \hspace*{-8em}
        \caption{Stream function contours for flow over a square cylinder for four different Reynolds numbers; Re=1, Re=15 and Re=30 using the GI preconditioned cascaded central moment LBM with $\gamma=0.5$.}
    \label{fig:strsquare}
\end{figure}

Next, in order to validate our preconditioned LB formulation for an external complex flow example, a two dimensional laminar flow over a square cylinder in a channel is studied. The geometry details and  the set up of the flow problem is provided in Fig.~\ref {fig:schsquare}. A fully developed velocity profile is considered at the inlet, and at the outlet, a convective boundary condition is used which is given by
\begin{align}
\partial_t u_i+U_{max}\partial_x u_i=0
\end{align}
where $U_{max}$ is the maximum velocity of the inflow profile. Computations were performed using $L=50D$, $H=8D$ and $L_1=12.5D$, where D is side of square the cylinder, L and H are the total  length and width of computation domain, respectively and the location of square cylinder from entrance is defined by $L_1$.
In order to visualize the general complex features and patterns of the flow, the streamlines plots at four different Reynolds numbers $Re=1$, $Re=15$, $Re=30$ and $Re=200 $ are presented in Fig.~\ref {fig:strsquare}. In Fig.~\ref {fig:strsquare}(a), as it may be expected, at a low Reynolds number, $Re=1$, where the fluid velocity is  relatively very slow and on the other hand, the viscosity is large, the fluid flow is creeping and symmetric without separation. However, with increasing Reynolds number an adverse pressure gradient is established which leads to the flow separation from the surface and a vortex pair regime is  formed (Fig.~\ref {fig:strsquare}(b)). As the Reynolds number is further increased further to $Re=30$, the size of the recirculation zone increases; besides the flow is still steady and symmetric about the horizontal centerline (Fig.~\ref {fig:strsquare}(c)). These general features and flow patterns are consistent with the prior benchmark results (e.g.~\cite{Breuer2000},~\cite{Guo2008}).

\begin{figure}[htbp]
\centering
\advance\leftskip-7cm
\advance\rightskip-7cm
    \subfloat[\label{subfig-2:dummy}]{
        \includegraphics[scale=0.5] {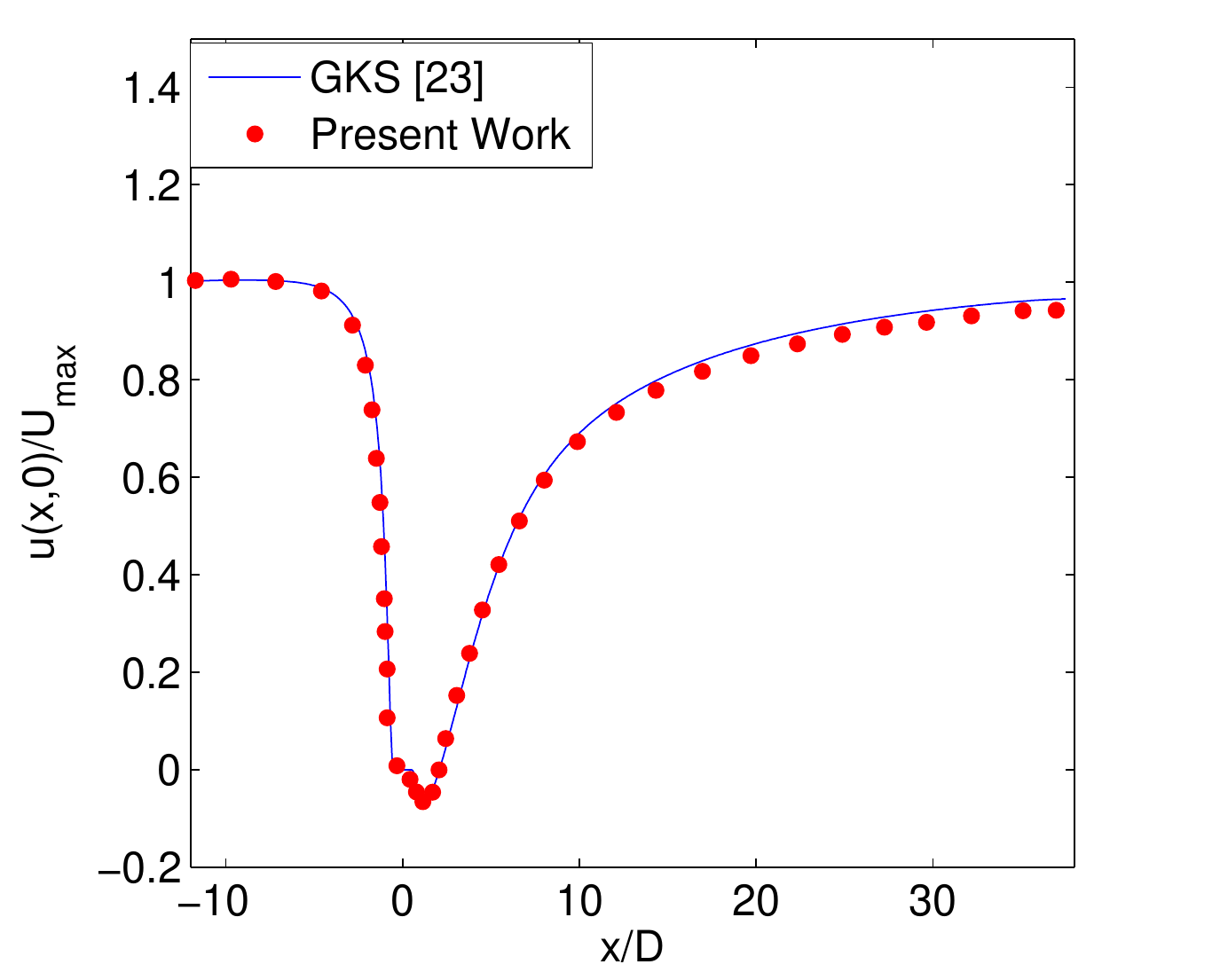}
        \label{fig:img1} } \hspace*{-2em}
    \subfloat[\label{subfig-2:dummy}]{
        \includegraphics[scale=0.5] {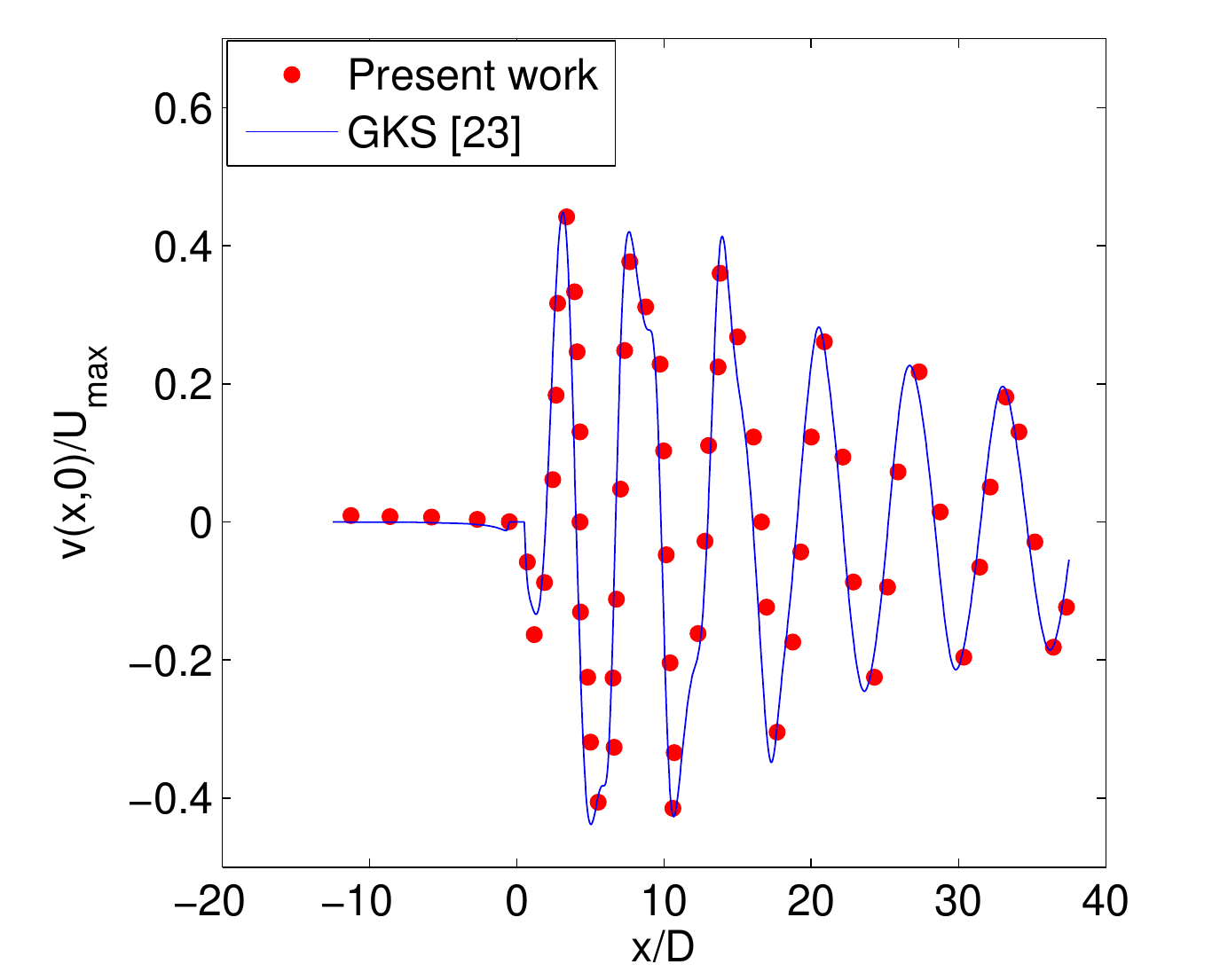}
        \label{fig:img2} } \\
        \advance\leftskip0cm
    \subfloat[\label{subfig-2:dummy}]{
        \includegraphics[scale=0.5] {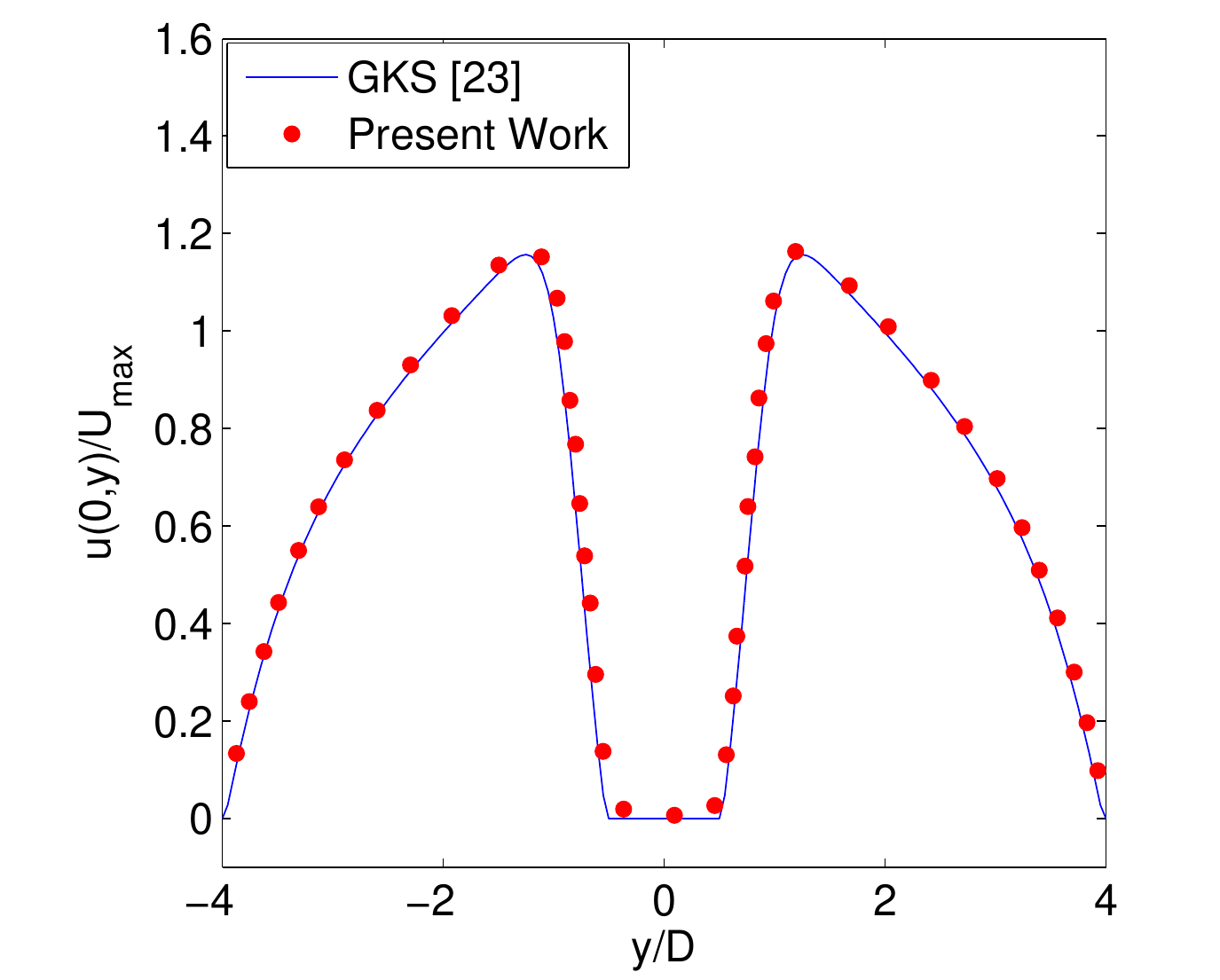}
        \label{fig:img3} } \hspace*{-2em}
     \subfloat[\label{subfig-2:dummy}]{
        \includegraphics[scale=0.5] {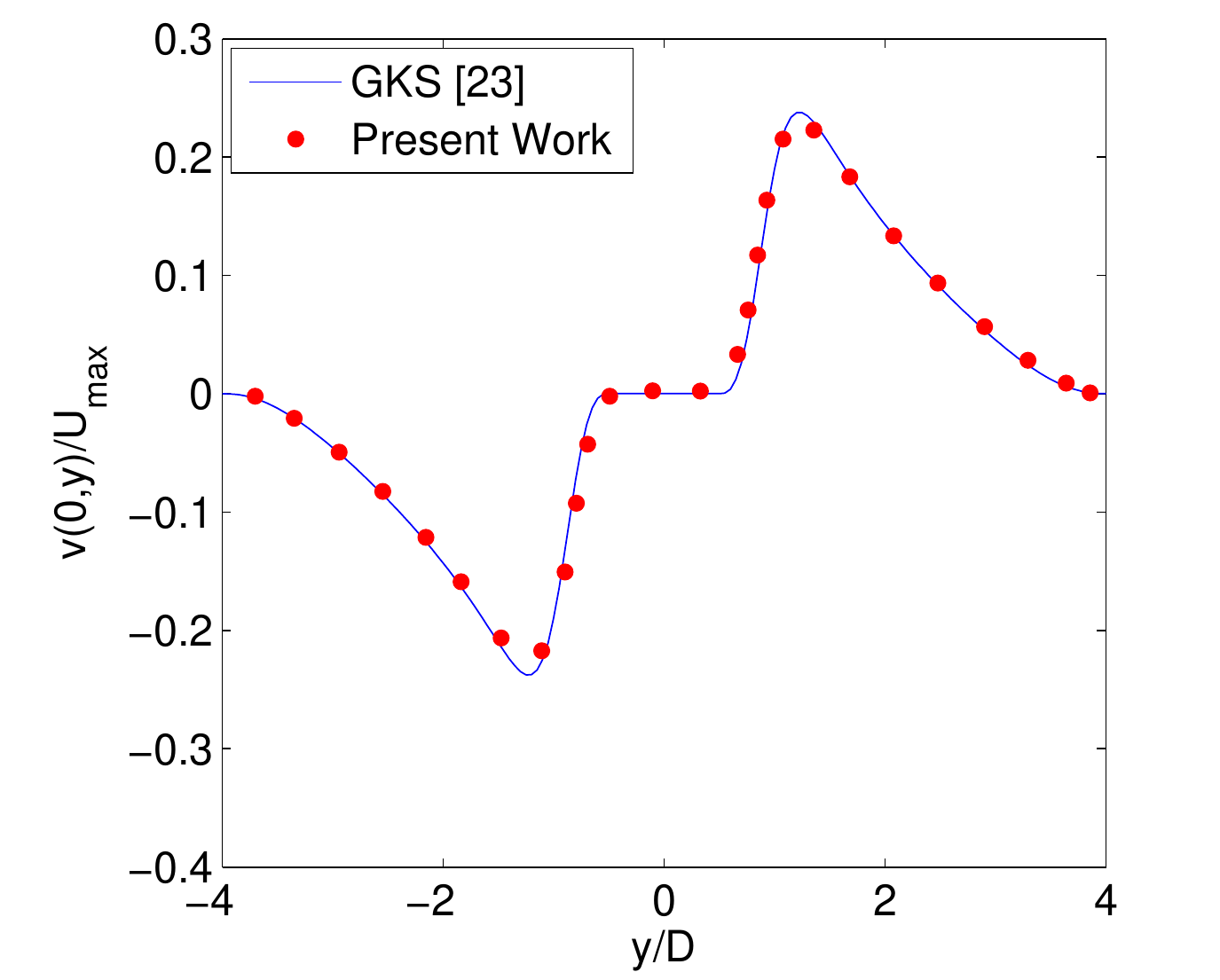}
        \label{fig:img4} }
    \caption{Comparison of the computed velocity profiles along and across the square cylinder along its centerline for both the horizontal $u$ and vertical $v$ velocity components obtained using the GI preconditioned cascaded central moment LBM with $\gamma=0.5$ for $Re=100$ with benchmark results obtained using the Gas Kinetic Scheme (GKS)~\cite{Guo2008}.}
    \label{fig:velsquare}
\end{figure}

\begin{figure}[htbp]
\centering
\advance\leftskip-7cm
\advance\rightskip-7cm
    \subfloat[\label{subfig-2:dummy-wakelength}]{
        \includegraphics[scale=0.6] {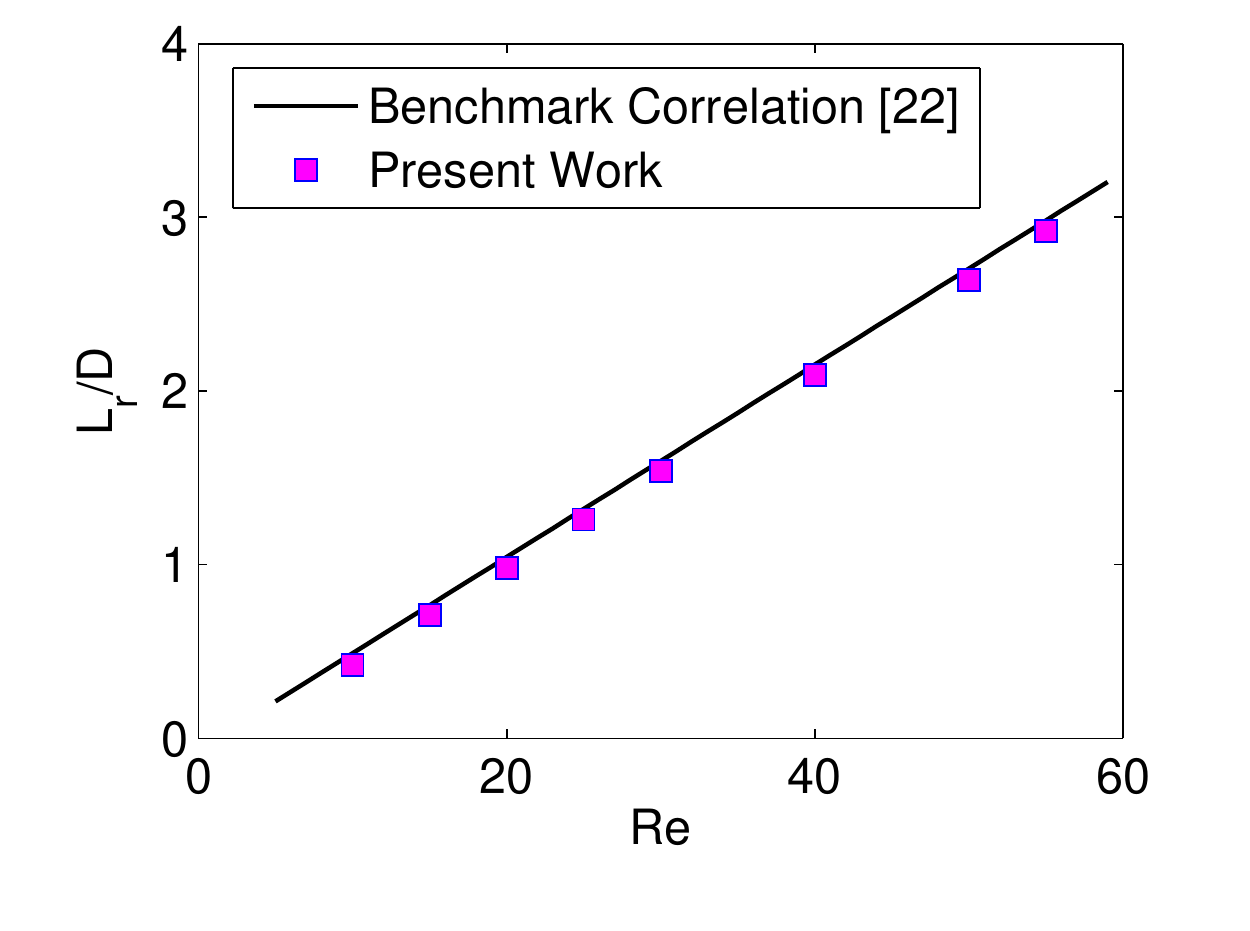}
       }
    \subfloat[\label{subfig-2:dummy1-drag}]{
        \includegraphics[scale=0.6] {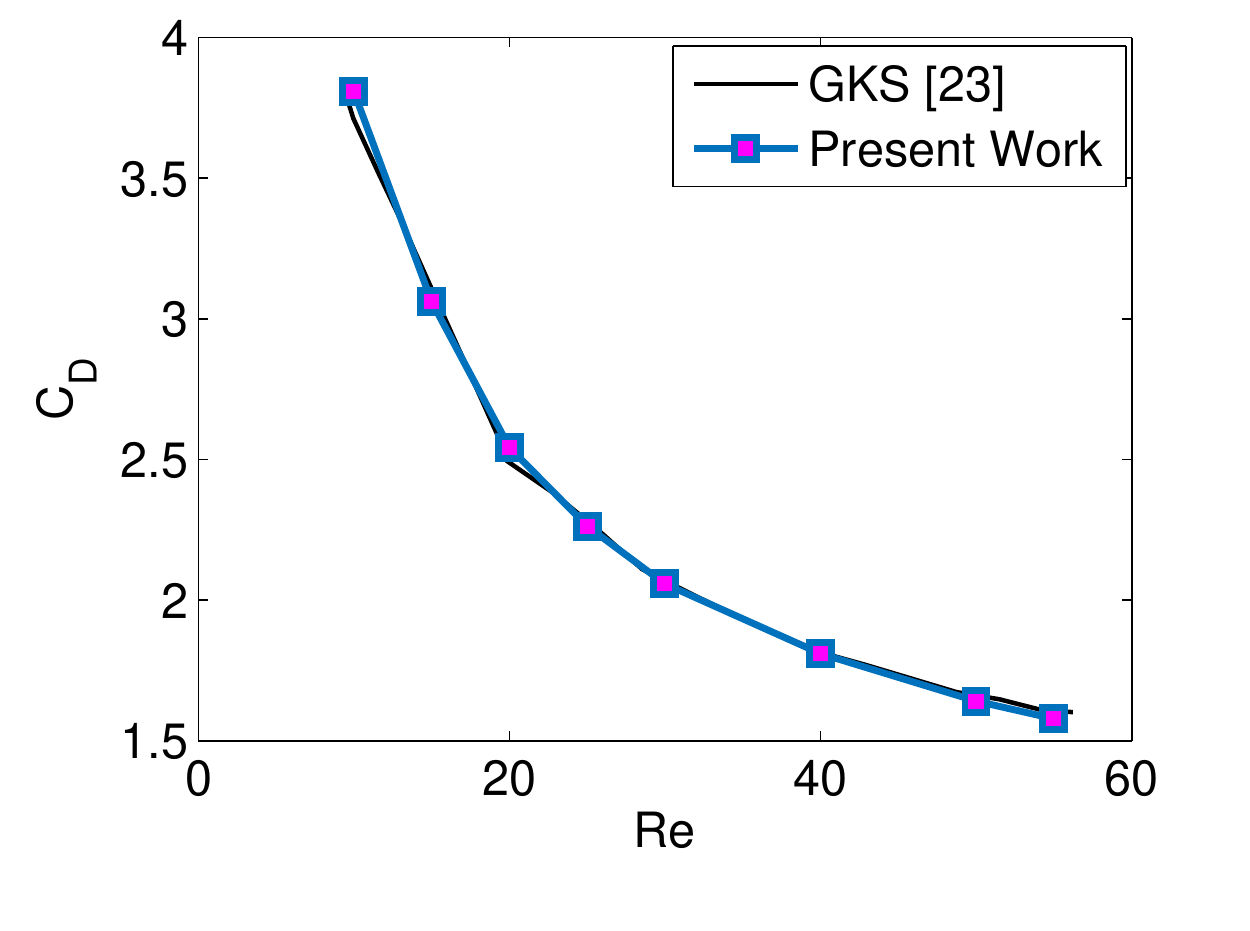}
        }
        \caption{Comparison of the computed Reynolds number dependence of the recirculating wake length $L_r$ on the left (a) and the Reynolds number dependence of the drag coefficient $C_D$ on the right with (b) benchmark correlation (Eq.~(\ref{eq:rec})) and GKS-based numerical results ~\cite{Guo2008} respectively.}
    \label{fig:his_cav}
\end{figure}

Then, we present the  velocity profiles  along the centerline at different sections  at $Re=100$ with a mesh resolution of $1000\times 320$. Figure.~\ref{fig:velsquare} illustrates the  horizontal and vertical components of the velocity profiles of $u$ and $v$, respectively. By comparing  the  present results against  the benchmark numerical results obtained using the Gas Kinetic scheme (GKS)~\cite{Guo2008}, a good agreement between the computational results is observed. An important global feature of the flow over a cylinder is the length of the recirculating flow pattern formed behind the cylinder. Quantitative characterization of this wake length $L_r$ and its dependence on the Reynolds number \mbox{Re} is a key element in the validation of numerical scheme. A widely used empirical correlation for the wake length $L_r$ as a linear function of the Reynolds number is given by~\cite{Breuer2000}
\begin{align}
\frac{L_r}{D}\approx -0.065+0.0554Re,\ for\ 5<Re<60.
\label{eq:rec}
\end{align}
As illustrated in Fig.~\ref{subfig-2:dummy-wakelength}, the computed results for the wake length $L_r$ obtained using the GI preconditioned cascaded central moment LBM are in very good agreement with the empirical correlation  presented in Eq.~(\ref{eq:rec}). As may be expected, for the steady 2D flow over a square cylinder  which, at relatively low $Re$ is characterized by symmetry, the lift force is zero and, as a result, a main  quantity of interest is the drag force or the drag coefficient $C_D$ in dimensionless form whose magnitude varies significantly with $Re$. We use the standard momentum exchange method to compute the drag force on the square cylinder in our preconditioned LB formulation.

\begin{figure}[htbp]
\centering
\advance\leftskip-4cm
\advance\rightskip-4cm
        \includegraphics[scale=0.5] {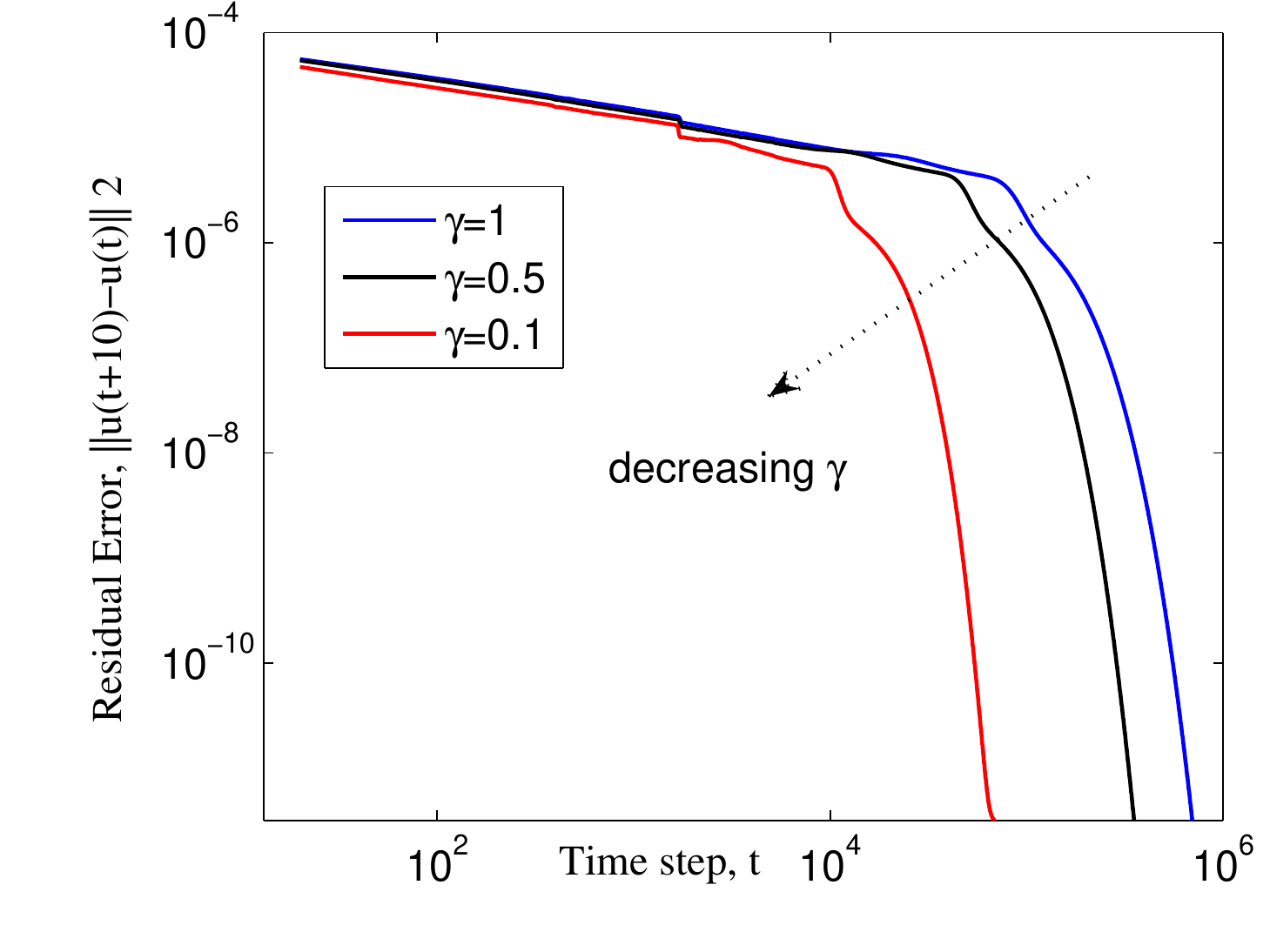}
        \caption{Convergence histories of  the GI preconditioned cascaded central moment LBM and the standard cascaded LBM ($\gamma=1$) for flow over the square cylinder for Re=30.}
        \label{fig:hissquare}
    \end{figure}

A comparison of the computed drag coefficient $C_D$ obtained using our GI preconditioned LB scheme with the GKS scheme~\cite{Guo2008} based benchmark results is presented in Fig.~\ref{subfig-2:dummy1-drag}. It can be observed that the obtained results agree well with the benchmark solutions.
Next, we analyze the influence of the precondition parameter $\gamma$ in our formulation on the steady state convergence of this complex flow problem. Figure~\ref{fig:hissquare} presents the convergence histories for $Re=30$. It can be seen that when compared to the  usual cascaded LBM without preconditioning $(\gamma=1)$, the preconditioned formulation (e.g. for $\gamma<0.1$) is able converge to the steady state significantly faster, with the residual error being reduced to the machine round off error by a factor of least 15 times more rapidly. Thus, the GI preconditioned cascaded central moments LBM exhibits significant convergence acceleration for complex flows.

\subsection{\label{sec:intro}Backward-Facing Step Flow}
\begin{figure}[htbp]
\centering
\advance\leftskip-4cm
\advance\rightskip-4cm
        \includegraphics[clip,trim=.0cm 8cm 0cm 0cm,width=0.6 \textwidth] {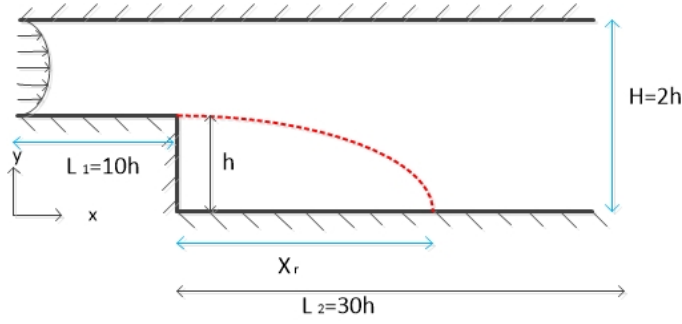}
        \caption{Schematic representation of the flow over a
backward-facing step in a 2D channel.}
        \label{fig:schstep}
    \end{figure}

\begin{figure}[htbp]
\centering
\advance\leftskip-5cm
\advance\rightskip-4cm
    \subfloat[Re=100\label{subfig-2:dummy}]{
        \includegraphics [scale=0.5] {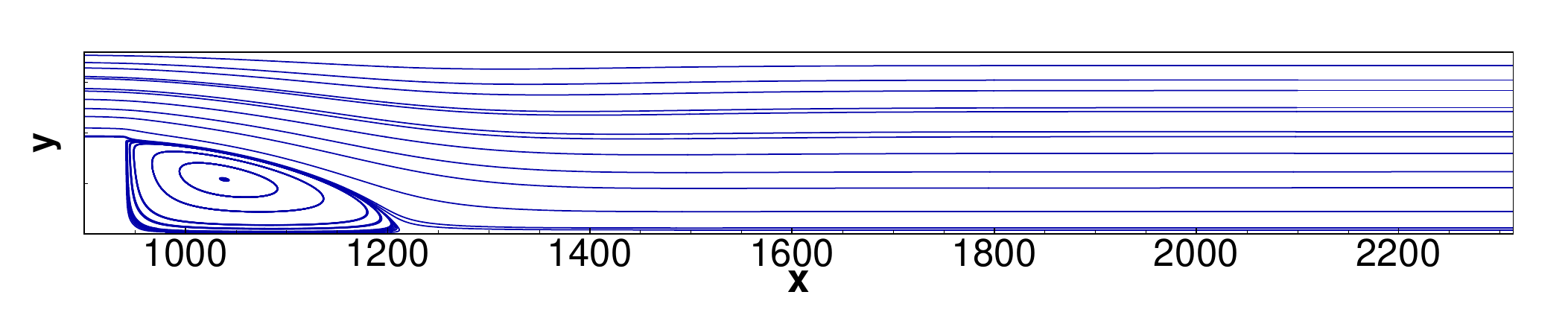}
        \label{fig:img1} }

    \subfloat[Re=300\label{subfig-2:dummy}]{
        \includegraphics [scale=0.5]  {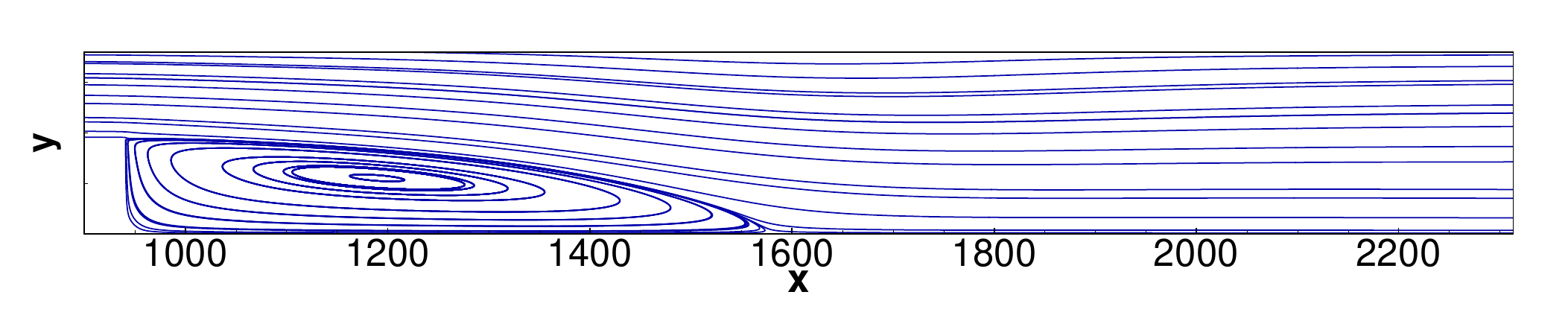}
        \label{fig:img2} }

    \subfloat[Re=500\label{subfig-2:dummy}]{
        \includegraphics [scale=0.5]  {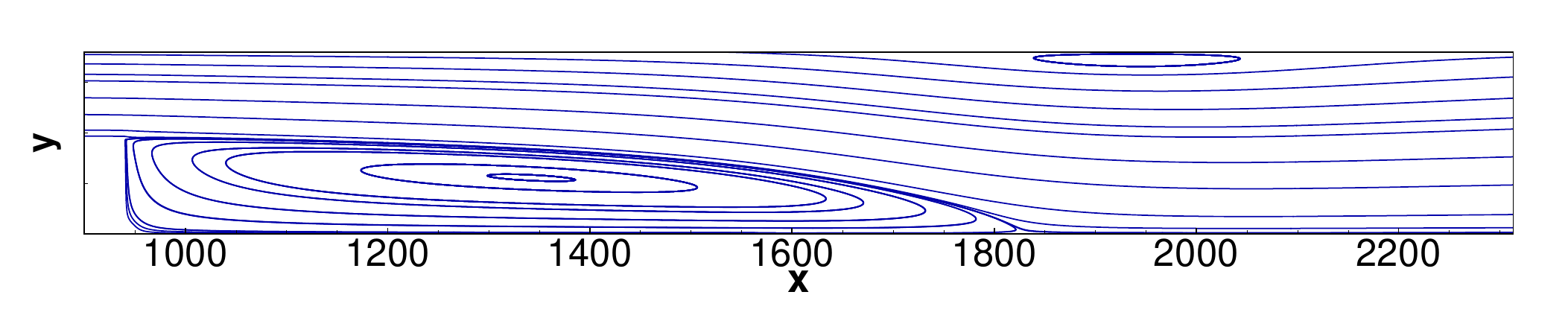}
        \label{fig:img3} }

    \subfloat[Re=800\label{subfig-2:dummy}]{
        \includegraphics [scale=0.5]  {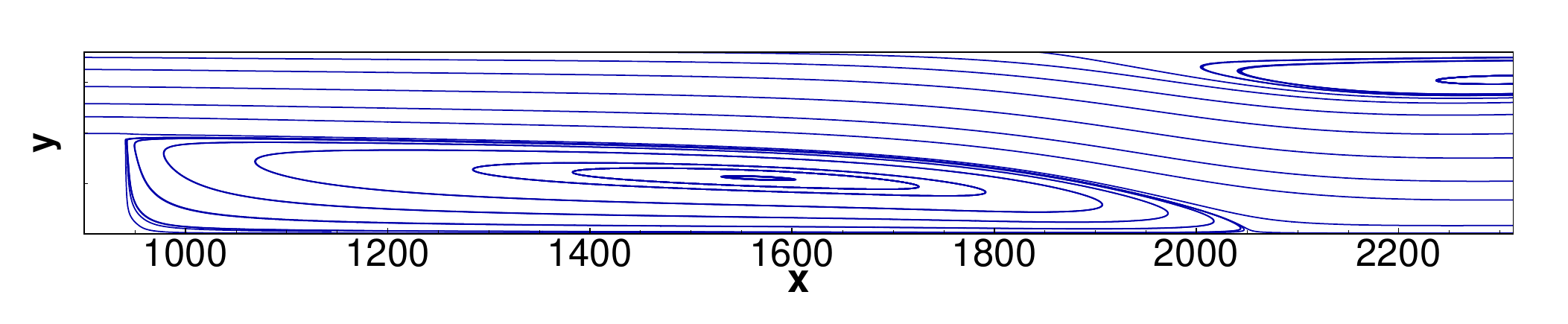}
        \label{fig:img4} }

    \caption{Streamline contours for flow over a backward-facing step at (a) Re = 100, (b) Re = 300, (c) Re = 500, (d) Re = 800 computed using  the GI preconditioned cascaded central moment LBM with $\gamma=0.3$.}
    \label{fig:strstep}
\end{figure}

\begin{figure}[htbp]
\centering
\advance\leftskip-4cm
\advance\rightskip-4cm
    {
        \includegraphics[scale=0.6] {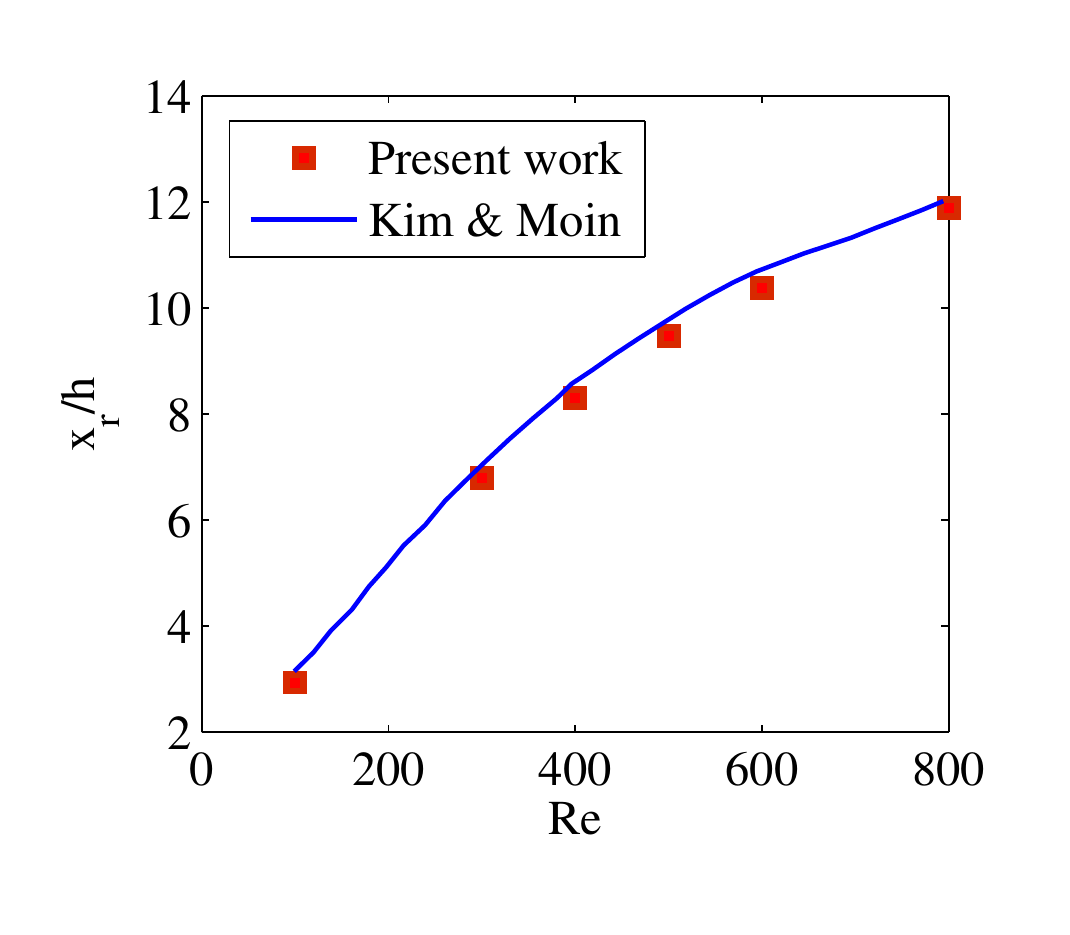}
                               \caption{Comparison of the  reattachment length as a function of the Reynolds number $Re$ computed using the  GI preconditioned cascaded central moment LBM with $\gamma=0.3$ (symbols) against the benchmark results of~\cite{Kim1985}.}
        \label{fig:reastep} }
\end{figure}

As the third flow benchmark flow problem involving complex separation and reattachment effects, we consider a two-dimensional laminar flow over a backward facing step, which is computed using the GI preconditioned central moment LBM. The geometry and boundary conditions for the simulation are shown in Fig.~\ref{fig:schstep}.  For a step of height $h$, the  flow entry is placed at $L_1=10h$ behind the step and the exit is located $L_2=30h$ downstream of the step, and the channel height is defined as $H=2h$. In this simulation, the number of nodes in resolving the step flow is defined by considering $h=94$. At the entrance, a parabolic profile,  and, at the outlet, a convective boundary condition are imposed, and,  finally, the half-way bounce-back scheme is utilized for the no-slip boundary condition at the walls. The computational results are then presented for Reynolds numbers up to 800, where the Reynolds number is defined as $Re=\frac{2hU_{max}}{{3\nu}}$. Here, $U_{max}$ is the maximum speed at the inlet channel.
For the purpose of investigating the flow behavior in the vicinity of the step, the distributions of streamlines are plotted  at four different Reynolds numbers in Fig.~\ref{fig:strstep}. Initially, a primary recirculation zone is created downstream of the step at $Re=100$ (Fig.~\ref{fig:strstep}(a)). However, it can be  seen from Fig.~\ref{fig:strstep}(a) to Fig.~\ref{fig:strstep}(d) that the Reynolds number has a remarkable effect on the structure recirculation regimes  and the length of this zone is seen to increase by increasing the Reynolds number. Furthermore, a second recirculation zone occurs along the top wall at the higher Reynolds number of $Re=500$ which becomes  more visible at $Re=800$. All these observed flow pattern are consistent with prior benchmark results.
In order to more precisely determine the quantitative  effect of the Reynolds number on the reattachment length in the primary recirculation zone, our computed results based on the GI preconditioned cascaded central moment LBM for different Reynolds numbers are computed with the numerical results of~\cite{Kim1985}, which are presented in Fig.~\ref{fig:reastep}. It can be observed that the agreement between the predictions based on our GI preconditioned LB scheme and the benchmark results is excellent. Moreover, it can be clearly seen that  by increasing the Reynolds number, the reattachment length increased, consistent with prior observations.

\subsection{Hartmann Flow}
In this section, in order to validate our preconditioned scheme for a problem involving a body force, the Hartmann flow of an incompressible fluid
bounded by two parallel plates is studied. An external uniform magnetic field $B_z=B_0$
is applied perpendicular to the plates. Since the body force varies spatially arising due to the interaction of the flow velocity and the induced magnetic field, i.e. the Lorentz force, it represents appropriate test problem for the present study. In our preconditioned LB model, the moments of the source terms at different orders are preconditioned differently to correctly recover the macroscopic with variable body forces. The relationship between the external magnetic field $B_0$ and an induced magnetic field $B_x(z)$ across the channel is given by $B_x(z)=\frac{F_bL}{B_0}\left[\frac{sinh\left(\mathrm{Ha}\frac{z}{L}\right)}{sinh(\mathrm{Ha})}-\frac{z}{L}\right]$, where $F_b$  and $L$ are driving force due to imposed pressure gradient and the half channel width, respectively, and \mbox{Ha} is the Hartmann number,which measures the ratio of the Lorentz force to viscous force.

\begin{figure}[htbp]
\centering
\advance\leftskip-5cm
\advance\rightskip-5cm
{
\includegraphics[scale=0.5] {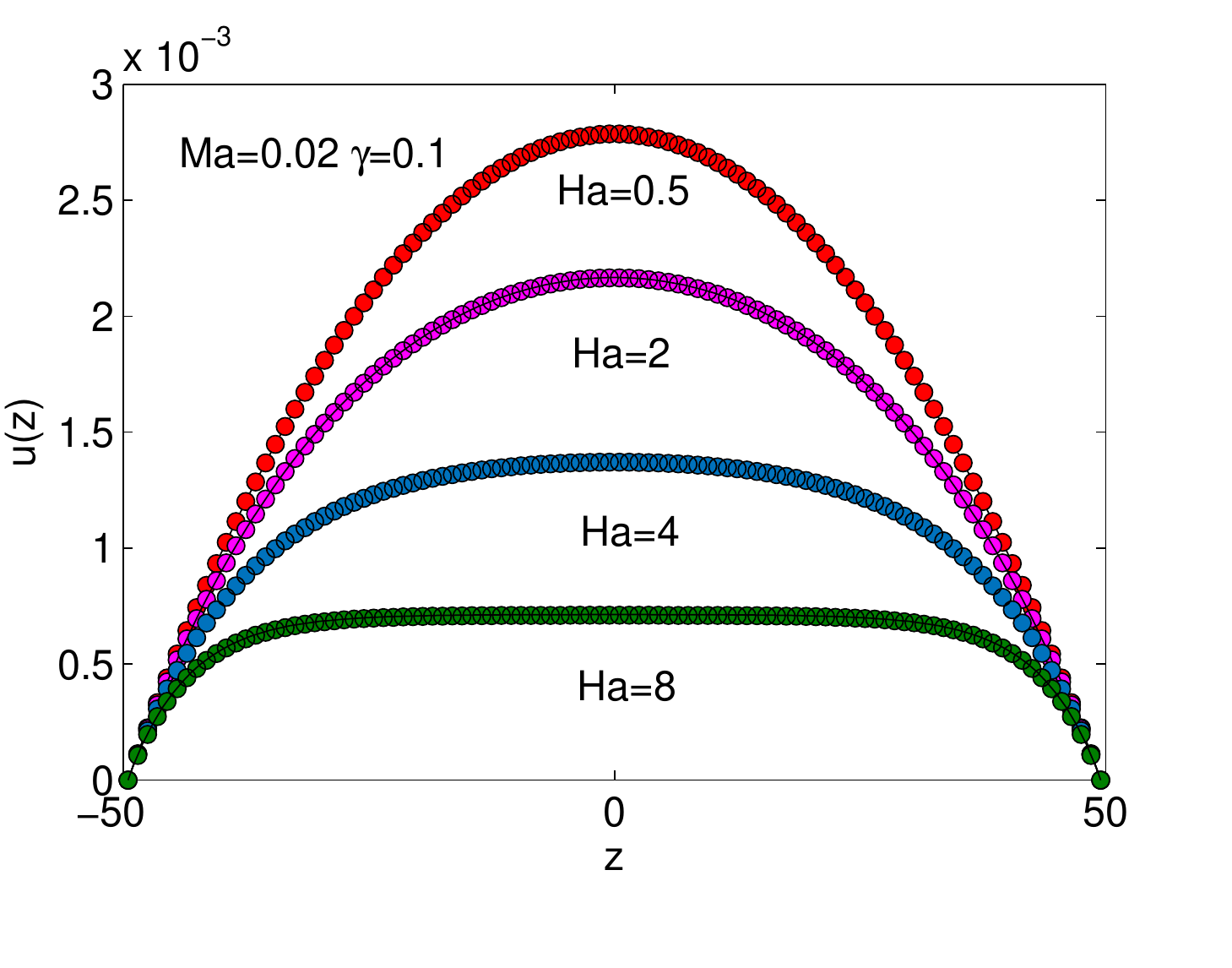}
\label{fig:img1} }
\caption{Comparison of the computed velocity profile using the preconditioned GI cascaded central moment LBM ($\gamma$ = 0.1) with the analytical solution for Hartmann flow for various $\mbox{Ha}$ at $\mbox{Ma} = 0.02$. The lines indicate analytical results, and the symbols are the solutions obtained by the GI preconditioned cascaded LBM.}
\label{fig:velhartmann}
\end{figure}

The Lorentz force component is then defined as $F_{mx}=B_0\frac{dB_x}{dz}$. In consequence, the effective  variable body force component is defined as $F_x=F_b+F_{mx}$. The analytical solution for the Hartmann flow has the following velocity profile  $u_x(z)=\frac{F_bL}{B_0}\sqrt{\frac{\eta}{\nu}}\mbox{coth}(\mathrm{Ha})\left[1-\frac{\mbox{cosh}\left(\mathrm{Ha}\frac{z}{L}\right)}{\mbox{cosh}(\mathrm{Ha})}\right]$, where $\eta$ is the magnetic resistivity given by $\eta={B_0}^2L^2/({\mbox{Ha}^2\nu})$.
Figure~\ref{fig:velhartmann} presents comparisons of the computed velocity profiles using the GI preconditioned cascaded LBM with $\gamma=0.1$ and Mach number $\mbox{Ma}=0.02$ against the exact solution for various values of \mbox{Ha}. It can be observed that the GI preconditioned cascaded central moment LBM is able to reproduce the benchmark solution very well. In particular, as  Ha is increased, the  resulting higher magnitudes of the Lorentz force causes significant flattering of the velocity profiles and this effect of \mbox{Ha} on the velocity profiles is represented by our preconditioned is model with very good accuracy.

\subsection{Four-rolls Mill Flow Problem: Comparison between GI Corrected and Uncorrected Preconditioned Cascaded LBM\label{sec:fourrollsmill}}
As seen in Sec.~\ref{3}, the GI errors for the LBM on the standard, tensor product lattices, such as the D2Q9 lattice, are generally related to the strain rates in the principal directions ($\partial_x u_x$ and $\partial_y u_y$). Hence, in order to compare the GI corrected formulation (Sec.~\ref{sec:5}), which is constructed to eliminate such errors, with the uncorrected formulation (Sec.~\ref{2}), we consider the
four-rolls mill flow problem, which is characterized by local extensional/compression strain rates (i.e. $\partial_x u_x \neq 0, \partial_y u_y \neq 0$),
and for which a well-defined analytical solution is available. It is a modified form of the classical Taylor-Green vortex flow driven by a local
body force, whose components are given by
\begin{equation*}
F_x(x,y)=2\nu u_0 \sin x \sin y, \quad F_y(x,y) = 2\nu u_0 \cos x \cos y
\end{equation*}
in a periodic square domain of side length $2\pi$ ($0 \leq x,y \leq 2\pi$), resulting in a steady vortical motion in the form of an array of
counterrotating vortices. Here, $\nu$ and $u_0$ are the kinematic viscosity and the velocity scale, respectively, and a unit reference density
is considered. The analytical solution of the velocity field, which follows from a simplification of the Navier-Stokes equations impressed by the
above body force, reads
\begin{equation*}
u_x(x,y)= u_0 \sin x \sin y, \quad F_y(x,y) = u_0 \cos x \cos y.
\end{equation*}
Clearly, the local flow field is subjected to local diagonal strain rates, i.e. $\partial_x u_x = - \partial_y u_y = u_0 \cos x \sin y$, and,
as a result, the uncorrected LB scheme induces additional GI errors, which should be annihilated by the corrected LB method; and thus, the difference
in the global flow fields against the analytical solution under a suitable norm in each case can be quantitatively studied and compared.

\begin{figure}[htbp]
\centering
\advance\leftskip-15cm
\advance\rightskip-2cm
    \subfloat{
         \includegraphics[clip, trim=.19cm .11cm 4cm .cm, width=.55\textwidth]{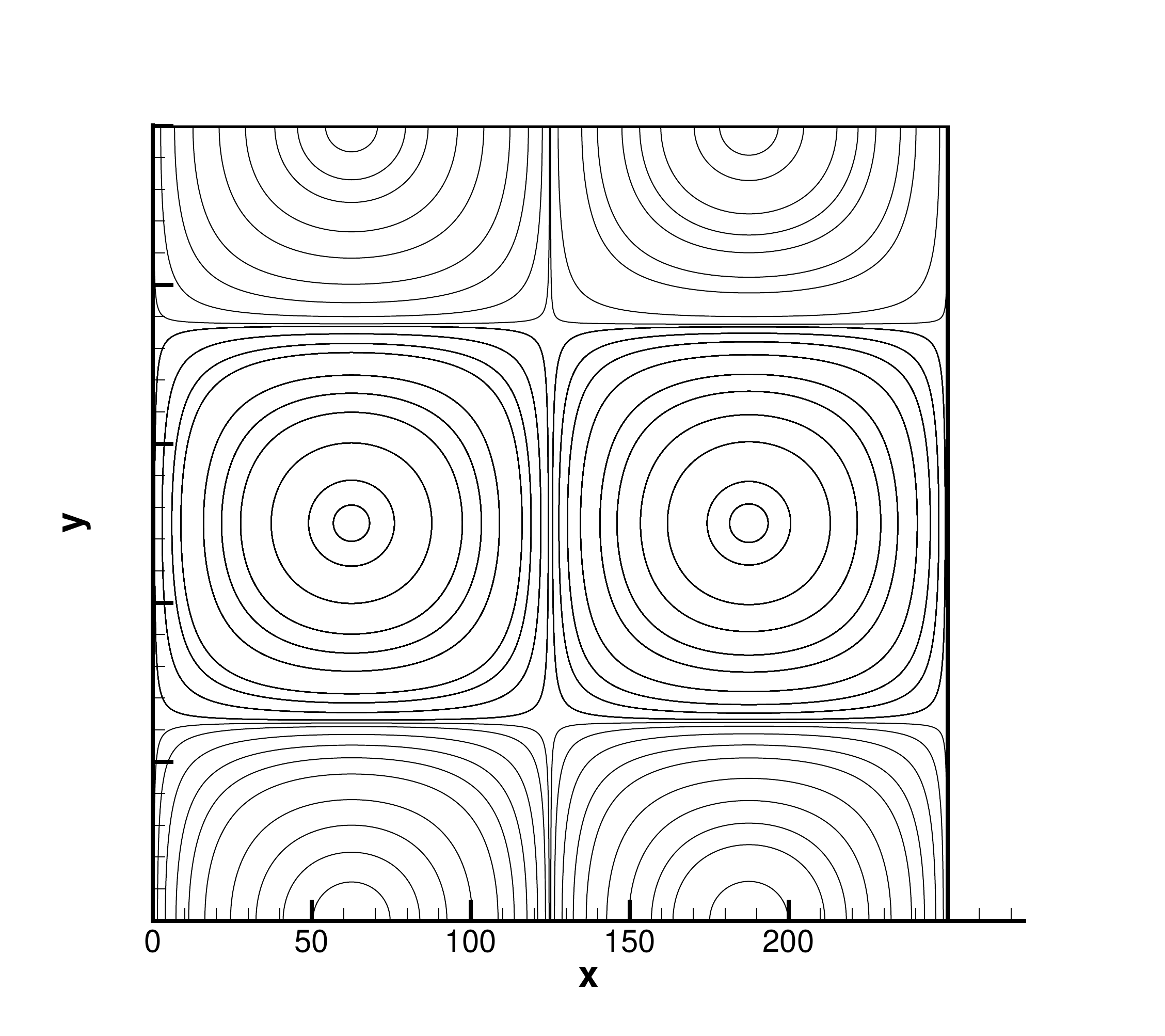}
        } \hspace*{-32em}
    \hfill
       \caption{Steady state streamline patterns for the four-rolls mill flow problem at $u_0=0.045$ and $\mbox{Re}=20$ computed using the GI corrected preconditioned cascaded LB scheme with $251\times 251$ grid nodes and $\gamma = 0.3$.}
    \label{fig:streamlinesfourrollsmill}
\end{figure}

\begin{figure}[htbp]
\centering
\advance\leftskip-1cm
\advance\rightskip-2cm
    \subfloat{
        \includegraphics[scale=0.45] {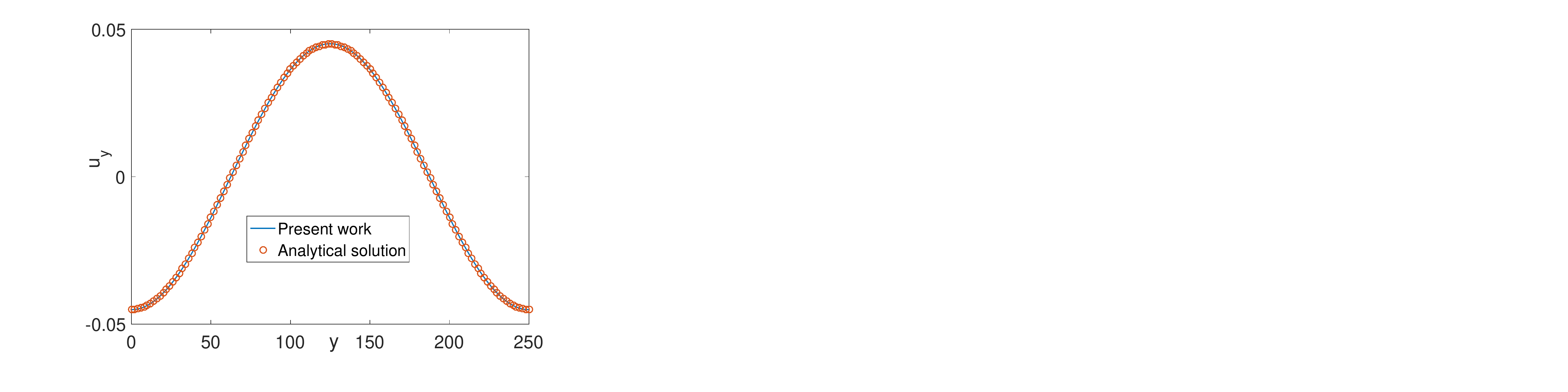}
    } \hspace*{-32em}
    \hfill
       \caption{Comparison of the computed and analytical vertical velocity profiles $u_y(x)$ at $y = \pi$ for the four-rolls mill flow problem at $\mbox{Re}=20$ obtained using the GI corrected preconditioned cascaded LB scheme with $251\times 251$ grid nodes, $u_0=0.45$ and $\gamma = 0.3$.}
    \label{fig:velocityprofilefourrollsmill}
\end{figure}

We performed computations on a square domain resolved by $251\times 251$ grid nodes with a velocity scale $u_0=0.045$ for a Reynolds number $\mbox{Re}=u_0L/\nu$, where $L=2\pi$, of 20. Figure~\ref{fig:streamlinesfourrollsmill} shows the streamline patterns at the steady state computed
using the GI corrected preconditioned LB scheme ($\gamma=0.3$), which manifest as a set of counterrotating vortices.
The computed velocity profile $u_y(x,y=\pi)$ obtained using the GI corrected LB scheme along the horizontal centerline of the domain presented in Fig.~\ref{fig:velocityprofilefourrollsmill} are compared against the analytical solution given above, which show good agreement.

\begin{figure}[htbp]
\centering
\advance\leftskip-10cm
\advance\rightskip-2cm
    \subfloat{
        \includegraphics[scale=0.3] {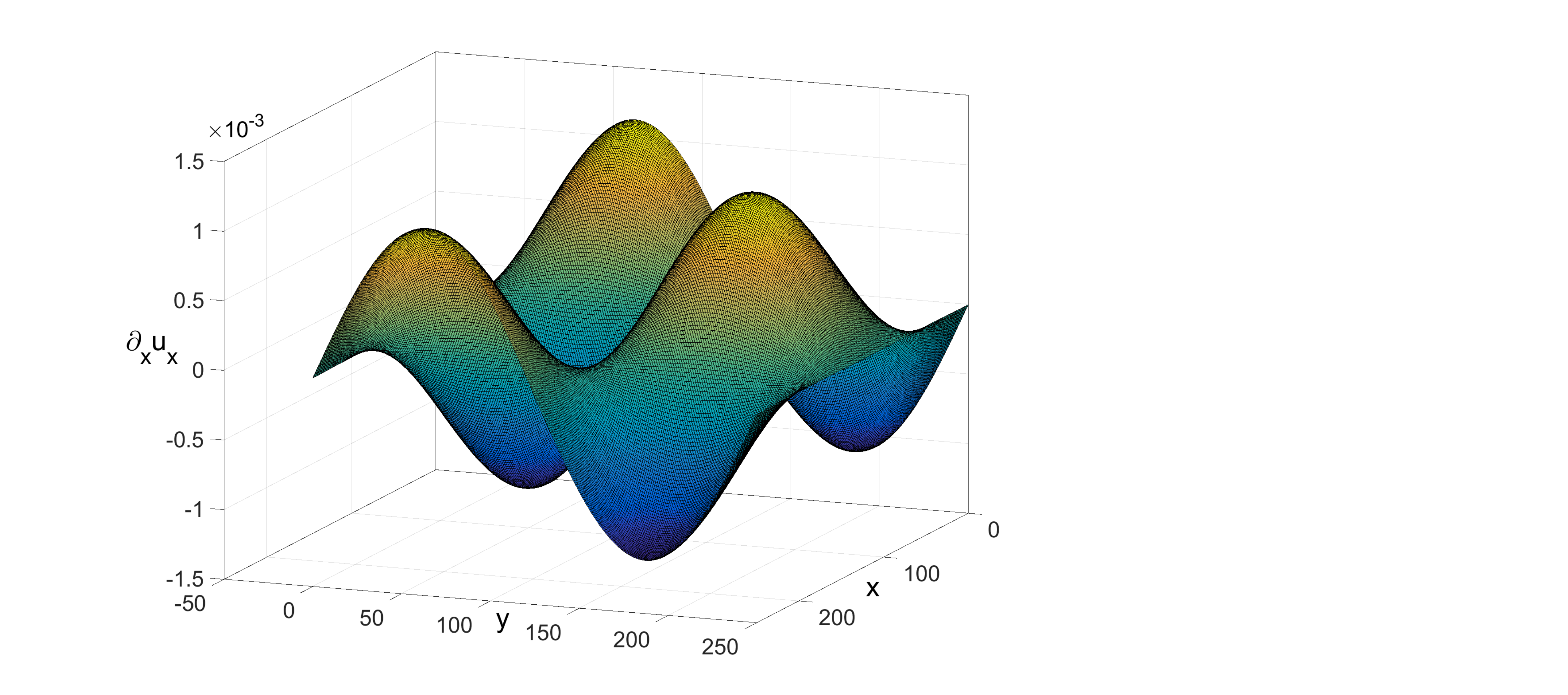}
    } \hspace*{-32em}
    \hfill
       \caption{Distribution of the diagonal strain rate component $\partial_x u_x=-\partial_y u_y$ for the four-rolls mill flow problem with $u_0=0.045$.}
    \label{fig:strainratefourrollsmill}
\end{figure}

Furthermore, Fig.~\ref{fig:strainratefourrollsmill} presents a surface plot of the diagonal strain rate component $\partial_x u_x$, which is seen to
have a significant local variation, due to which quantitative differences in the solutions between the GI corrected and uncorrected preconditioned
LB schemes can be expected, which will now be demonstrated in the following.

In order to make a quantitative comparison between the solutions obtained using the two different LB methods, we first define the global relative errors for the velocity field $||\mbox{GRE}_u^{GI}||_2$ and $||\mbox{GRE}_v^{GI}||_2$ between the components of the solution obtained using the GI corrected preconditioned LB scheme (i.e. ($u_c,v_c$)) and the analytical solution (i.e. ($u_a,v_a$)) under a discrete $\ell_2$ norm; and similarly $||\mbox{GRE}_u||_2$ and $||\mbox{GRE}_v||_2$ between the uncorrected preconditioned LB scheme (i.e. ($u_{uc},v_{uc}$)) and the analytical solution. These
are written as follows:
\begin{equation*}
||\mbox{GRE}_u^{GI}||_2=\sqrt{\frac{\sum (u_c-u_a)^2}{\sum u_a^2}}, \quad ||\mbox{GRE}_v^{GI}||_2=\sqrt{\frac{\sum (v_c-v_a)^2}{\sum v_a^2}},
\end{equation*}
\begin{equation*}
||\mbox{GRE}_u||_2=\sqrt{\frac{\sum (u_{uc}-u_a)^2}{\sum u_a^2}}, \quad ||\mbox{GRE}_v||_2=\sqrt{\frac{\sum (v_{uc}-v_a)^2}{\sum v_a^2}},
\end{equation*}
where the summations in the above are carried out for the whole computational domain. Table~\ref{tab:GREfourrollsmill} presents the above global
relative errors for the velocity field components for both the preconditioned cascaded LB formulations for different values of the preconditioning parameter ($\gamma=0.2,0.3,0.4$ and $0.5$). It can be seen that significant improvements in accuracy is achieved by the GI corrected preconditioned LB
scheme. In particular, the errors relative to the analytical solution are reduced by about a factor of two with the corrected preconditioned LB scheme
for the conditions considered for the computation of this problem. Such improvements are consistent with the fact that the corrected LB
scheme eliminates the additional GI errors arising in this flow subjected to the local variations of the diagonal (compression/extension) strain rates, which are present in the uncorrected LB scheme.
\begin{table}[htbp]
\centering
\caption{Comparison between the global relative errors in the computed solutions for the velocity field using the GI corrected preconditioned cascaded LB scheme and the uncorrected preconditioned cascaded LB scheme for the four-rolls mill flow problem at $\mbox{Re}=20$, $u_0=0.045$ and a grid resolution of $251\times 251$.}
\label{tab:GREfourrollsmill}
\begin{tabular}{|c|c|c||c|c|}\hline
Preconditioning        & GI corrected $u$ error & Uncorrected $u$ error & GI corrected $v$ error & Uncorrected $v$  error\\
parameter $\gamma$ &$||\mbox{GRE}_u^{GI}||_2$ & $||\mbox{GRE}_u||_2$  & $||\mbox{GRE}_v^{GI}||_2$ & $||\mbox{GRE}_v||_2$\\ \hline \hline
0.2      & $3.386\times 10^{-3}$    & $6.662\times 10^{-3}$ & $3.377\times 10^{-3}$     & $6.665 \times 10^{-3}$\\ \hline
0.3      & $1.850\times 10^{-3}$    & $4.104\times 10^{-3}$ & $1.854\times 10^{-3}$     & $4.126\times 10^{-3}$\\ \hline
0.4      & $1.384\times 10^{-3}$    & $2.851\times 10^{-3}$ & $1.389\times 10^{-3}$     & $2.865\times 10^{-3}$ \\ \hline
0.5      & $1.135\times 10^{-3}$    & $2.113\times 10^{-3}$ & $1.140\times 10^{-3}$     & $2.123\times 10^{-3}$\\  \hline
\end{tabular}
\end{table}

Finally, we now obtain an estimate for the additional computational cost associated with including the GI corrections. For the flow condition employed ($u_0=0.045$, $\mbox{Re}=20$, and $251\times 251$ grid nodes), with $\gamma = 0.3$, the uncorrected preconditioned LB scheme for 6000 iterations
incurs a CPU time of $356.1$ secs on a standard Dell workstation, while the GI corrected preconditioned LB scheme takes $390.9$ secs. Thus, the
additional computational overhead of applying the GI corrections is about $9.7\%$. These involved computations of the GI correction terms related to the velocity gradients using non-equilibrium moments and the finite-difference (FD) calculations of the density gradients in our present 2D simulations, with the latter taking $16.1$ secs out of the total overhead of $34.8$ secs. Also, it was found that there were negligible differences in the accuracy variations between using a isotropic FD scheme or a standard central difference FD scheme for the density gradients in the GI correction terms. Thus, especially in extensions to 3D, it may be more efficient to adopt the simpler standard FD schemes for density gradient calculations in the GI corrections terms. In summary, a significant improvement in accuracy was achieved with the use of the GI corrected preconditioned LB scheme when compared to the uncorrected preconditioning formulation with a relatively minor additional computational effort.

\section{\label{sec:7}Summary and Conclusions}
Lattice Boltzmann schemes on standard tensor product lattices can result in cubic-velocity errors in Galilean invariance (GI)
as the third-order diagonal moments are not independently supported and degenerates to the first-order moments. Recent
investigations have presented corrections to the collision operator to yield schemes free of these errors for the
representation of the standard Navier-Stokes (NS) equations. Convergence acceleration of simulations of steady state flows
can be achieved by solving the preconditioned NS equations involving a preconditioning parameter $\gamma$ to tune the
pseudo-sound speed thereby alleviating the numerical stiffness. In our prior work, we devised a modified central moment
based cascaded LBM to represent such preconditioned NS equations, which may be referred to as a specific example of an
extended or generalized NS equations containing a free parameter, here the preconditioning parameter $\gamma$. In this
work, we have presented a new preconditioned central moment based cascaded LB scheme that eliminates such non-GI
cubic-velocity and parameter dependent errors for the simulation of steady state flows. A detailed analysis
based on the Chapman-Enskog expansion reveals the structure of the non-GI truncation errors that appear
in the second-order non-equilibrium moment components, which are related to the viscous stress.
Subsequently, we prescribe an extended second-order moment equilibria that restores GI free of cubic-velocity errors for the
preconditioned LB model on the standard D2Q9 lattice. The following are among the main findings arising from our analysis:
\begin{itemize}
\item In general, the use of central moments in a LB scheme provides a natural setting to partially restore GI for the
third-order off-diagonal moments. In particular, by setting the third-order \emph{central} moment equilibria of the off-diagonal
components to zero (e.g. $\widehat{\kappa}^{eq}_{xxy}=0$), one naturally arrives at the precise forms of the corresponding
\emph{raw} moment equilibria (e.g. $\widehat{\kappa}^{eq'}_{xxy}=c_s^2\rho u_y+\rho u_x^2u_y$) that restores GI of such
components in the representation of the standard NS equations. On the other hand, in the preconditioned LB scheme,
the cubic-velocity terms appearing in the third-order, off-diagonal moment equilibria needs to be scaled by $\gamma^2$
(e.g. $\widehat{\kappa}^{eq'}_{xxy}=c_s^2\rho u_y+\rho u_x^2u_y/\gamma^2$) to fully eliminate the spurious cubic-velocity
cross-derivative terms (e.g. $u_xu_y\partial_yu_x,u_yu_x\partial_xu_y$) appearing in the derivation of the preconditioned
macroscopic equations.

\item In order to effectively eliminate the non-GI, diagonal velocity gradient terms (e.g. $u_x^2\partial_xu_x$), the
second-order, diagonal moment equilibria needs additional corrections in both the velocity and density gradients when
$\gamma \ne 1$, which are prescribed via extended moment equilibria. The velocity gradients can be locally and efficiently
obtained using the non-equilibrium second order moment components; on the other hand, the density gradients can be computed
using a finite-difference approximation.

\item Unlike that for the standard NS equations, the representation of the preconditioned NS equations using a LB scheme
results in additional, non-GI, cross-coupling velocity terms (e.g. $u_y^2\partial_xu_x$), which are also eliminated by
our GI-corrected preconditioned LB scheme.

\item For the second-order, off-diagonal moment equilibria, additional gradient velocity correction terms are needed to
restore GI for these components when $\gamma \ne 1$. However, for incompressible flows ($\bm{\nabla}\cdot\bm{u}=0$), they
vanish regardless of the value of $\gamma$. Such a situation is unique to the representation of the preconditioned NS
equations using LB schemes, as the non-GI corrections are generally restricted only to the diagonal components of the
second-order equilibria for the representation of the standard NS equations.

\item In general, the prefactors in GI defect terms exhibit dramatically different behaviors for the asymptotic limit cases:
For example, $\gamma \rightarrow 1$ (No preconditioning):$\left(\frac{4}{\gamma^2}-\frac{1}{\gamma}\right) \sim 3$ and
$\gamma \rightarrow 0$ (Strong preconditioning):$\left(\frac{4}{\gamma^2}-\frac{1}{\gamma}\right) \sim \frac{4}{\gamma^2}$.

\item When $\gamma = 1$, i.e. when the present LB model is used to simulate flows represented by standard NS equations as
a special case, all our results for the GI defect terms and corrections become identical with those derived by~\cite{Dellar2014}
and~\cite{Geier2015}.

\item Finally, the results of our present analysis can be extended to three-dimensions (e.g. D3Q27 lattice) and other collision
models for the simulation of the preconditioned NS equations.
\end{itemize}

In addition, we have presented numerical validation of our new GI preconditioned LB scheme based on central moments against
several complex flow benchmark problems including the lid-driven cavity flow, flow over a square cylinder, the backward
facing step flow, the Hartmann flow and the four-roll mills flor problem. Comparison against prior numerical solutions show
good agreement for the modified preconditioned scheme. In addition, it is demonstrated that our GI corrected preconditioned cascaded
LB scheme results in significant convergence acceleration of complex flow simulations, and a quantitative improvement in accuracy
when compared to the uncorrected preconditioned LB scheme. Finally, it may be noted that our analysis of non-GI aspects
for the preconditioned LB scheme has implications for LB schemes for other situations such as the porous media flows. For example,
there is a formal analogy between the preconditioned NS equations and the Brinkman-Forchheimer-Darcy equations, where the porosity
serves as a free parameter (e.g.~\cite{hsu1990thermal,nithiarasu1997natural}). LB models constructed for such flows (e.g.~\cite{guo2002lattice})
can be further improved by the approach presented in this work. Investigations involving such flow problems will be reported in
our future studies.

\section*{Acknowledgements}
The authors would like to acknowledge the support of the US National Science Foundation (NSF) under Grant CBET-1705630.


\end{document}